\documentclass{article}

\usepackage{arxiv}

\usepackage[utf8]{inputenc} 
\usepackage[T1]{fontenc}    

\usepackage{url}            
\usepackage{booktabs}       
\usepackage{amsfonts}       
\usepackage{nicefrac}       
\usepackage{microtype}      

\usepackage{mathpazo} 

\usepackage{amsmath}

\usepackage{lipsum}
\usepackage{graphicx}
\usepackage{multirow}
\usepackage{colortbl}
\usepackage{xcolor}
\usepackage{caption}
\usepackage{subcaption}
\usepackage{longtable}
\usepackage{placeins}
\usepackage{setspace}
\usepackage{float}

\usepackage{multirow}
\usepackage[normalem]{ulem}
\useunder{\uline}{\ul}{}
\graphicspath{ {./img/} } 
\usepackage{hyperref}
\hypersetup{
    colorlinks=true,
    linkcolor=blue,
    urlcolor=blue,
    linktoc=all,
    citecolor=cyan
}

\usepackage{natbib}
\setcitestyle{open={(},close={)}}

\title{The Narrow Depth and Breadth of Corporate Responsible AI Research}

\author
{ Nur Ahmed~$^{1}$\thanks{Corresponding author: \href{mailto:nura@uark.edu}{nura@uark.edu}}, Amit Das~$^{2}$, Kirsten Martin~$^{3}$, Kawshik Banerjee~$^{4}$  \vspace{1em} \\
\normalfont{\small $^{1}$Sam M. Walton College of Business, University of Arkansas}\\
\normalfont{\small $^{2}$Lundquist College of Business, University of Oregon}\\
\normalfont{\small $^{3}$Mendoza College of Business, University of Notre Dame}\\
\normalfont{\small $^{4}$School of Computing, Southern Illinois University}\\
}

\makeatletter
\input{size12.clo}
\makeatother

\begin{document}
\maketitle

\begin{abstract}

The transformative potential of AI presents remarkable opportunities, but also significant risks, underscoring the importance of responsible AI development and deployment. Despite a growing emphasis on this area, there is limited understanding of industry's engagement in responsible AI research, i.e., the systematic examination of AI's ethical, social, and legal dimensions. To address this gap, we analyzed over 6 million peer-reviewed articles and 32 million patent citations using multiple methods across five distinct datasets to quantify industry's engagement. Our analysis reveals notable heterogeneity between industry's substantial presence in conventional AI research and its comparatively modest engagement in responsible AI. Leading AI firms exhibit significantly lower output in responsible AI research compared to their conventional AI research and the contributions of leading academic institutions. Our linguistic analysis reveals a more concentrated scope of responsible AI research within industry, with fewer distinct key topics addressed.  Our large-scale patent citation analysis uncovers limited linkage between responsible AI research and the commercialization of AI technologies, suggesting that industry patents infrequently draw upon insights from the responsible AI literature. These patterns raise important questions about the integration of responsible AI considerations into commercialization practices, with potential implications for the alignment of AI development with broader societal objectives. Our results highlight the need for industry to publicly engage in responsible AI research to absorb academic knowledge, cultivate public trust, and proactively address the societal dimensions of AI development.

\end{abstract}

\section{Introduction}
\vspace{-0.8em}

As AI continues to mediate our lives and relationships \citep*{1, wagner2021measuring}, it brings a host of benefits \citep*{ai-force-for-good}, such as automation of routine tasks \citep*{machine-trans-intl-trade}, improved medical services \citep*{classification-cancer-deep-learning}, and increased productivity \citep*{noy2023experimental}. At the same time, studies have documented ethical challenges associated with AI deployment, including algorithmic recommendations of harmful content \citep*{5}, discriminatory predictions \citep*{7,3, predicting-recidivism}, and the spread of disinformation \citep*{4,2,6}. Consequently, public discourse around AI has increasingly focused on questions of accountability and oversight \citep*{8,haenlein2022guest}. Regulators, including the US Congress, the White House, and the European Parliament, have called for greater transparency and research engagement from AI firms (organizations that own one or more AI patents) on these societal dimensions \citep*{9,10}.

In this article, we use the term ``responsible AI'' to denote the examination of the ethical and societal implications of AI. This includes improving its design, development and deployment, and assessing the appropriateness of AI use in specific contexts \citep*{barocas2020not}. Research into responsible AI contributes to aligning AI models with societal values \citep*{van2020embedding}, respecting fundamental human rights \citep*{aizenberg2020humanrights}, and understanding how to maximize societal benefits while addressing potential harms \citep*{hoffmann2023adding, brad-smith-res-ai-microsoft}, as well as establishing the trustworthiness of AI technologies \citep*{sanction-authority-public-trust}. Firms creating these technologies are positioned to critically evaluate their AI systems, as the outcomes of AI depend substantially on how it is developed and deployed \citep*{acemoglue-harms-of-ai}.

AI firms are in a unique position in shaping the trajectory of AI development, possessing the resources, talent, and capability to drive innovation in this field \citep*{18, frank_wang_cebrian_iyad}. This position affords opportunities for these organizations to contribute to the socially beneficial development and deployment of AI \citep*{14,15,16}. Engagement in responsible AI research may enable firms to develop absorptive capacity \citep*{cohen1990absorptive, 26,28}, foster transparency and accountability \citep*{brundage_toner_fong_et_al_2020, mittelstadt_2019}, and navigate the trade-offs inherent in aligning AI systems with societal values \citep*{korinek_avital_2022}. Public engagement and deliberation are essential in navigating the responsible development of AI, as the substantial conceptual ambiguities surrounding this field require input from diverse stakeholders \citep*{manish_2023}. Given that AI firms bear accountability for the societal consequences of their products, engagement in responsible AI research offers a pathway toward understanding the limitations and risks associated with their technologies \citep*{holstein2019improving}. Overall, prioritizing public engagement in responsible AI development not only benefits society as a whole but also provides tangible financial benefits to AI firms themselves by building trust among regulators and customers.

While research into responsible AI by firms could enable them to develop and deploy the technology in a socially beneficial manner \citep*{cooperation-res-ai}, the extent and nature of industry involvement in responsible AI research remain underexplored. The literature presents divergent perspectives on this topic. One strand of scholarship examines the relationship between industry participation and the independence of responsible AI research \citep*{29,38,39}. For example, Baker and Hanna \citeyearpar{40} highlight the role of funding for independent, grassroots-led research. Related work analyzes how industry's substantial presence in AI research may shape the trajectory of the field \citep*{39, democratization}. Other scholars have examined industry collaborations with academics on responsible AI research in relation to regulatory dynamics \citep*{bigtech-manipulate-media}, exploring the various motivations that may underlie such engagement \citep*{ai-good-society-sandra-wachter}.

At the same time, emerging findings present a more nuanced view of industry participation in responsible AI research. Recent evidence indicates that certain firms have made notable progress in this area \citep*{ai-index-2022, ai-index-2023, ahmed&jia}, advancing beyond initial efforts \citep*{companies-committed-res-ai}. These findings suggest that industry engagement in responsible AI research exhibits considerable heterogeneity, with some organizations engaged in socially beneficial AI development. This heterogeneity motivates systematic empirical investigation into the nature and extent of industry's engagement in responsible AI research. 
\vspace{0.8em}

In this study, we ask: 

\begin{itemize}    
    \item To what extent does industry engage in responsible AI research?
    \item How do the priorities of responsible AI research differ between industry and academia
    \item To what extent does industry integrate responsible AI research into its commercial inventions?
\end{itemize}

To answer these questions, we conducted a systematic analysis of more than 6 million peer-reviewed papers (2010-22) sourced from Scopus and 32 million USPTO patent citations (1947-2022) \citep*{33}. Our comprehensive dataset encompasses a wide range of conference and journal articles, adopting an expansive definition of responsible AI research. We employed multiple methods, including supervised machine learning and keyword search, to classify papers as responsible AI research. Leveraging five distinct datasets (see Table \ref{table:3_distinct_data}), we evaluated the extent of industry’s engagement in this important subfield. Our analyses reveal considerable heterogeneity in industry's engagement with responsible AI research, with most firms exhibiting limited depth and breadth of participation relative to their investments in conventional AI development.

\vspace{0.8em}

Our \textbf{key contributions} include: 
\begin{itemize}

    \item First, our comprehensive analysis of industry-authored AI papers documents patterns of industry engagement in responsible AI research. We employ multiple complementary methods, including a supervised machine learning classifier, a keyword-based approach, and large language model classification (Gemini), to identify responsible AI papers. Among AI research firms (those with one or more AI publications), we observe that 34.5\% engage in responsible AI research. These findings provide an empirical baseline for understanding the current landscape of industry participation in this emerging subfield.

    \item Second, comparing industry to academic institutions by analyzing over 5.9 million papers, our study documents differences in the production of responsible AI research. Leading AI firms exhibit lower output in responsible AI research relative to their conventional AI research and relative to the contributions of leading academic institutions. At the same time, we observe that some leading firms have published responsible AI research at levels comparable to elite academic institutions, suggesting heterogeneity across firms.

    \item Third, analyzing industry presence at AI conferences reveals differential patterns: while industry has substantial presence at conventional AI conferences, its participation at responsible AI conferences is comparatively limited.

    \item Fourth, by employing various natural language processing (NLP) methods on paper abstracts, we find that AI firms' responsible AI research tends to focus on different topics compared to academic researchers. Areas such as moral considerations, environmental concerns, and ethical consequences appear less frequently in industry papers.

    \item Finally, to the best of our knowledge, this work provides the first large-scale empirical documentation of the relationship between responsible AI research and industry's commercial inventions. Our analysis of over 32 million patent citations reveals that industry patents infrequently cite responsible AI research, suggesting limited integration of this research stream into commercial development.
\end{itemize}

Our findings indicate that AI's rapid commercialization is proceeding with modest attention to its potential consequences, despite mounting pressure from regulators \citep*{White_House_2023}, civil society \citep*{activist_facct_24}, and employees \citep*{ahmed&jia, emp_activism_2020} for greater attention to responsible development. Given industry's leading role in AI development, understanding the extent of corporate engagement in responsible AI research has important implications for both scholarship and policy. Our analysis indicates that AI firms have comparatively limited involvement in responsible AI research and development, as measured by publications and patent citations. These patterns raise questions about whether current market incentives are sufficient to align corporate AI development with broader societal considerations.

In the following sections, we discuss: related literature on industry engagement in responsible AI research (section \ref{section:literature}), why industry should engage in responsible AI research (section \ref{section:reason-for-res-ai}), what extent industry engages in responsible AI research (section \ref{section:narrow-depth}), research priorities in responsible AI between industry and academia (section \ref{section:topic-modeling}), to what extent industry incorporates responsible AI research into their commercial inventions (section \ref{section:citation-analysis}), and finally, we discuss the potential policy implications of our findings (section \ref{section:discussion}).

\setcounter{footnote}{0}

\begin{longtable}{| p{1.3cm}|p{2.7cm}|p{2.3cm}|p{1.85cm}|p{1.65cm}| p{3.9cm}| }
\caption{Datasets used to analyze industry engagement in responsible AI}
\label{table:3_distinct_data} \\
\hline
Dataset &  Source &  Number of \newline Observations & Period & Reference Figure & Sampling Strategy \\ \hline
\multirow{6}{*}{1} & Industry & 679,919 & 2010-22 & \ref{fig:fig1}, \ref{fig:fig3}, \ref{fig:patent-analysis}, \ref{fig:keyword-appendix} & All peer-reviewed \newline papers published by firms holding at least one AI patent \\ \cline{2-6} 
 & Academia & 5,265,419 & 2010-22 & \ref{fig:fig1}, \ref{fig:fig3}, \ref{fig:patent-analysis}, \ref{fig:keyword-appendix} & All peer-reviewed \newline papers published by leading 100 US \newline universities \\ \hline
\multirow{8}{*}{2} & Leading \newline conventional AI conferences & 106,012 & 2010-22 & \ref{fig:fig2}  & Leading conferences in conventional AI \newline (see Table \ref{tab:conf-list} in the \newline appendix) \\ \cline{2-6} 
 & Leading \newline responsible AI conferences & 851 & 2018-22 & \ref{fig:fig2}, \ref{fig:patent-analysis} & Leading conferences in responsible AI (see Table \ref{tab:conf-list} in the \newline appendix) \\ \hline
\multirow{2}{*}{3} & Responsible AI journal and conference papers\footnote{This comprehensive sample encompasses papers from a diverse array of sources, including various journals and conferences in addition to top-tier outlets.} & 36,022 & 2010-22 & \ref{fig:fig2}, \ref{fig:patent-analysis} &  Papers identified using expert-suggested keywords (see Table \ref{table:oldA6} in the appendix) \\ \hline
\multirow{2}{*}{4} & Patents-to-paper citations\footnote{This citation data comes from over 2 million USPTO patents between 1947 and 2022 citing over 5 million papers between 1800 and 2022.} & 32,698,465 & 1947-2022 & \ref{fig:patent-analysis} & Reliance on Science \newline data \citep*{33} \\ \hline
\multirow{1}{*}{5} & AI patents & 141,770 & 1985-2018 & \ref{fig:patent-analysis} & AI patents data \newline \citep*{34} \\ \hline

\end{longtable}

\section{Related Literature on Industry's Engagement in Responsible AI Research}
\label{section:literature}
\vspace{-0.8em}

Recent research has documented industry's growing presence at the frontiers of AI research \citep*{frank_wang_cebrian_iyad, koch2021reduced, 18}. Studies indicate that industry not only maintains an increased presence but also shapes the trajectory of AI research through developing the majority of large AI models \citep*{benkler2019don, jurowetzki2021privatization, 18}. Scholars have examined the implications of industry's influence over the direction of AI research \citep*{39, democratization}, including patterns of hiring academics \citep*{18, 37, 43} and funding AI research at universities \citep*{29, 38, 39}. This body of work has raised questions about the evolving division of labor between academia and industry in AI research \citep*{democratization, jurowetzki2021privatization}.

However, the literature presents mixed findings regarding industry's engagement in responsible AI research. One stream of research suggests that such engagement has increased notably \citep*{ai-index-2023, ai-index-2024}. The Stanford AI Index 2022 reports a 71\% year-over-year increase in industry co-authored publications at leading responsible AI conferences \citep*{ai-index-2022}. Research has also documented cases where AI firms have substantively engaged with responsible AI practices \citep*{companies-committed-res-ai}. Building on these observations, some scholars have examined whether AI firms might shape the future research agenda in responsible AI \citep*{38,39,40,41}. Related work has explored the relationship between industry-academic collaboration in responsible AI research and regulatory processes \citep*{bigtech-manipulate-media, ai-good-society-sandra-wachter}. Patent citation analysis suggests that firms' responsible AI publications may serve regulatory or reputational functions rather than directly informing product development \citep*{ahmed2026}. Some scholars have also examined how varying levels of industry involvement may interact with the independence of responsible AI research initiatives \citep*{29,38,39,40}.

A separate stream of literature examines whether industry may have limited incentives to engage substantively in responsible AI development \citep*{cooperation-res-ai}. Some scholars have characterized certain forms of industry engagement as primarily symbolic rather than substantive, a phenomenon sometimes termed "ethics washing" \citep*{companies-committed-res-ai, practitioner-perspective, falco_gupta_hart_et_al_2021}. This perspective suggests that observed industry engagement in responsible AI may vary in depth and substance, presenting a different interpretation than the literature documenting increased engagement.

Scholars have also examined the topical breadth of industry's responsible AI research. Prior studies, while offering valuable insights, have typically focused on specific venues or time periods. Recent textual analyses indicate that certain areas of responsible AI have gained traction over time \citep*{4yearsfacct}, and that the thematic priorities of mainstream AI research may differ from topics emphasized in academic responsible AI scholarship \citep*{19}. One limitation of prior studies is their focus on the most prestigious responsible AI conferences, which may not fully capture the breadth of research activity in the field. Our study addresses this gap by employing a more comprehensive sampling strategy across multiple data sources.

Industry engagement in responsible AI research exhibits notable heterogeneity, with certain firms producing scholarship at levels comparable to leading academic institutions. For example, Anthropic, founded in 2021 with safety as a central organizational mission, has published work on "Constitutional AI," a method for training AI systems using explicit principles rather than solely human feedback \citep*{bai2022constitutional}. Anthropic’s other works have contributed to interpretability research examining neural network representations through dictionary learning methods \citep*{Templeton2024}. Similarly, Google DeepMind maintains one of the larger industry research portfolios in this area, with published work on frontier safety frameworks for managing risks from advanced AI systems, holistic safety evaluation methodologies, and interpretability tools such as Gemma Scope \citep*{lieberumGemmaScopeOpen2024}. These examples indicate that substantive responsible AI research from industry does occur. However, less is known about the prevalence and characteristics of such engagement across the broader population of AI firms, a gap our empirical analysis seeks to address.

\section{Motivations for Industry to Engage in Responsible AI Research}
\label{section:reason-for-res-ai}
\vspace{-0.8em}

AI firms, as primary drivers of technological development and deployment, hold a distinctive position in responsible AI research given their expertise, resources, and potential societal impact. Several theoretical perspectives suggest reasons why these organizations may benefit from publicly engaging in responsible AI research. We outline these perspectives below, recognizing that the actual motivations and engagement patterns of firms remain empirical questions.

\textit{Unique position and capability.} AI firms occupy a unique position to shape the trajectory of AI development due to their central role in developing and deploying this technology \citep*{34, 18}. The decisions these firms make during the design and deployment phases have far-reaching implications for society \citep*{14,15,16}. Furthermore, their privileged vantage point allows them to identify potential shortcomings and risks associated with their AI systems that may not be apparent to external researchers and take necessary measures to address those \citep*{role-of-workers-ai-ethics}. This unique position imposes a heightened responsibility on AI firms to ensure that the technology they develop is socially beneficial and morally sound \citep*{brad-smith-res-ai-microsoft}.
 
\textit{Access to required resources.} The development of cutting-edge AI requires vast amounts of data, computational power, and talent, resources that are largely concentrated within AI firms \citep*{18, 46, democratization}. Industry actors develop the majority of state-of-the-art AI models \citep*{18, ai-index-2024}, which are often difficult for external researchers to reproduce and audit \citep*{bommasani2023foundation}. Moreover, AI firms attract a substantial share of top AI talent, with nearly 70\% of AI PhDs opting for industry roles over academic positions \citep*{18}. Research suggests that certain limitations of AI models, such as toxicity in language, may only become evident at larger scales \citep*{21, 46}. This concentration of resources suggests that industry engagement in responsible AI research may offer access to settings and scale that academic researchers face challenges replicating independently.

\textit{Cultivation of absorptive capacity.} Engaging in responsible AI research is crucial for firms to develop their ``absorptive capacity''--the ability to identify, assimilate, and apply external knowledge \citep*{cohen1990absorptive, 26,28}. This process is complex and costly, requiring firms to establish specific routines and organizational procedures \citep*{26}. Recent research suggests that active participation in conferences is essential for firms to effectively learn from other researchers \citep*{28}. By actively participating in responsible AI research and presenting their findings at leading conferences, AI firms can enhance their capacity to absorb and integrate the knowledge generated by the broader research community \citep*{epstein2018closing, microsoft_Transparency_Report_24}.

\textit{Transparency and accountability.} Prior literature suggests that transparency and accountability in AI development may contribute to the trustworthiness of AI systems \citep*{brundage_toner_fong_et_al_2020, martin2019ethical,mittelstadt_2019}. Martin \citeyearpar{martin2019ethical} argues that algorithms are value-laden rather than neutral, creating moral consequences and influencing stakeholder rights, which grounds firm accountability regardless of algorithmic complexity. Engagement in public deliberations and consultation with external stakeholders represents one mechanism through which firms may navigate critical decisions about design choices during AI development. Several dimensions of this relationship have been explored in the literature.

First, participating in peer-review processes facilitates interaction with various stakeholders, particularly those in academia and the non-profit sectors. Research indicates that mitigating bias and reducing unfairness in AI models often require making choices that conflict with each other \citep*{manish_2023}. Furthermore, the complexity of designing technology for diverse user needs, with varying incentives and often unclear user preferences, may benefit from collaborative decision-making \citep*{sucheta2023value, kleinberg2023challenge}.

Second, decisions regarding what to commercialize, how to commercialize, and for whom have been identified as candidates for public deliberation \citep*{barocas2020not}. Scholars have also emphasized \textit{what not to develop or commercialize}, as some AI technologies may have unintended negative consequences or raise significant ethical concerns \citep*{bengio_science_24}. This perspective reflects concerns that the repercussions of commercialization may disproportionately impact marginalized groups \citep*{raji2019actionable, bruckner2018promise}. Transparency regarding the values encoded and reflected in AI models \citep*{19} may enable stakeholders to identify potential biases or limitations in AI systems. Furthermore, the substantial conceptual ambiguities and debates surrounding responsible AI \citep*{ethics-of-ethics-ai} highlight the potential value of public deliberation with diverse stakeholders, including those from the Global South \citep*{global-south, subjective-global-opinion}. Recent scholarship emphasizes the role of diverse stakeholders in determining the values that AI systems align with and the risks and harms to be addressed \citep*{luke_2024_facct}. A sociotechnical approach emphasizes that no single group, especially technologists alone, is positioned to unilaterally make these critical decisions \citep*{korinek_avital_2022}. Some scholars argue that public debates on these questions contribute to AI safety \citep*{ai_safety_terms_2023}. Involving a range of perspectives and expertise may support the development of more robust solutions to complex challenges \citep*{gina-neff-cambridge}. Overall, research engagement in responsible AI may facilitate communication within the AI research community around transparency and accountability.

Third, publishing in peer-reviewed forums facilitates greater scrutiny of the decision-making processes of AI firms. It also enables outsiders to observe how firms balance their business models and financial incentives with the direction of technology development \citep*{zuboff2015big}. Such scrutiny may support transparency and accountability in AI development \citep*{brundage_toner_fong_et_al_2020, mittelstadt_2019}.

\textit{Risk reduction.} AI is a tool that lacks autonomy in decision-making. Drawing an analogy to automobile manufacturers, who are held liable for accidents due to vehicle defects and therefore undertake thorough pre-market testing, some scholars have argued that AI companies may similarly bear responsibility for the consequences of their products \citep*{martin2019designing}. From this perspective, engagement in responsible AI research may help firms better understand the limitations of their technology and products \citep*{holstein2019improving}.

\textit{Potential benefits.} Literature suggests that investment in responsible AI research may yield benefits for firms. By engaging in responsible AI research, firms may improve the quality of AI products and services by incorporating ethical considerations and developing AI systems that are fair, unbiased, and trustworthy. Moreover, by developing internal expertise in this area, firms may be better positioned to navigate the complex challenges that arise as the technology advances \citep*{renieris2022responsible}. Collaborating with the broader responsible AI community and sharing knowledge may help industry leaders stay attuned to evolving societal concerns and adapt accordingly. Research in this area suggests that firms with demonstrated commitment to responsible AI may benefit from increased trust among regulators and customers \citep*{renieris2022responsible}. Additionally, firms that demonstrate commitment to responsible AI principles may be better positioned to attract and retain top AI talent, who increasingly prioritize ethical considerations when selecting employers \citep*{ahmed&jia}.

\section{The Narrow Depth of Industry's Responsible AI Research}
\label{section:narrow-depth}
\vspace{-0.8em}

\subsection{Comparing Responsible AI Research Output Across Industry and Academia}
\label{section:scibert}
In this section, we provide a comprehensive analysis of responsible AI research output from industry and academia. We begin by examining the extent to which industry publishes responsible AI research and compare this output to their involvement in conventional AI research. This comparison allows us to characterize the composition of responsible AI research within industry's research portfolio. Subsequently, we compare industry's responsible AI research output with academia's research output in this domain. We consider both publication counts and publication quality to document the current distribution of responsible AI research across these two sectors.

\subsubsection{Data Description} To identify AI firms and assess industry engagement in responsible AI research, we focused on companies holding at least one AI patent registered with the United States Patent and Trademark Office (USPTO), yielding a list of 1,771 AI firms. Patents signal a firm's ability to actively research and commercialize technology in a field \citep*{30}. Our sample included firms with patents under the Cooperative Patent Classification (CPC) class ``computer systems based on specific computational models,'' excluding quantum computing, per consultations with USPTO examiners. Afterward, we collected each AI firm's peer-reviewed publications from Scopus, including publications from both conferences and journals. Our sampling strategy ensures that we have the \textit{full} research portfolio of AI firms. This process resulted in 679,919 papers from 2010 to 2022, which included AI papers, non-AI papers, and responsible AI papers (\hyperref[table:3_distinct_data]{dataset 1}). 

To compare industry research engagement with academia, we first collected data from Scopus for the leading 100 universities in computer science research. We selected these universities based on their research productivity between 2010 and 2022, as reported by CSRankings.org\footnote{CSRankings (\href{https://csrankings.org/}{csrankings.org}) is a well-cited resource in the literature \citep*{18,42,43} that ranks universities based on their research output in computer science.}. The complete list of universities can be found in Table \ref{table:oldA2} in the appendix. For each of the selected universities, we downloaded their publication data from Scopus, covering the years 2010 to 2022. This process resulted in a comprehensive dataset (\hyperref[table:3_distinct_data]{dataset 1}) consisting of 5,265,419 peer-reviewed publications. This dataset serves as the basis for our analysis of academic research engagement.

\subsubsection{Classification Method} 
\label{section:scibert-subsection}

The classification of responsible AI papers presents a significant challenge due to the lack of a universally accepted definition of responsible AI \citep*{44}. Consequently, researchers may employ varying definitions and classification approaches, leading to inconsistencies in the identification and categorization of responsible AI research. For our analysis, we conducted a supervised machine learning method using an ensemble classification \citep*{dietterich2000ensemble} that combined a transformer-based pre-trained SciBERT model \citep*{31} and a boosting-based XGBoost classifier \citep*{47}. Such ensembled classifiers are generally used to gain better results than those from the constituent classifiers alone \citep*{ensemble-better-performance, decision_fusion}. The intuition here is that our SciBERT model, combined with a separate classifier and being trained on research articles, would be able to learn the patterns from the data and identify responsible AI and conventional AI papers (see Fig. \ref{fig:flowchartSML} in the appendix for our overall workflow).

Papers from the three leading responsible AI conferences were initially labeled as responsible AI papers for our ensemble model's training data. The goal was to maintain a broad operationalization of responsible AI and reduce human intervention, thus mitigating selection bias and making the classification process more reproducible. Additionally, to classify conventional AI and non-AI papers that are unrelated to responsible AI, we added data from reputed journals and leading conventional AI conferences that were not about responsible AI to our training data. We manually validated the titles and abstracts of these articles and labeled them as either conventional AI or non-AI papers. Our training data contained 7,355 papers comprising 930 responsible AI papers and 6,425 papers from leading AI conferences and journals (see Tables \ref{table:ethics_TrainingData} and \ref{table:AITrainingData} in the appendix for the details of the model's training data).

Transformer-based models outperform other neural network models on text processing tasks \citep*{transformer_original}. Among them, BERT \citep*{bert_original} and other BERT-based models, like RoBERTa \citep*{roberta_original}, have demonstrated state-of-the-art performance on different NLP tasks. For our analysis, we evaluated multiple pre-trained transformer-based models on a hold-out test sample. Our findings show the SciBERT model, originally trained on over a million research articles, outperformed other models, yielding better accuracy (see Table \ref{tbl:bert-experiment}).\footnote{As part of the ensemble model, we fine-tuned the SciBERT model by setting 50 epochs with 10-fold cross-validation. Additionally, we enabled early stopping with a delta of $1e^{-4}$, a patience value of 3 to avoid overfitting, and set the training batch size to 50.}

\begin{table}[htb]
\centering
\caption{Transformer-based supervised ML models' accuracy test results}

\begin{tabular}{ l c c c }
\hline
\textbf{Model} & \textbf{Accuracy} & \textbf{AUC} & \textbf{F$_1$ Score} \\ \hline
SciBERT & \textbf{0.993} & \textbf{0.919} & \textbf{0.882} \\ 
RoBERTa & 0.992 & 0.915 & 0.844 \\ 
XLNet & 0.992 & 0.880 & 0.827 \\ 
BERT & 0.993 & 0.902 & 0.856 \\ 
ELECTRA & 0.993 & 0.917 & 0.862 \\ 
ALBERT & 0.991 & 0.869 & 0.807 \\ 
\hline
\end{tabular}
\label{tbl:bert-experiment}
\end{table}

To complement the SciBERT model, we used the XGBoost classifier as it has demonstrated better performance on diverse data types when ensembled with a neural network-based model \citep*{xgboost-dnn, xgboost-cnn}.

Fine-tuning the ensemble model also included manual validation of the training dataset and feature ablation on it. Ablation studies are helpful in identifying and quantifying the actual impact of the constituent model components \citep*{autoablation}. Following prior literature, we performed ablations on features \citep*{randomized-feature-ablation, feature-hierarchy-ablation, beyond-pagerank-abation} and training data subsets \citep*{unreal-engine-data-ablation, autoablation}. We tested multiple models by using varied subsets of features (e.g., abstract, title and abstract, abstract and publication outlet title). This careful ablation led us to a robust model which included paper title, paper abstract, and publication outlet title as the most effective features. 

In our ensemble model, we utilized paper abstracts as a feature for the SciBERT model and paper titles and publication outlet titles (e.g., conference or journal names) as features for our XGBoost classifier. Both of these classifiers independently resulted in prediction values indicating whether the paper is a responsible AI paper, a conventional AI paper, or neither. Afterward, we used a weighted summation of the predicted values to obtain the final prediction. Our rationale is that abstracts convey a more comprehensive representation of a paper's content and theme than titles or outlet names. Hence we assigned a weight of 70\% to the prediction from the SciBERT model, which used abstracts as features, and 30\% to the prediction from the XGBoost classifier. This weighted summation, if greater than 50\%, resulted in a final predicted ``positive'' class (responsible AI paper or conventional AI paper, measured separately); otherwise, the paper was classified as ``negative'' (not a responsible AI paper).

After the ensemble model's classification, we manually validated our findings on 500 randomly selected classified papers. This analysis showed that our classification approach was reliable, achieving 92.4\% accuracy. To ensure further robustness of our approach, we conducted data ablation afterward with multiple different training and testing datasets (see Section \ref{section:scibert-robustness}).

\subsubsection{Citation-Weighted Publication Count} 
\label{section:citation-weighted}

Here we used a citation-weighted count to account for the quality of the publications. For each of the $n$ papers, we first measured its age as the difference between the current year $y_{cur}$ and the paper's publication year $y_{pub}$. Then we measured its age-weighted citation $w$ by dividing its total number of citations $c$ by its age. 
\begin{equation}
\label{eq:weight_formula}
    w = \frac{c}{(y_{cur} - y_{pub}) + 1}
\end{equation}

Afterward, to assess the overall impact of a firm or university's research publications in the current year, we aggregated the age-weighted citations for all of the $n$ papers published by that firm or university in that year. The sum of these age-weighted citations, denoted as $I$, represents the citation impact of that firm or university's research outcomes for the current year.
\begin{equation}
\label{eq:weighted_citations}
    I = \sum_{i=1}^{n} w_i
\end{equation}

\subsubsection{Findings}
\label{section:scibert-findings}

\begin{figure}
    \centering
    \textbf{Industry’s limited engagement in responsible AI research (a comparison with academia)}\par\medskip
    \includegraphics[width=\textwidth]{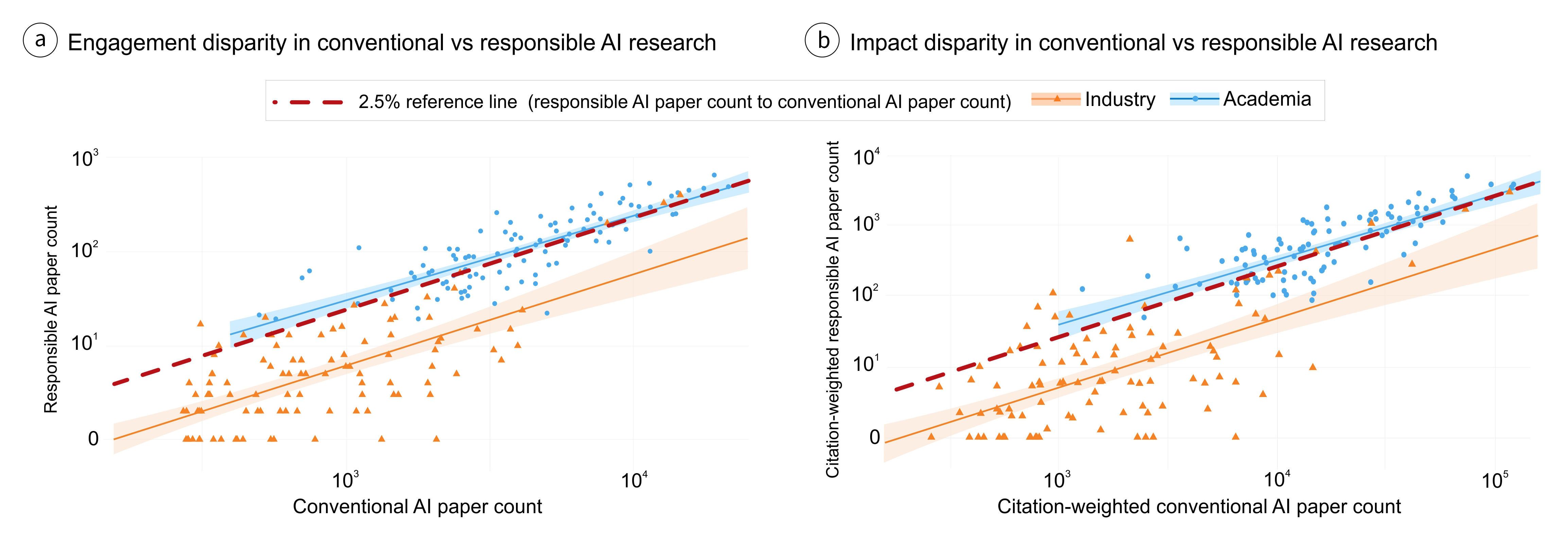}
    \caption{{\fontsize{10}{10}\selectfont Here, \textbf{Fig. \ref{fig:fig1}a} showcases the distribution of the leading 100 AI firms’ (n = 506,017 papers, 2010-22) and universities’ (n = 5,265,419 papers, 2010-22) participation in conventional AI research compared to their engagement in responsible AI research. \textbf{Fig. \ref{fig:fig1}b} presents a citation-weighted paper count for the same data. A trend line reflects the participation trend within each group with 95\% confidence intervals. In both Fig. \ref{fig:fig1}a and \ref{fig:fig1}b, the dashed lines indicate a reference line where the proportion of responsible AI papers to conventional AI papers is 2.5\%.}}
    \label{fig:fig1} 
\end{figure}

We begin by examining patterns of responsible AI research engagement among AI research firms (those with at least one AI publication). Among 519 such firms, we observe considerable heterogeneity in publication activity: 34.5\% engaged in responsible AI research during the study period (2010–2022), while 65.5\% had no responsible AI publications. Among firms that did engage, output varied substantially, with a subset demonstrating sustained commitment through five or more publications. This variation may reflect differences in firm capabilities, strategic priorities, or organizational factors that shape research investment decisions. These patterns provide a baseline for examining how industry engagement in responsible AI compares to conventional AI research output.

We then compared industry’s proportion of responsible AI research to its conventional AI research with the corresponding proportion in academia. We calculated the total number of AI papers and responsible AI papers for each university and firm. Then, for each university, we used AI paper counts as the x-axis and responsible AI papers count as the y-axis, and we fitted a regression line with a 95\% confidence interval. Finally, we introduced a 2.5\% reference line indicating the projected ratio of responsible AI papers to total AI papers.

Here, we compared the 100 leading academic research institutes with the 100 leading AI firms (for a comparison with our full sample of AI firms, see Fig. \ref{fig:ex1} in the appendix). We selected the leading AI research institutes and firms based on their number of conventional AI papers. This selection ensured a fair comparison in terms of research capability and reputation. Notably, the leading 100 AI firms have significantly greater resources than even the most elite universities \citep*{18}. Fig. \ref{fig:fig1}a indicates that industry's participation in conventional AI research is comparable to that of leading academic institutions. However, on average, leading AI firms exhibit lower rates of responsible AI research output than comparable leading academic institutions. This difference is reflected in the slope of industry's regression line, which is flatter than that of academia. Notably, we observe considerable heterogeneity across firms: several organizations, including IBM, Google, and Microsoft, demonstrate responsible AI publication rates comparable to elite research universities, and the number of firms with active responsible AI research programs has increased since 2018. These patterns suggest variation in how different firms allocate research effort between conventional and responsible AI topics.

Finally, to account for the quality of publications, we used a citation-weighted publication count (see Section \ref{section:citation-weighted} for details). Fig \ref{fig:fig1}b suggests that after considering publication quality, the difference between industry and academia remains similar in magnitude. This pattern is observable in the slope of industry's regression line compared with academia's regression line. Notably, all papers in our analysis are peer-reviewed, ensuring comparable quality standards across sectors. The performance of some AI firms \textit{improved} when quality was taken into account, suggesting that these firms produce high-impact responsible AI research, albeit in smaller quantities relative to their conventional AI output. Indeed, several industry-authored responsible AI papers rank among the most highly cited in the field, and academic researchers increasingly cite industry responsible AI work (see Fig. \ref{fig:citation}), indicating that the research quality of active industry contributors is recognized within the broader scholarly community.

Consistent with the findings of de Laat \citeyearpar{companies-committed-res-ai}, we observed that a subset of AI firms publish responsible AI research. However, their responsible AI output remains lower in proportion to their conventional AI research. When considering the quality of their contributions, we note that a smaller number of firms surpassed the 2.5\% threshold. It is plausible that firms selectively disclose their high-quality papers, which may partially explain the observed improvement in performance when quality is accounted for. Overall, the difference between industry and academia in terms of responsible AI research output is persistent across our measures. This pattern suggests that variation exists in how market incentives shape research portfolio composition across organizational types \citep{ahmed2026}.

\subsubsection{Robustness of Classification Method}
\label{section:scibert-robustness}

We tested in three different ways to validate whether the patterns observed above are robust to changes in the selection of dataset or methodology. First, we performed data ablation by varying the training and testing data. Second, we classified the papers with a different method.

To conduct data ablation, we tested three different models by (a) training and testing exclusively on journal articles, (b) training and testing exclusively on conference papers, and finally (c) training on prior ACM FAccT conference papers and testing exclusively on conference papers (see Fig. \ref{fig:validation-scibert} in the appendix). For this, we prepared a subset of the data from the same sample as our prior analysis. This data ablation produced similar results to our findings from the main model (as described in Section \ref{section:scibert-findings}), which increases confidence that our classification method is robust to changes in the underlying data.

We also disaggregated the publications into conferences and journals (see Fig. \ref{fig:ex3}a and Fig. \ref{fig:ex3}b in the appendix, respectively). This analysis suggests that industry has a lower proportion of responsible AI research in both journal and conference publications; however, the difference is larger in journal publications.

\subsubsection{Validating Results Using Expert-Suggested Keywords}

To validate the findings from our ensemble model, we classified the papers using an alternate approach: by searching for expert-recommended keywords in the paper titles and abstracts. Here, we used a predefined list (i.e., dictionary) consisting of specific terms that indicated the category of interest, i.e., whether an observation matches the themes of responsible AI or not. However, a keyword-based search approach can be challenging in the absence of a relevant and representative dictionary. To address this concern, we utilized an extended list of keywords for responsible AI as suggested by experts in this field (see Table \ref{table:oldA6} in the appendix). Similarly, when classifying conventional AI papers, we utilized keywords from the extant literature \citep*{conventional-ai-patent-keyword, ai_key_based_Baruffaldi} and added additional words (see section \ref{list:AIkeys} in the appendix).

It is important to note that keywords alone are not enough to categorize responsible AI papers. For instance, the word ``fairness'' in the search query would include papers such as ``Achieving proportional fairness in WiFi networks via bandit convex optimization.''\footnote{Famitafreshi, G., \& Cano, C. (2022). Achieving proportional fairness in WiFi networks via bandit convex optimization. \textit{Annals of Telecommunications}, 77(5-6), 281-295.} This study is clearly not a responsible AI paper. Thus, our manual validation process, combined with machine learning-based analyses, indicates that data collection solely dependent on keywords tends to \textit{overestimate} the extent of responsible AI research.  

We used the same dataset employed in our prior analysis of industry output using the machine learning model: 679,919 papers between 2010 and 2022 co-authored by researchers affiliated with 1,771 leading AI firms, and 5,265,419 papers (2010-22) co-authored by researchers affiliated with 100 leading universities. Our findings from this keyword-based search on paper titles and abstracts, as shown in Fig. \ref{fig:keyword-appendix} in the appendix, align with our prior findings using the machine learning model and indicate that AI firms exhibit lower rates of responsible AI research output relative to academia. This methodological triangulation \citep*{triangulation}, employing varied approaches, yields consistent results regarding the difference in responsible AI research between industry and academia.

\subsubsection{Validating Results Using LLM-based Classification}

To assess the robustness of our findings, we employed an alternative classification approach using large language models (LLMs) with in-context learning (ICL). This method leverages an LLM’s capacity to perform classification tasks based on annotated examples without fine-tuning model parameters \citep*{xu2024context, tay2022transformer}. We evaluated multiple proprietary LLMs, including Gemini 1.5 Flash, GPT-4o, OpenAI o1, and Claude 3.5 Sonnet (versions as of October 2024). For each model, we prepared sets of 47 and 44 manually annotated samples to classify papers as conventional AI, responsible AI, or non-AI. We assessed model performance through data ablation across multiple hold-out subsets and validated results both manually and against our ensemble model. On the basis of this evaluation, we selected Gemini 1.5 Flash for subsequent validation.
Finally, using LLM as a classifier produced results consistent with our prior analyses. As shown in Fig. \ref{fig:LLM-appendix} in the appendix, the patterns of industry and academic engagement in responsible AI research observed using LLM classification correspond with those obtained from both our supervised machine learning model and keyword-based approach. This convergence across three distinct classification methods confirms the reliability of our findings.

\subsubsection{Quality of Industry's Responsible AI research}

Prior literature has raised questions about the nature and depth of industry involvement in responsible AI research \citep*{gerrit_de_vynck_oremus_2023, owning-ethics}. To examine this dimension, we analyzed the 50 most-cited papers from leading responsible AI conferences. We found that half of these highly cited papers (2018–2022, leading three responsible AI conferences) have authors with industry affiliations, and six are exclusively industry-authored. This substantial representation among the field's most influential work suggests that industry researchers have contributed meaningfully to shaping the responsible AI research agenda. Google and Microsoft are the most represented firms, with ten and nine papers, respectively, reflecting sustained institutional investment in this area. Several other firms, including IBM, Meta, and DeepMind, also appear among these highly cited contributions. These findings indicate that while industry's responsible AI output is lower in volume than academia's, the work produced by active industry researchers has achieved considerable scholarly impact and recognition within the field.

\subsection{Assessing Industry Engagement in Responsible and Conventional AI Research}
\label{section:conference-participation}
\vspace{-0.8em}

To assess the variation in industry participation between responsible AI and conventional AI research, we conducted three distinct analyses using multiple datasets. First, we compared industry engagement in leading responsible AI conferences with its presence at the leading conventional AI conferences. This comparison was performed using two approaches: counting the unique number of firms and examining their co-authorship behavior. Second, we analyzed the growth of industry engagement in responsible AI research over time using a more comprehensive sample that extends beyond top-tier articles.

\subsubsection{Data and Method} The underlying data for our analysis of industry's engagement in responsible AI and conventional AI research come from three different datasets, consisting of (a) papers from three leading responsible AI conferences, (b) papers from ten leading conventional AI conferences, and (c) a unique sample of journal and conference papers containing expert-suggested responsible AI keywords in their abstracts or titles.

First, we selected three leading responsible AI conferences--ACM FAccT, AAAI/ACM AIES, and ACM EAAMO--based on our consultation with scholars researching responsible AI across multiple disciplines (computer science, information science, business, and philosophy). We found 851 papers in total from 2018 to 2022, of which 748 were downloaded from Scopus and 103 were downloaded from their respective websites (\hyperref[table:3_distinct_data]{dataset 2}). Then we manually classified all of these papers’ affiliations. We counted a paper as an industry paper if at least one of its co-authors had any industry affiliation, following prior literature \citep*{frank_wang_cebrian_iyad}. We then manually reviewed these papers' affiliations to compile a list of unique firms represented in the authorship. Henceforth, this process of measuring industry engagement was applied separately for conferences and journals, allowing us to quantify and compare the level of industry engagement across different categories. 

Second, we selected the ten leading AI conferences (see Table \ref{tab:conf-list} in the appendix) from CSRankings.org, and collected 106,012 papers from Scopus for these conferences from 2010 to 2022 (\hyperref[table:3_distinct_data]{dataset 2}).

Third, we collected a unique dataset of 36,022 papers from between 2010 and 2022 on responsible AI (\hyperref[table:3_distinct_data]{dataset 3}) using an extensive list of over 170 keywords (see Table \ref{table:oldA6} in the appendix) from Scopus. To select those keywords, we consulted eight leading experts in responsible AI research. These experts have co-authored papers at the three leading responsible AI conferences or written highly cited responsible AI papers. Our search was confined to English-language, peer-reviewed publications available on Scopus. We included papers in our dataset if their titles or abstracts contained any of the keywords from our list. 

\textbf{Keywords-based distinct sample.} This distinct sample is a key strength of our study, setting it apart from previous works \citep*{ai-index-2024, 19} that focused solely on leading conferences. We adopted a more comprehensive definition of responsible AI, incorporating papers from a wider range of outlets beyond the top-tier conferences. This allows us to examine industry participation in less prominent outlets and provide a more comprehensive assessment.

To classify the papers' affiliations, we first employed string-matching algorithms and an extensive set of regular expressions to systematically categorize them. Our strategy involved leveraging an extensive list of firm names containing 1,771 firm entities and variations of their names with a list of keywords that includes, but is not limited to, ``ltd,'' ``llc,'' ``inc,'' ``limited,'' ``consult,'' ``industries,'' ``llp,'' ``gmbh,'' ``corp,'' ``incorporated,'' ``incorporation,'' ``corporation,'' and ``company.'' Then, we manually reviewed all the unclassified affiliations to classify them and minimize misclassification.

\subsubsection{Findings}

\begin{figure}
    \centering
    \textbf{Industry’s limited engagement in responsible AI research}\par\medskip
    \includegraphics[width=0.90\textwidth]{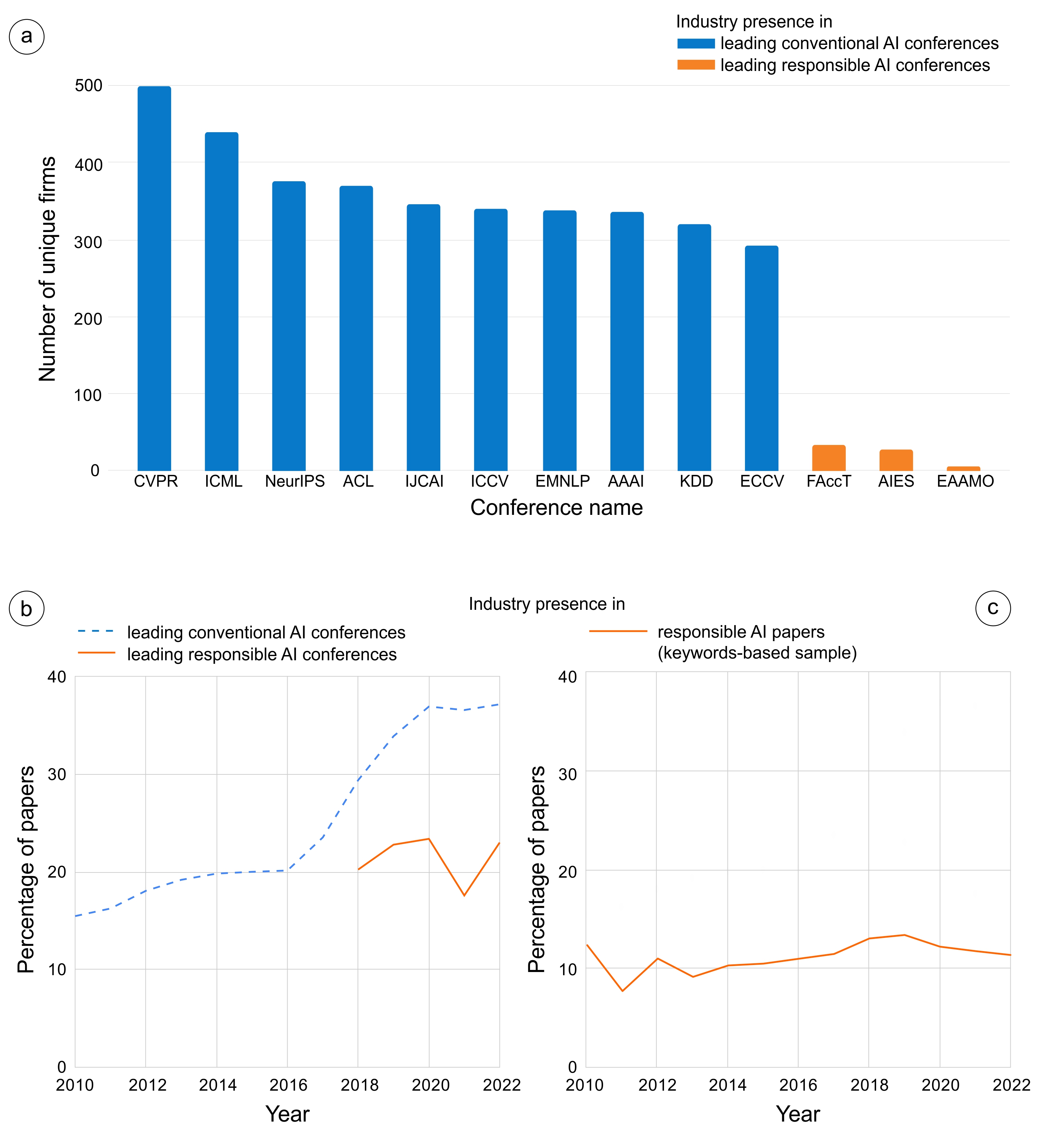}
    \caption{{\fontsize{10}{10}\selectfont Here, \textbf{Fig. \ref{fig:fig2}a} shows the distribution of industry participation by measuring firms’ presence (based on co-authors’ industry affiliation) in ten leading conventional AI conferences (n = 63,526 papers) and three leading responsible AI conferences (n = 851 papers) between 2018 and 2022. \textbf{Fig. \ref{fig:fig2}b} shows the percentage of papers with at least one co-author having industry affiliation in conventional AI (n = 106,012; 2010-22), and responsible AI conferences (n = 851; 2018-22) over the years. This result is robust to changes in methods of co-authorship count. \textbf{Fig. \ref{fig:fig2}c} represents the percentage of industry-papers (n = 36,022; 2010-22) classified as responsible AI papers (a distinct sample based on expert-suggested keywords) from both journals and conferences. }}
    \label{fig:fig2}
\end{figure}

First, we examined the unique number of firms that have presented research at leading conventional AI conferences and responsible AI conferences. To measure this, we took the subset of papers where at least one co-author was affiliated with industry and manually curated the list of unique firms from co-authors' affiliations. Fig. \ref{fig:fig2}a presents the distribution of firm participation across these venues.
For conventional AI conferences (see Table \ref{tab:conf-list} in the appendix for the list of conferences), more than 250 firms participated in each of all 10 venues. In contrast, participation at responsible AI conferences (FAccT, AIES, and EAAMO) was more concentrated, with 33, 28, and six firms participating, respectively (see Fig. \ref{fig:fig2}a). This variation may reflect differences in conference size, disciplinary focus, or the relative maturity of these venues within the broader AI research ecosystem.

Subsequently, we examined industry co-authorship in the three major responsible AI conferences. Fig. \ref{fig:fig2}b plots the proportion of industry co-authored papers in leading responsible AI conferences compared to the proportion of industry papers in conventional AI conferences (the dashed line). We calculated this annual proportion by dividing the total number of industry papers in each year for each conference by the total annual publications of each conference. The trends shown in Fig. \ref{fig:fig2}b suggests that industry participation rates differ between responsible AI conferences and conventional AI conferences, with participation growing in the latter. From 2018 to 2022, industry's participation in responsible AI conferences remained relatively stable (the percentage varies roughly between 20\% and 23\%).\footnote{Our supplementary analysis of ACM FAccT 2023 (not included in this study) reveals a similar trend in industry participation.} By comparison, during the same period, industry increased its participation from 29\% to 37\% at the leading conventional AI conferences. Our results are consistent with recent research \citep*{18} indicating industry has increased its presence in conventional AI research.

Finally, using a separate sample based on expert-suggested keywords (\hyperref[table:3_distinct_data]{dataset 3}), we found that industry proportion has largely stayed the same in recent years (see Fig. \ref{fig:fig2}c), which is consistent with our earlier findings. This suggests that, among all the peer-reviewed responsible AI publications over the years, annually, roughly 15\% or less of them had a co-author affiliated with industry. This proportion has remained stable over the observation period. These patterns are consistent across different samples and classification methods.

Overall, our results document different trajectories in industry participation between conventional AI and responsible AI research. Different interpretations of the criteria defining an industry paper produce similar outcomes.

\subsubsection{Alternative Measurements}
\label{section:keywords-sampling-details}

\textbf{Author-weighted affiliation (AWA) count:} To further validate our findings, we used an alternate method of attributing industry co-authorship. To more precisely measure industry engagement from author affiliations, we first allocated an equal contribution weight $ \frac{1}{n} $ to each of the $n$ co-authors of a paper, assuming their equal contribution to it. We then calculated each co-author's `industry-weight,' by taking a fraction of their affiliations that are from industry.

To illustrate, if one of the co-authors has $a$ affiliations in total, and among them $\hat{a}$ are from industry, then we defined industry-weight to be $\frac{\hat{a}}{a}$. Thus, we calculated any co-author's final weight in a paper by multiplying the equal contribution weight by industry-weight. Finally, the total author-weighted affiliation count of industry for a paper, $AWA$, was defined by summing the final weights of all its $n$ co-authors:

\begin{equation}
    \label{eq:awa}
    AWA = \frac{1}{n} \times \sum_{i=1}^{n} \left( \frac{\hat{a}_i}{a_i} \right)
\end{equation}

Where, $AWA$ is the author-weighted affiliation count of industry for a paper, $n$ is the total number of co-authors in the paper, $a_i$ represents the total number of affiliations, and $\hat{a_i}$ represents the number of industry affiliations of the $i$-th author in the paper.

This approach provides a more accurate picture of industry engagement in responsible AI by addressing the authors' multiple affiliations, instead of counting only industry affiliations. Even with this counting approach, we found that industry participation has doubled in conventional AI research but has largely stayed the same in responsible AI (see Fig. \ref{fig:author_weighted} for the author-weighted count of leading conferences \& \ref{fig:author_weighted_Keywords} for the responsible AI papers in the appendix).

\textbf{Responsible AI research at leading conventional AI conferences:} Recent research indicates the growing presence of responsible AI research at conventional AI conferences \citep*{ai-index-2024}. This trend raises the question of whether industry is more focused on engaging in responsible AI research at these conferences. To investigate that possibility, we analyzed industry's contribution to responsible AI research presented at leading conventional AI conferences. Our findings reveal that industry's share of responsible AI research at these conferences has largely stayed the same in recent years (see Fig. \ref{fig:industry-rai-paper-cai-conf} in the appendix). 

\subsection{Summary of Findings}

In this section, by analyzing nearly 6 million papers from industry and academia, we documented patterns of industry engagement in responsible AI research. Comparing leading AI firms with leading academic institutions, we observe that industry's responsible AI research output is lower in both quantity and impact relative to academia, though considerable heterogeneity exists across firms, with several organizations demonstrating publication rates comparable to elite research universities. By examining industry presence in leading conferences on conventional and responsible AI, we observe that industry participation rates are lower in responsible AI conferences than in conventional AI conferences. While industry's participation in responsible AI conferences has remained relatively stable, their presence in conventional AI conferences has increased over the years. Additionally, analysis of highly cited papers indicates that industry researchers have contributed substantially to influential work in the field. Overall, our analysis shows that responsible AI research constitutes a smaller share of industry's research portfolio relative to academia's, though the quality and impact of active industry contributors is recognized within the scholarly community.

\section{The Narrow Breadth of Industry's Responsible AI Research}
\label{section:topic-modeling}
\vspace{-0.5em}

To gain a comprehensive understanding of the breadth of industry's responsible AI research, and for meaningful comparisons with academia, we employed a multi-pronged approach that leverages various NLP techniques. These methods consisted of clustering algorithms to identify thematic patterns, different topic modeling analyses to uncover latent themes and their prevalence, and a series of word frequency analyses to highlight key concepts. 

\subsection{Data Description}

For our linguistic analysis, we used 10,799 paper abstracts, as abstracts provide a more comprehensive representation of a paper's content, published between 2010 and 2022. This dataset included papers classified as responsible AI papers by our supervised machine learning model (see section \ref{section:scibert-subsection}). Among the papers, 10,408 were authored by researchers affiliated solely with academia (from the leading 100 universities) and 391 were authored by researchers affiliated exclusively with AI firms.\footnote{To ensure the robustness of our findings, we also conducted an analysis using a randomly selected subset containing 400 responsible AI papers with academic affiliations and compared them to industry papers. The results of this additional analysis were consistent with our main findings.} We focused on papers with sole authorship from either academia or industry to better capture the differences in language usage between these two groups without the potential confounding influence of industry-academia collaboration. By omitting papers with mixed affiliations, we aimed to minimize bias in the analysis that could arise from differing research priorities, potential conflicts of interest, or the influence of joint funding on the framing and language of the research.

\subsection{K-means Clustering}

To analyze the research themes of industry and academia separately in responsible AI research, we used k-means clustering \citep*{k-means} on our data. K-means clustering is used to organize large volumes of unstructured text data into meaningful clusters, providing a foundation for an intuitive understanding.

We performed a sensitivity analysis by varying the number of clusters (k=6, 8, 10), but for this discussion, we will focus on the insights from the ten-cluster analysis, which provides a more granular understanding of the research themes within responsible AI. After preprocessing and vectorizing all the abstracts together (regardless of affiliation) with term frequency-inverse document frequency (TF-IDF), each paper was assigned to one of the ten clusters based on semantic similarity by the k-means algorithm. These ten clusters correspond to ten different themes of responsible AI.

To assess the extent to which industry and academia contributed to these different themes, we calculated the percentage of papers from each group that belonged to each cluster. For academia, we determined the percentage of academia-only papers belonging to each cluster by dividing the number of academia-only papers in a specific cluster by the total number of academia-only papers in our dataset and multiplying the result by 100. Similarly, for industry, we calculated the percentage of industry-only papers belonging to each cluster by dividing the number of industry-only papers in a specific cluster by the total number of industry-only papers in our dataset and multiplying the result by 100.

This clustering, shown in Fig. \ref{fig:fig3}a, illustrates the topical distribution of industry's responsible AI research. Here we quantify the distribution of industry and academia research across different topics in responsible AI research. Our results suggest that industry research concentrates on issues like bias,'' explainable AI,'' human-AI interaction,'' and algorithmic fairness,'' with less representation in topics addressing broader societal implications. Academia, along with bias,'' shows greater concentration in human-centric AI,'' ethical and moral concerns,'' and equitable AI.'' Both sectors, however, converge on privacy'' and data governance'' (see Table \ref{table:3AClusters} in the appendix for associated terms).

We conducted sensitivity analyses by varying the number of clusters, and our findings remained consistent (see Fig. \ref{fig:ex9} and Fig. \ref{fig:ex10} in the appendix). Additionally, our temporal analysis suggests industry's thematic distribution has become more concentrated over time, and industry tends to publish on certain topics after academia introduces them (see Fig. \ref{fig:ex_fig3} in the appendix).

\textbf{Comparison with conventional AI conferences.} We compared the research themes of industry and academia by examining their distribution in both conventional AI and responsible AI conferences. We performed separate clustering analyses on industry-only and academia-only papers from the leading ten conventional AI conferences (see Fig. \ref{fig:CAI_topic_cluster} in the appendix). We found that the research themes of industry and academia in these conventional AI conferences are largely similar. In contrast, the thematic distribution differs between industry and academia in responsible AI research, with industry research distributed across fewer topic clusters than academia.

\subsection{Structural Topic Modeling}

Next, we used structural topic modeling (STM) \citep*{stm-main} to identify how different research topics are distributed across industry and academia. STM builds on probabilistic topic modeling \citep*{prob-topic-model} and is widely used in the literature \citep*{lucas2015computer, mulder2021recommend, 4yearsfacct}. This method allows additional information, such as affiliations, to be incorporated along with the target text to be analyzed to estimate topic proportions more precisely. This let us analyze the relative prevalence of specific topics across our target groups (i.e., industry and academia). In our analysis, we used the papers' authors' affiliations (i.e., industry or academia) as additional metadata along with abstracts. We used the same dataset here as we did in our earlier k-means clustering analysis.

\begin{figure}
    \centering
    \textbf{Linguistic analysis suggests that industry has limited engagement in key issues in responsible AI}\par\medskip
    \includegraphics[width=0.98\textwidth]{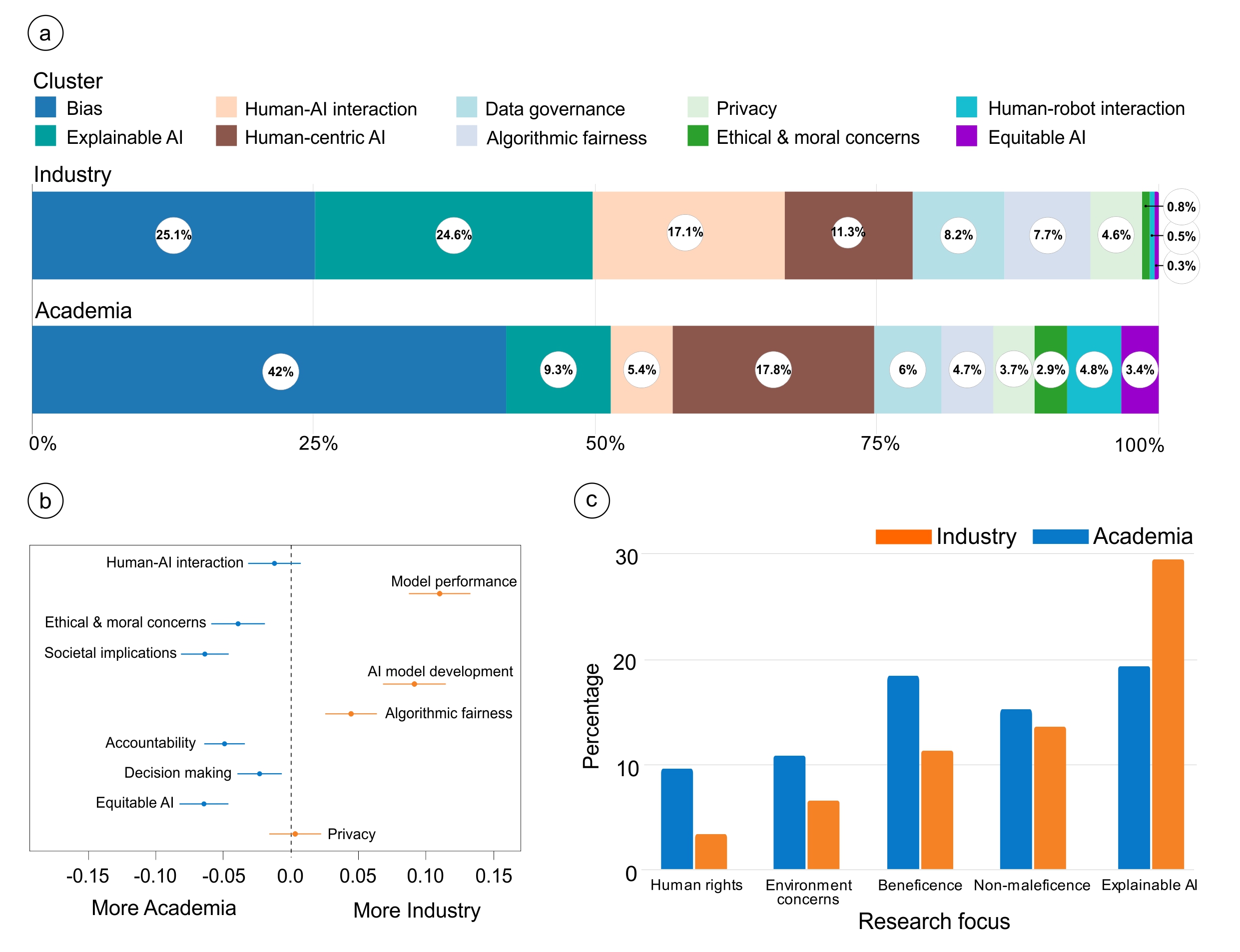}
    \caption{{\fontsize{10}{10}\selectfont Here, the first figure \textbf{Fig. \ref{fig:fig3}a} quantifies the key focus areas in responsible AI research between industry and academia (n = 10,799 papers in total; 2010-22) using k-means clustering on paper abstracts. \textbf{Fig. \ref{fig:fig3}b} shows the structural topic modeling estimates using the same data. Topics toward the right-hand side are more prevalent in abstracts from industry, while those toward the left are more prevalent in abstracts from academia. Both analyses indicate that industry research is more concentrated in technical topics, while academia shows greater concentration in ethical, moral, and societal topics. Subsequently, by conducting a frequency analysis of relevant terms on the same data, \textbf{Fig. \ref{fig:fig3}c} shows the relative frequency of topic-related terms, with industry research showing lower frequencies for terms related to ``Human rights'' and ``Environmental concerns,'' and higher frequencies for terms related to ``Explainable AI''.}}
    \label{fig:fig3}
\end{figure}

Our analysis covered different numbers of topics (k=5, 10, 15), but in this section, we focus on the results from the ten-topic model, as illustrated in Fig. \ref{fig:fig3}b. This highlights the contrasting priorities between industry and academia. Specifically, industry pays more attention to technical improvement (for example, ``AI model development,'' ``model accuracy,'' and ``algorithmic fairness'') while ignoring their applicability or societal needs. In contrast, academia seems to place more emphasis on ``ethical and moral concerns,'' ``accountability,'' ``societal implication,'' and ``equitable AI.'' On the other hand, ``privacy'' appears to be of mutual interest to academia and industry.
 
Our STM analysis, by analyzing the dominant topics present in both groups' work, suggests a concerning difference between the research priorities of industry and academia. This shows that industry heavily emphasizes research on developing and improving AI technology. In contrast, they pay limited attention to moral and ethical issues, accountability, and the societal implications of their deployed technologies. Even though there is a growing recognition within computing research that computing tools have an important role to play in addressing social issues \citep*{role-computing-social-change}, firms' engagement in responsible AI research seems to overlook such normative issues. Overall, our results suggest that industry is more focused on technical rather than sociotechnical aspects of responsible AI.

 \subsection{Frequency count}

Additionally, we searched the abstracts of the same 10,799 papers used in our prior k-means clustering analysis. We used regular expressions (regex) to identify relevant papers and counted the total number of papers on each topic from industry and academia. To get an intuitive measure of industry's topic distribution, we calculated the percentage of papers on each topic by dividing the total number of papers from industry on each one (identified using regex patterns) by the total number of papers from industry on responsible AI (classified by our earlier model described in Section \ref{section:scibert-subsection}).

Repeating the same calculations for academia, and comparing the results with those of industry, we observed differences in topic distribution. We found that industry research shows lower frequencies for terms related to human rights,'' environmental concerns,'' and beneficence'' (Fig. \ref{fig:fig3}c) and higher frequencies for explainable AI.'' This pattern is consistent with theoretical accounts suggesting that market-facing organizations may concentrate research efforts on areas with clearer paths to commercialization \citep*{rosenberg1990firms, cohen1990absorptive}. For example, the increasing computational power required for AI systems can contribute to environmental concerns, such as high energy consumption and carbon emissions, a topic that appears less frequently in industry research than in academic research. Overall, while both industry and academia show similar frequencies for terms related to ``non-maleficence,'' their topic distributions differ for other dimensions of responsible AI research.

\subsection{Additional Linguistic Analyses}

Furthermore, we employed complementary approaches on a subset of the dataset previously analyzed, specifically focusing on papers from three leading conferences in the field of responsible AI. First, we used latent Dirichlet allocation (LDA) topic modeling \citep*{lda}. The results mirrored our prior findings from STM analysis as previously shown in Fig. \ref{fig:fig3}b. The patterns held even when we varied the number of topics. In Fig. \ref{fig:ex11} in the appendix, we present LDA topic modeling results for ten topics, and we list the top keywords in Table \ref{table:oldA8} in the appendix.

Finally, we used a bigram frequency analysis, which requires fewer assumptions than topic modeling and clustering approaches. Fig. \ref{fig:ex12} in the appendix shows the top 20 keywords for industry-only and academia-only bigrams using a different dataset containing the abstracts of papers from the three leading responsible AI conferences. This analysis suggests that industry research shows lower frequencies for terms like criminal justice,'' implicit bias,'' and fairness metrics.'' These results are consistent with our earlier findings indicating that industry and academia differ in their topic distributions, with industry research more concentrated in technical topics. Industry's responsible AI research shows higher frequencies in areas such as explainable AI'' and ``model performance,'' which align with market-oriented applications. Firms may have incentives to develop and launch products that are explainable, as this can help satisfy regulatory requirements \citep*{ahmed2026} and gain market acceptance.

\subsection{Summary of Findings}

Our linguistic analysis, employing multiple methods across different datasets, documents variation in the thematic composition of responsible AI research between industry and academia. Industry research concentrates in technical aspects of responsible AI, such as explainability, privacy, and algorithmic fairness, areas that align closely with product development and market applications. Academia, by contrast, places greater emphasis on human rights considerations, ethical and moral dimensions, and the broader societal implications of AI, consistent with the sociotechnical approach discussed in recent literature \citep*{ai_safety_terms_2023, 19}. These patterns hold across STM topic modeling, k-means clustering, frequency analysis, and LDA, suggesting they are robust to methodological choices. Taken together, the evidence indicates that industry and academia exhibit distinct thematic priorities within the responsible AI research landscape.

\section{Limited Adoption of Responsible AI Research in Commercialization: Patent Citation Analysis}
\label{section:citation-analysis}
\vspace{-0.8em}

\begin{figure}
    \centering
    \textbf{Example of a patent that cited responsible AI research}\par\medskip
    \includegraphics[width=0.9\textwidth]{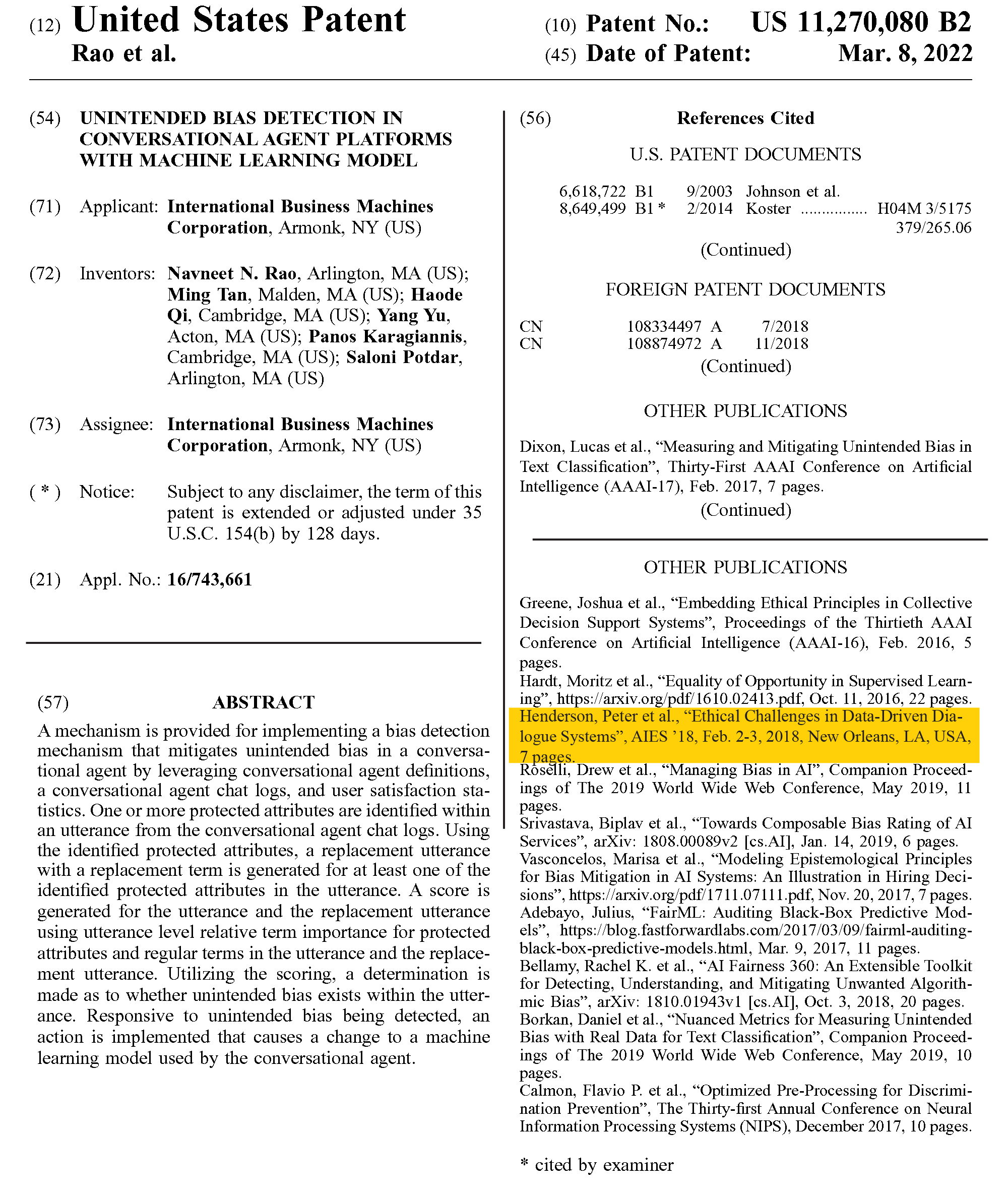}
    \caption{{\fontsize{10}{10}\selectfont A patent by IBM, ``Unintended Bias Detection in Conversational Agent Platforms with Machine Learning Models,'' is an illustrative case that cited multiple responsible AI papers, including ``Ethical Challenges in Data-Driven Dialogue Systems,'' which was published in one of the leading responsible AI conferences--AIES. This highlights the importance of responsible AI research in commercial invention.}}
    \label{fig:patent-cite-res-ai}
\end{figure}

To assess the extent to which industry integrates responsible AI research into its product commercialization processes, we employed patent citations as a measure of how industry builds upon academic research in its products and services, following established practices in prior literature \citep*{patent-citation-hci, jaffe2004patent}. The intuition is that if AI firms incorporate responsible AI research in their products, it will likely be reflected in their patent citations \citep*{jaffe2004patent}. This can be seen in the example provided in Fig. \ref{fig:patent-cite-res-ai}, which illustrates how a patent citation can reveal the impact of responsible AI research on firms' innovations.

\subsection{Data and Methodology:}

We used two distinct datasets to capture the research paper references in patents. First, for generic patents, we used the Reliance on Science data (\hyperref[table:3_distinct_data]{dataset 4}) by Marx \& Fuegi \citeyearpar{33}. This dataset is continually updated and maintained by the authors, and we have used the latest data available as of May 2024.\footnote{Marx, M. (2023). Reliance on Science (V62). Zenodo. \href{https://doi.org/10.5281/zenodo.10215169}{doi.org/10.5281/zenodo.10215169}} The dataset includes 32,698,465 patents-to-paper citations data from 2,905,718 USPTO patents (1947-2022) and 5,210,014 scientific publications (1800-2022). Second, we used 141,770 AI patents between 1985 and 2018 (\hyperref[table:3_distinct_data]{dataset 5}) from Miric et al. \citeyearpar{34}. After matching the AI patents data with the Reliance on Science data, we identified 726,712 AI patent-to-paper citations to research papers from 1985 to 2018.

Prior research suggests that industry patents tend to cite industry research at higher rates and cite their own research more in their commercial innovations \citep*{made-in-academia, arora-corporate-investment-in-research}. To account for this citation pattern, we conducted our patent citation analysis for industry and academia co-authored papers separately. We compiled two new, distinct datasets of responsible AI papers, separating papers from industry and academia

For industry co-authored papers, the dataset comprised 683,390 papers (dating from 2010 to 2022) including: (a) Scopus-indexed papers from 1,771 AI firms (n = 679,919; 2010-22), (b) industry co-authored papers from three leading responsible AI conferences (n = 160; 2018-22), and (c) a sample of industry co-authored papers containing expert-suggested responsible AI keywords in their abstracts or titles (n = 3,311; 2010-22). For academia co-authored papers, our dataset comprised 5,296,547 papers (dating from 2010 to 2022) including: (a) Scopus-indexed papers from leading 100 universities (n = 5,265,419; 2010-22), (b) academia co-authored papers from leading responsible AI conferences (n = 802; 2018-22), and (c) a sample of academia co-authored papers containing expert-suggested responsible AI keywords (n = 30,326; 2010-22). 

To observe \textit{all} the USPTO patents citing any industry co-authored papers between 2010 and 2022, we first preprocessed our industry paper data and then cross-referenced their DOIs with the Reliance on Science dataset. The matched distinct industry papers (n = 34,590; 2010-22; cited by 96,376 patents) were then disaggregated into three groups--responsible AI, conventional AI, and non-AI papers--by matching their DOIs with our previously classified papers' DOIs (using our ensemble classification model; see Section \ref{section:scibert}). For the sake of simplicity, we present only two categories--responsible AI and conventional AI--in the output plots shown in Fig. \ref{fig:patent-analysis}. For USPTO patents citing academic papers, we followed a similar approach to the one we used for industry papers and obtained 156,140 distinct academia papers (2010-22; cited by 203,105 patents). Likewise, we disaggregated them into the same groups (responsible AI, conventional AI, and non-AI papers) using the DOIs of the papers classified by our ensemble model (see Fig. \ref{fig:flowchart1} and Fig. \ref{fig:flowchart3} in the appendix for the complete process). 
 
Additionally, we matched paper DOIs exclusively with AI patents for both industry and academia co-authored papers separately. For this, we first matched the DOIs of our AI patent dataset with the Reliance on Science dataset. Afterward, we compared this matched data with our industry papers, which resulted in a dataset of 3,051 patents citing 2,349 distinct industry papers. Using the same process for academia papers, we obtained 4,509 distinct papers cited by 4,152 patents. Utilizing this publication data cited in patents for both groups, we disaggregated them into three groups as we had done previously for generic patents (non-AI, conventional AI, and responsible AI papers) by matching the DOIs of the papers classified by the ensemble classifier (see Fig. \ref{fig:flowchart2} and Fig. \ref{fig:flowchart4} in the appendix for the complete process). 

\subsection{Responsible AI Research in Generic Patents}

\begin{figure}
    \centering
    \textbf{Patent citation analyses illustrate that industry patents rarely cite responsible AI research}\par\medskip
    \includegraphics[width=0.9\textwidth]{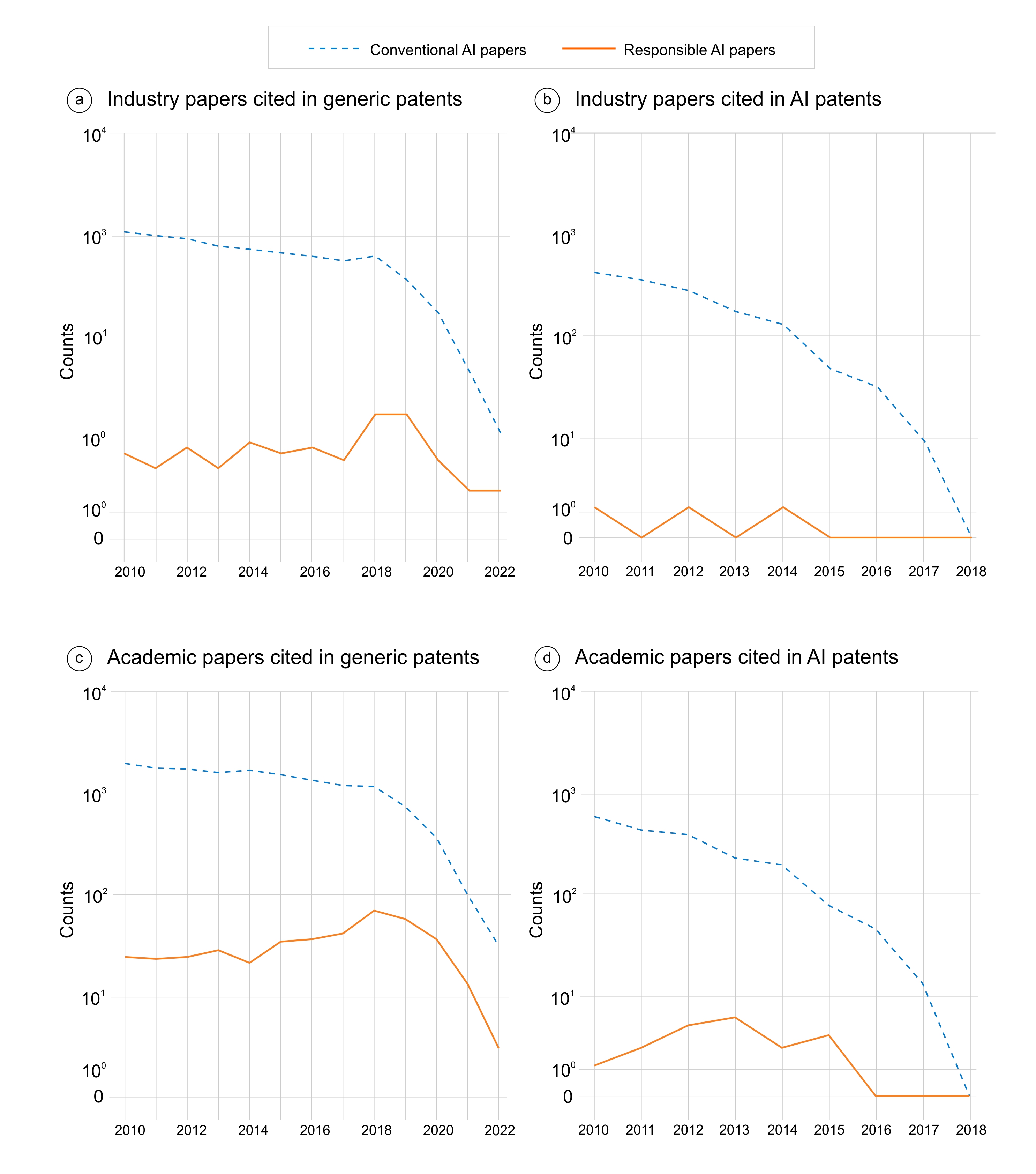}
    \caption{{\fontsize{10}{10}\selectfont This figure analyzes industry and academic papers’ citations in USPTO patents. By matching against a comprehensive list of USPTO patent-paper citation data (n = 32,698,465 citations; 1947-2022), we show in \textbf{Fig. \ref{fig:patent-analysis}a} and \textbf{\ref{fig:patent-analysis}c} that 88 and 396 responsible AI papers from industry and academia, respectively, have been cited in generic patents, while 15,236 academia and 7,532 industry conventional AI papers have been cited in patents between 2010-22. Using a separate dataset of AI patents (n = 141,770 patents; 1985-2018), \textbf{Fig. \ref{fig:patent-analysis}b} and \textbf{\ref{fig:patent-analysis}d} illustrate that three responsible AI papers from industry and 17 from academia have been cited in the AI patents between 2010-18, compared to higher citation counts for conventional AI research. }}
    \label{fig:patent-analysis}
\end{figure}

Our analysis of over 32 million patent citations in USPTO patents shows that responsible AI research represents a small share of industry patent citations. We found that 88 industry-authored responsible AI papers were cited in these generic patents from 2010 to 2022, as presented in Fig. \ref{fig:patent-analysis}a. By comparison, generic patents cited 7,532 conventional AI papers and 26,970 non-AI papers from industry over the same period.\footnote{A notable trend is the decreasing total number of citations across both categories--conventional AI and responsible AI (see Fig. \ref{fig:citation-cumulative} in the appendix for the cumulative counts). There could be two potential explanations for this trend. First, it takes time for basic knowledge to diffuse to commercial patents \citep*{mansfield1991academic, patent-citation-hci}. Also, older papers have more time to be discovered by others. Therefore, patents are likely to cite older papers more than newer papers. The second reason is the time-consuming patent application process; our data considers only the approved patents, not patent applications still under review, contributing to the reduced citation count in recent years.} We found a total of 246 patents cited 88 responsible AI papers. We then examined the top patent assignees that cited industry-authored responsible AI research: IBM, Microsoft, and Google were the top patent assignees. These are the same firms that had higher rates of responsible AI research output. The correlation between firms' responsible AI research output and their patent citations is consistent with absorptive capacity theory \citep*{cohen1990absorptive}, which suggests that research participation facilitates the translation of knowledge into commercial inventions.

We observed a similar pattern when we examined industry patents' citations of academic papers. By matching over 5 million academic papers from Scopus (2010-22) with data on over 32 million patent-paper-citations (1947-2022), we observed that industry cited 15,236 conventional AI and 140,508 non-AI academic papers, and 396 responsible AI papers from academia (see Fig. \ref{fig:patent-analysis}c). Upon examining the patents' citations we found a total of 983 patents cited academic responsible AI papers. The top patent assignees citing such academic papers were IBM, Microsoft, and Google.\footnote{When we manually checked these responsible AI papers we found a slight over-count of responsible AI papers in industry and academia because of the keywords sampled responsible AI papers.}

\subsection{Responsible AI Research in AI Patents}

We further examined a subset of patents--specifically AI patents using a different source. Our result, presented in Fig. \ref{fig:patent-analysis}b shows that out of 141,770 AI patents (1985-2018), three responsible AI papers from industry were cited. During the same period, the patents cited 1,438 conventional AI papers and 908 non-AI papers.

Similarly, we analyzed academic papers cited in AI patents from industry. This analysis yielded comparable results--17 responsible AI papers from academia were cited in AI patents (see Fig. \ref{fig:patent-analysis}d). By comparison, 1,949 conventional AI and 2,541 non-AI papers were cited by those patents. This pattern indicates that citation rates for responsible AI research differ from those for conventional AI research in patent filings. In Fig. \ref{fig:citation}, we present evidence that AI firms' responsible AI research builds on academic responsible AI research.

We then analyzed the patents that cite the responsible AI papers from industry and academia. We find that the 17 university-authored responsible AI papers were cited by a total of 22 patents and 3 industry-authored responsible AI papers received citations from 5 patents.

\subsection{Alternative Channels for Responsible AI Integration}

Patent citations represent one channel through which responsible AI research may inform commercial development, though firms may integrate responsible AI principles through additional mechanisms not captured by bibliometric analysis. Several firms have implemented responsible AI frameworks through governance structures, open-source tools, and certification processes. For example, IBM developed and open-sourced the AI Fairness 360 toolkit, which provides over 70 fairness metrics and bias mitigation algorithms for practitioners \citep*{bellamy2019ai}. Salesforce established its Office of Ethical and Humane Use in 2018, developed an AI Ethics Maturity Model to guide organizational practices, and implemented consequence scanning methodologies across product teams \citep*{salesforce_2023}. SAP, the first European technology firm to publish AI ethics principles, has achieved ISO 42001 certification for its AI governance framework covering Joule and SAP AI Core, and maintains both an internal Global AI Ethics Steering Committee and an external Global AI Ethics Advisory Panel \citep*{sap_responsible_2025}. These examples illustrate that some firms operationalize responsible AI principles through product design, governance frameworks, and certification processes, activities that may not be reflected in traditional bibliometric measures of research output or patent citation patterns.

\subsection{Summary of Findings}

Taken together, our large-scale patent-to-paper citation analysis is consistent with our prior results. The analysis documents that industry patents cite responsible AI research at lower rates than conventional AI research, suggesting differences in how these research streams are integrated into commercial innovation. This pattern could reflect several factors: variation in firms' research priorities, differences in the applicability of responsible AI findings to patentable innovations, or variation in absorptive capacity \citep*{cohen1990absorptive, 26} necessary to assimilate and apply knowledge from this emerging field. Notably, firms with greater engagement in responsible AI research, such as IBM, Microsoft, and Google, are also the most frequent citers of responsible AI papers in their patents, consistent with the view that research engagement facilitates knowledge integration.

\vspace{-0.8em}
\section{Discussion}
\label{section:discussion}
\vspace{-0.8em}

Our results indicate that the rapid commercialization of AI  \citep*{ai-index-2023} is occurring concurrently with comparatively limited industry engagement in responsible AI research. This pattern has implications for the trajectory of AI development, as AI firms are currently at the forefront of research and innovation, shaping the field's direction. Our findings document that firms exhibit lower engagement with responsible AI in both research (as measured by publications) and development (as measured by patent citations of responsible AI research). These patterns raise questions about whether the current trajectory of AI development adequately incorporates considerations of societal implications.

These patterns of engagement also have implications for firms' absorptive capacity. Innovation research suggests that firms benefit from participating in research to develop their capacity to absorb external knowledge \citep*{cohen1990absorptive, 26}. Recent reports indicate that some AI firms have reduced their responsible AI research teams \citep*{gerrit_de_vynck_oremus_2023}, with several leading technology firms restructuring positions in this area. \citep*{field_vanian_2023} \footnote{Prior research further reveals that employees who prioritize ethical considerations often encounter obstacles, as firms tend to prioritize product launches over investing resources in addressing ethical concerns \citep*{walking-the-walk-ai-ethics}.} If industry participation in responsible AI research remains limited, firms' ability to adopt the latest external research may be constrained. This dynamic could affect the extent of critical examination of social benefits and costs in the early stages of design and may influence the diffusion of responsible AI knowledge from academia to industry.

While industry engagement with the public represents one avenue for advancing responsible AI, scholars have noted that such involvement functions most effectively when it supplements, rather than replaces, public responsible AI research. Prior work has raised questions about the potential for industry presence to shape the trajectory of AI research \citep*{39, democratization, benkler2019don} and influence the future research agenda in the field of responsible AI \citep*{38,39,40,41}. These considerations suggest that industry's increased presence may be most beneficial when complementary to existing responsible AI research efforts.

Additionally, the responsible AI research community plays an important role in making its work more accessible and applicable for practitioners. Prior studies \citep*{holstein2019improving, walking-the-walk-ai-ethics, fairMeetsProAttribute} have documented challenges practitioners face when attempting to apply and implement research findings in real-world settings. Moreover, a substantial portion of academic responsible AI research may not be immediately applicable or easily integrated into commercial products, potentially contributing to the observed patterns of industry engagement in this area. Strengthening mechanisms for industry-academia collaboration could facilitate the translation of academic responsible AI research into commercial innovation.

Our research raises important questions for both innovation and responsible AI research. For example, future studies could explore the factors that motivate certain firms to publicly engage in responsible AI research. Similarly, examining the perceived benefits and costs of such engagement, including enhanced trust, talent acquisition, innovation outcomes, and regulatory compliance, could provide valuable insights for both managers and policymakers.

\vspace{-0.8em}
\subsection{Policy Implications}

Our systematic analysis documents patterns of industry involvement in responsible AI research that have implications for AI governance. The development and commercialization of AI technologies intersects with questions about societal implications, which policymakers and scholars have increasingly sought to address. These findings may inform discussions about measures to support responsible development and deployment of AI systems.

Our results suggest that patterns of participation in responsible AI research may be associated with firms' capacity to anticipate challenges arising from AI deployment. Limited research engagement may constrain firms' ability to identify and address potential issues with their tools during the design phase. Documented incidents illustrate this dynamic: in 2016, Microsoft launched an AI chatbot on Twitter targeting 18-to-24-year-olds, which within 24 hours of its launch began posting harmful content \citep*{microsoft-racist-chatbot}. Years later, in 2023, Snapchat launched its AI chatbot, My AI, which reportedly generated inappropriate content involving a minor \citep*{snapchat-chatbot}. These cases, spanning several years, illustrate how firms may encounter challenges that could potentially have been anticipated through greater engagement with responsible AI research. Active engagement in responsible AI research may enable firms to develop a more comprehensive understanding of the potential risks and ethical implications of their technologies, potentially allowing them to address these considerations during the design and development process.

Notwithstanding increased attention from regulators \citep*{White_House_2023}, civil society \citep*{activist_facct_24}, and employees \citep*{ahmed&jia, emp_activism_2020}, alongside a growing number of documented AI incidents \citep*{ai-index-2024}, industry participation in responsible AI research remains comparatively limited. This pattern suggests that current incentive structures may not fully align with the development of responsible AI \citep*{cooperation-res-ai}. Prior research suggests that competitive pressures may accelerate the commercialization process, potentially affecting investment in responsible AI development \citep*{cooperation-res-ai}. Regulatory frameworks represent one mechanism that could encourage AI firms to demonstrate engagement in responsible AI research as part of product development and deployment processes.

Transparency and public accountability represent important dimensions of the AI development process \citep*{brundage_toner_fong_et_al_2020, mittelstadt_2019}. To foster trust and confidence in the AI ecosystem, policymakers may consider mechanisms for AI firms to disclose their responsible AI research efforts, ethical frameworks, and decision-making processes related to the development and deployment of AI systems \citep*{companies-committed-res-ai, deAlmeida2021-DEAAIR}. Such disclosures could be complemented by independent audits and public review processes, supporting alignment between AI development and societal values and interests \citep*{mokander2021ethics}.

Furthermore, allocation of funding and resources toward independent, public-sector responsible AI research initiatives represents another policy consideration. While industry participation contributes to the field, our findings highlight the value of maintaining a robust research ecosystem that can examine AI technologies and their societal impacts from perspectives independent of commercial considerations \citep*{40}.

Industry-academia research collaborations in responsible AI represent one mechanism that could help bridge observed gaps. Such partnerships may promote knowledge exchange, foster the integration of academic insights into industry practices, and support a culture of responsible AI development within firms. By leveraging the combined expertise and resources of academia and industry, such collaborations could contribute to the advancement of responsible AI research and its practical implementation.

\subsection{Implications for the Direction of AI Development} 

The direction of AI research has implications for power, politics, and social dynamics in society \citep*{acemoglue-harms-of-ai}. The path-dependent nature of technological progress \citep*{nature-of-technology} implies that decisions made during the design and implementation phases of AI will have enduring societal implications once the technology is widely adopted \citep*{bicycles-bakelites-bulbs, winner2017artifacts}. The effects of AI systems can be difficult to reverse \citep*{algorithmic-imprint}, highlighting the value of considering potential consequences and guiding AI development toward beneficial outcomes during early stages.

To the extent that market forces do not fully direct AI innovation toward socially beneficial outcomes, public policies and democratic processes may play a role in addressing potential challenges and supporting alignment between AI and broader societal goals \citep*{nathan1969techChange, acemoglue-harms-of-ai}. Just as subsidies have been introduced to encourage innovation in renewable energy \citep*{Acemoglu2012EnvTechChange}, similar measures could be considered to address challenges associated with AI and other digital technologies. As AI progresses, governments can provide guidance on which validated AI systems are appropriate to deploy in high-stakes public contexts, considering factors like efficacy, safety, fairness, and transparency. By establishing policy frameworks, societies can guide the development of human-complementary AI in directions that enhance human potential and benefit society \citep*{12}. Thoughtfully designed policies can support alignment between AI progress and societal values.

\subsection{Limitations}
Our study is not without limitations. First, our focus on peer-reviewed responsible AI publications may not capture the full extent of AI firms' research engagement. While our research questions led us to examine only peer-reviewed responsible AI publications, AI firms may engage in research that they do not publish. Additionally, firms may have papers that are not peer-reviewed. That said, prior research suggests that research engagement with academics increases firms' capacity to adopt responsible AI research \citep*{28, cohen1990absorptive}. Additionally, peer-reviewed research and engagement with academic outlets are important for building trust and working through important and ambiguous design choices \citep*{sucheta2023value, kleinberg2023challenge}.

Second, since the diffusion of knowledge from basic science to product creation requires time \citep*{mansfield1991academic}, our patent citation analysis may \textit{underestimate} the actual extent to which responsible AI research informs commercial innovation. Moreover, not all elements of responsible AI research are amenable to patenting, nor are they always reflected in patent citations \citep*{why-firms-patent}. However, we observe that some AI patents cite recent academic papers on conventional AI and non-AI, suggesting that industry can, to some extent, build upon academic work in this area.

Third, the training data for our supervised machine learning model, sourced from leading responsible AI conferences, might not fully represent all industry research activities. While these conferences address a wide range of issues, they may still miss some industry research topics. To address this, we implemented measures such as data ablation, using different subsets of training datasets to classify industry papers (see Fig. \ref{fig:validation-scibert} in the appendix), and conducted manual validation. Additionally, we validated our findings using two alternative classification methods, an expert-consulted keyword search (see \ref{fig:keyword-appendix}) and LLM-based classification (see \ref{fig:LLM-appendix}), both of which produced consistent results.

Notwithstanding these limitations, our findings document patterns of industry engagement in responsible AI research and provide an empirical baseline for future studies examining the relationship between research activity and commercial innovation in this field.
\vspace{-0.8em}
\section{Conclusion}
\vspace{-0.8em}

AI is poised to transform society, and like all technologies, its development trajectory carries important implications \citep*{acemoglue-harms-of-ai, 1, acemoglu2023regulation, 12}. Our large-scale study documents patterns of industry engagement in responsible AI research. We find that among AI research firms, which are shaping the research agenda and driving innovation \citep*{18, frank_wang_cebrian_iyad, koch2021reduced},  34.5\% publish responsible AI papers, and that the ratio of responsible AI to conventional AI research is lower in industry than in academia. We also observe that industry's responsible AI research concentrates in a narrower range of topics compared to academic research, with greater emphasis on technical dimensions such as explainability and algorithmic fairness. These patterns are consistent across different datasets and measurements. Our large-scale patent analysis further documents that industry patents cite responsible AI research at lower rates than conventional AI research, though firms with active responsible AI research programs are also the most frequent citers. These findings provide an empirical baseline for understanding how responsible AI research is distributed across sectors and integrated into commercial innovation. As AI continues to advance, ongoing measurement of these patterns may inform discussions among researchers, practitioners, and policymakers regarding the development and deployment of AI systems.

\section*{Acknowledgements}
\vspace{-0.8em}

We would like to thank Nick Berente, Michael Park, Julian Barg, Nan Jia, Ranjit Singh, Maja Susanna Svanberg, Yian Yin, Ahmed Abbasi, Cameron Kormylo, Ana Trisovic, Mayur P. Joshi, Milan Miric, and Industry colleagues from DeepMind, Spotify, Microsoft, Google, and seminar participants at AOM, Harvard LISH, AIM at USC, NBER Productivity Seminar for their feedback and suggestions. All errors are our own.

\newpage
\bibliographystyle{apalike}
\bibliography{references}

\newpage

\renewcommand{\thesection}{S\arabic{section}}
\setcounter{section}{0}
\renewcommand{\thetable}{S\arabic{table}}
\setcounter{table}{0}
\renewcommand{\thefigure}{S\arabic{figure}}
\setcounter{figure}{0}

\section*{Supplementary Materials}

\section{Additional Analyses on Engagement Analysis}

\subsection{List of leading conferences: \hyperref[table:3_distinct_data]{Dataset 2}}

\begin{table}[H]
\centering
\caption{ List of leading responsible and conventional AI conferences}
\label{tab:conf-list}
\begin{tabular}{c c c}
\hline
\textbf{Conference Type} & \textbf{Conference Name} & \textbf{Paper Count} \\ \hline

\multirow{2}{*}{Conventional AI} & ICML & 16,070 \\
\multirow{2}{*}{(2010-2022)}& CVPR & 15,226 \\
 & ACL & 14,260 \\
 & AAAI & 12,338 \\
 & NeurIPS & 11,969 \\
 & ICCV & 9,901 \\  
  & IJCAI & 7,513 \\ 
  & ECCV & 7,459 \\
 & EMNLP & 6,067 \\
 & KDD & 5,209 \\\hline
 
 \multirow{2}{*}{Responsible AI} & AIES & 383 \\ 
  \multirow{2}{*}{(2018-2022)}& FAccT & 383 \\ 
 & EAAMO & 85 \\ \hline
\end{tabular}
\end{table}

\subsection{Supervised Machine Learning Model Details}
\label{section:scibert-detail}

\begin{figure}[h]
    \centering
    \textbf{Process for constructing Fig. \ref{fig:fig1}a and \ref{fig:fig1}b}
    \includegraphics[width=0.95\textwidth]{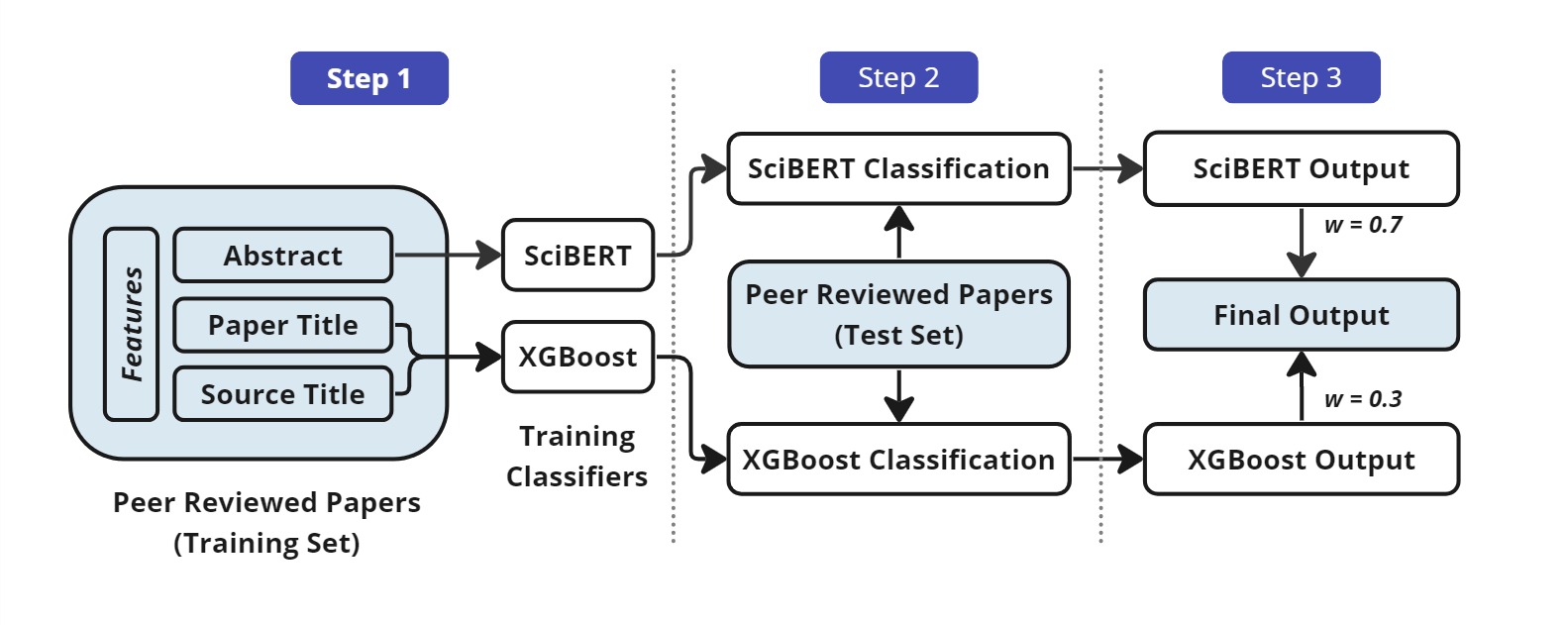}
    \caption{{\fontsize{10}{10}\selectfont This figure illustrates the overall workflow for classifying conventional AI and responsible AI papers using XGBoost and SciBERT classifiers. }}
    \label{fig:flowchartSML}
\end{figure}

\newpage
\begin{figure}[h]
    \centering
    \textbf{Industry’s limited engagement in responsible AI research (all 1,771 AI firms)}
    \includegraphics[width=\textwidth]{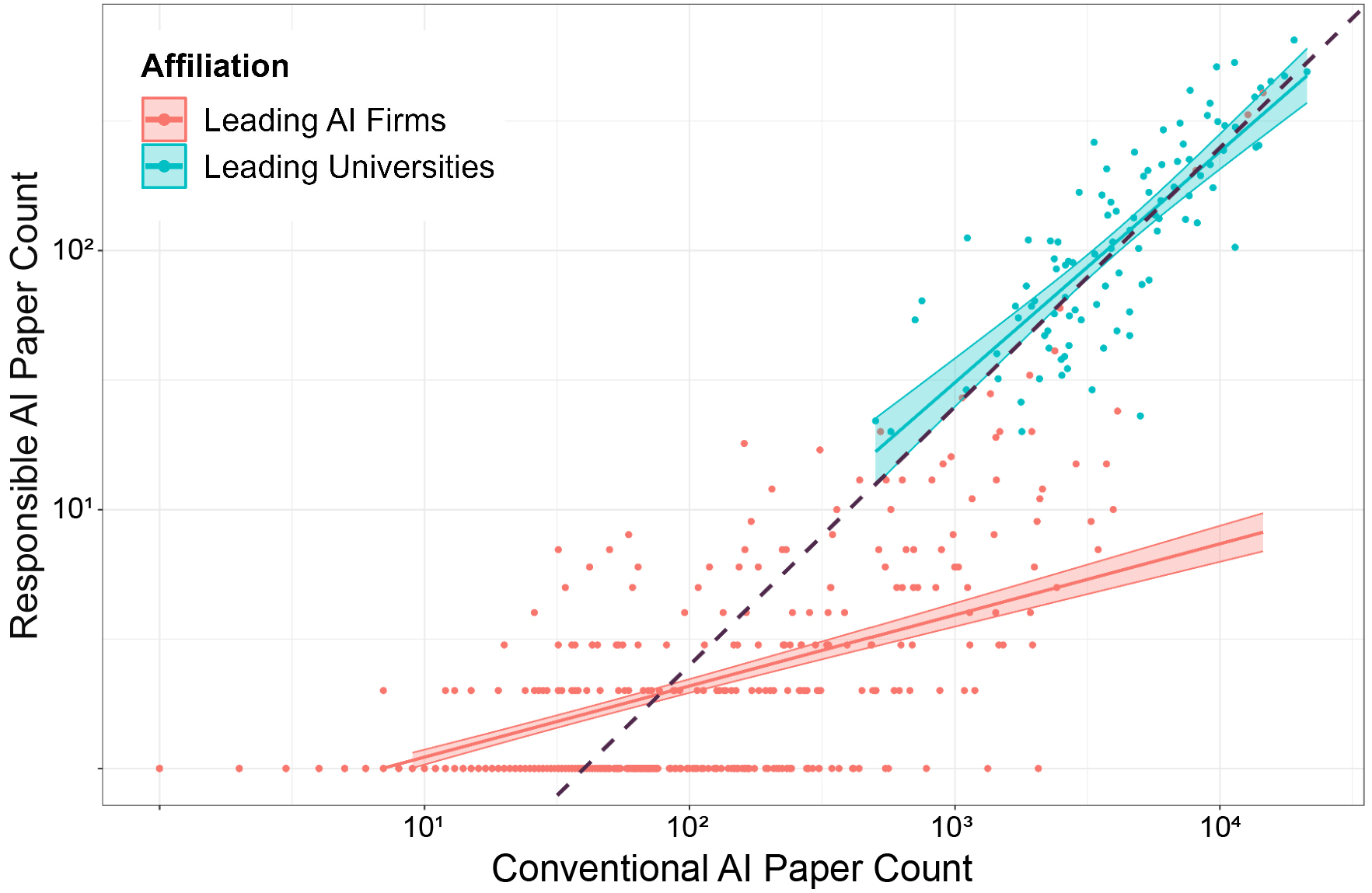}
    \caption{{\fontsize{10}{10}\selectfont This figure showcases industry and academia’s participation in conventional AI research compared to their responsible AI research. This time we present all 1,771 AI firms’ (n = 679,919; 2010-22) research instead of only the leading 100 firms. Here, individual organizations are symbolized by discrete dots. A trend line reflects the participation trend within each group (industry and academia), and the shaded bands correspond to 95\% confidence intervals. Additionally, the black dashed line indicates a reference line where the proportion of responsible AI papers to conventional AI papers is 2.5\%. }}
    \label{fig:ex1}
\end{figure}

\newpage
\begin{figure}[h]
    \centering
    \textbf{Industry’s limited engagement in responsible AI research (conference and journal separately)}
    \includegraphics[width=\textwidth]{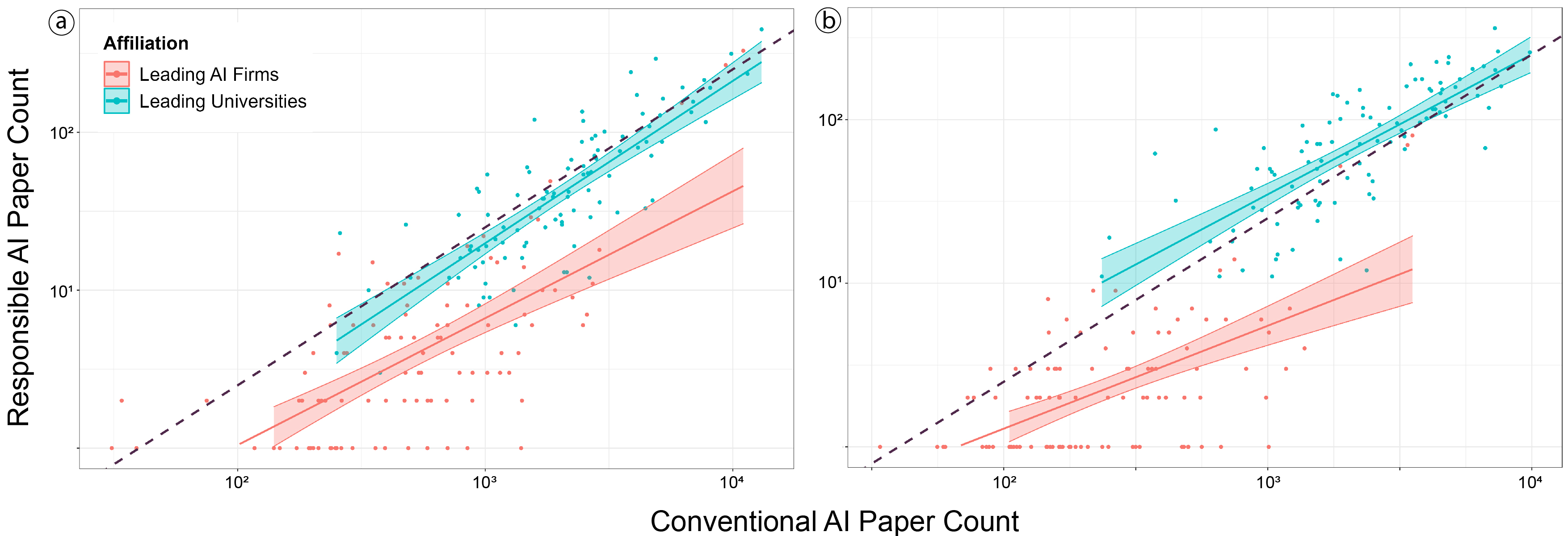}
    \caption{{\fontsize{10}{10}\selectfont This figure showcases industry and academia’s conference (Fig. \ref{fig:ex3}a) and journal (Fig. \ref{fig:ex3}b) publications, respectively, on responsible AI research compared to their conventional AI research. Here, individual organizations are symbolized by discrete dots. A trend line reflects the participation trend within each group, and the shaded bands correspond to 95\% confidence intervals. Additionally, the black dashed line indicates a reference line where the proportion of responsible AI papers to conventional AI papers is 2.5\%. }}
    \label{fig:ex3}
\end{figure}

\newpage
\begin{figure}[h]
    \centering
    \textbf{Industry's limited engagement in responsible AI research (classification model is robust to changes in training data)}
    \includegraphics[width=\textwidth]{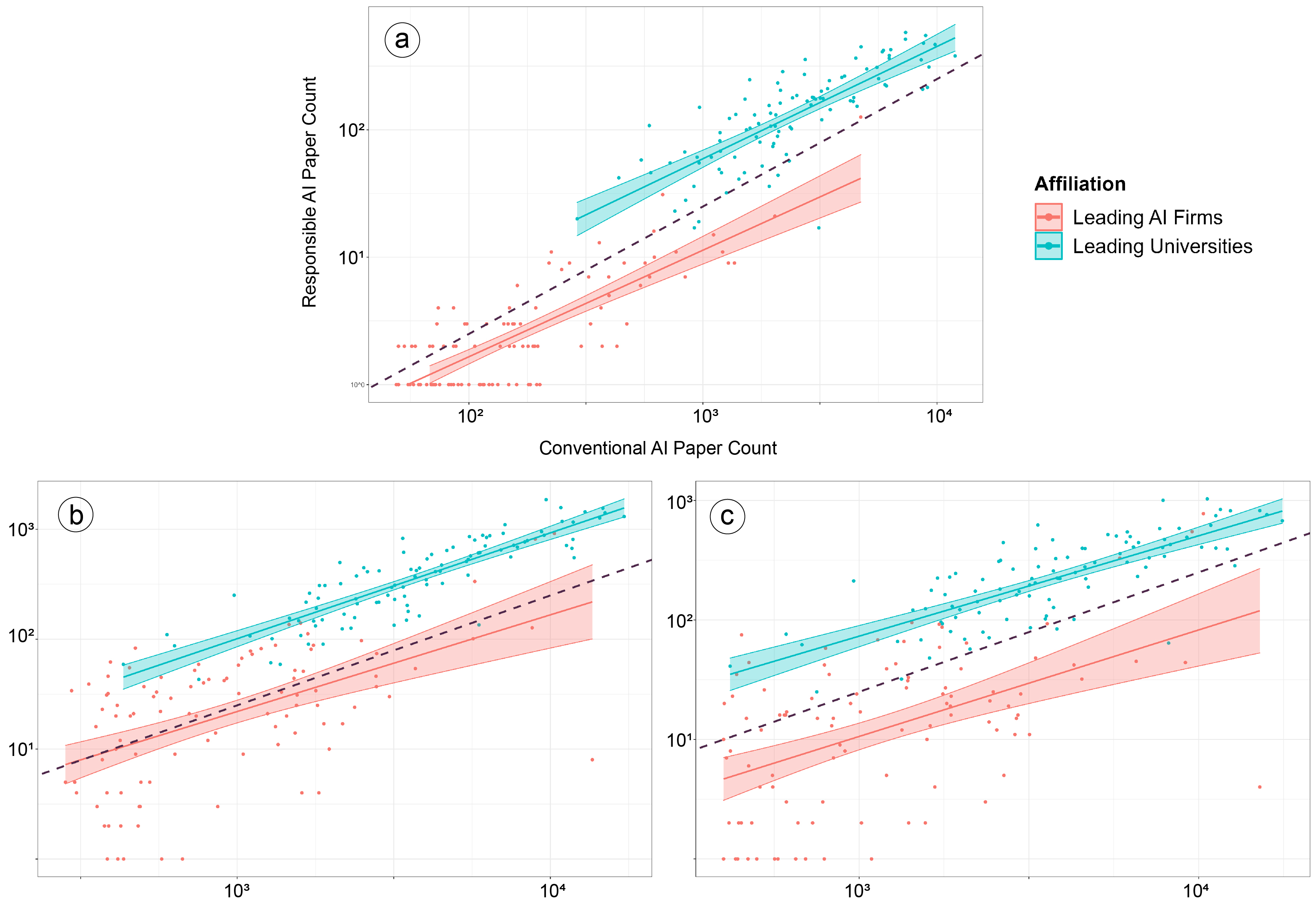}
    \caption{{\fontsize{10}{10}\selectfont This figure showcases industry and academia’s engagement in responsible AI research compared to their conventional AI research for each group's leading 100 organizations. Here, individual organizations are symbolized by discrete dots. A trend line reflects the participation trend within each group (industry and academia), and the shaded bands correspond to 95\% confidence intervals. Additionally, the black dashed line indicates a reference line where the proportion of responsible AI papers to conventional AI papers is 2.5\%. In this graph, we have trained the model with data from (a) only journals (and tested them on journal papers), (b) only conferences (and tested them on conference papers) (see Table \ref{table: Confonly_data}), and (c) previous ACM FAccT conferences for positive samples along with negative samples from other conferences (and tested them on conference papers) (see Table \ref{table: Facct_Confonly_data}). Even with varying training and testing data, the model produces results consistent with our prior analysis shown in Fig. \ref{fig:fig1}. }}
    \label{fig:validation-scibert}
\end{figure}

\newpage
\begin{figure}[h]
    \centering
    \textbf{Industry’s limited engagement in responsible AI conferences (author-weighted count)}
    \includegraphics[width=\textwidth]{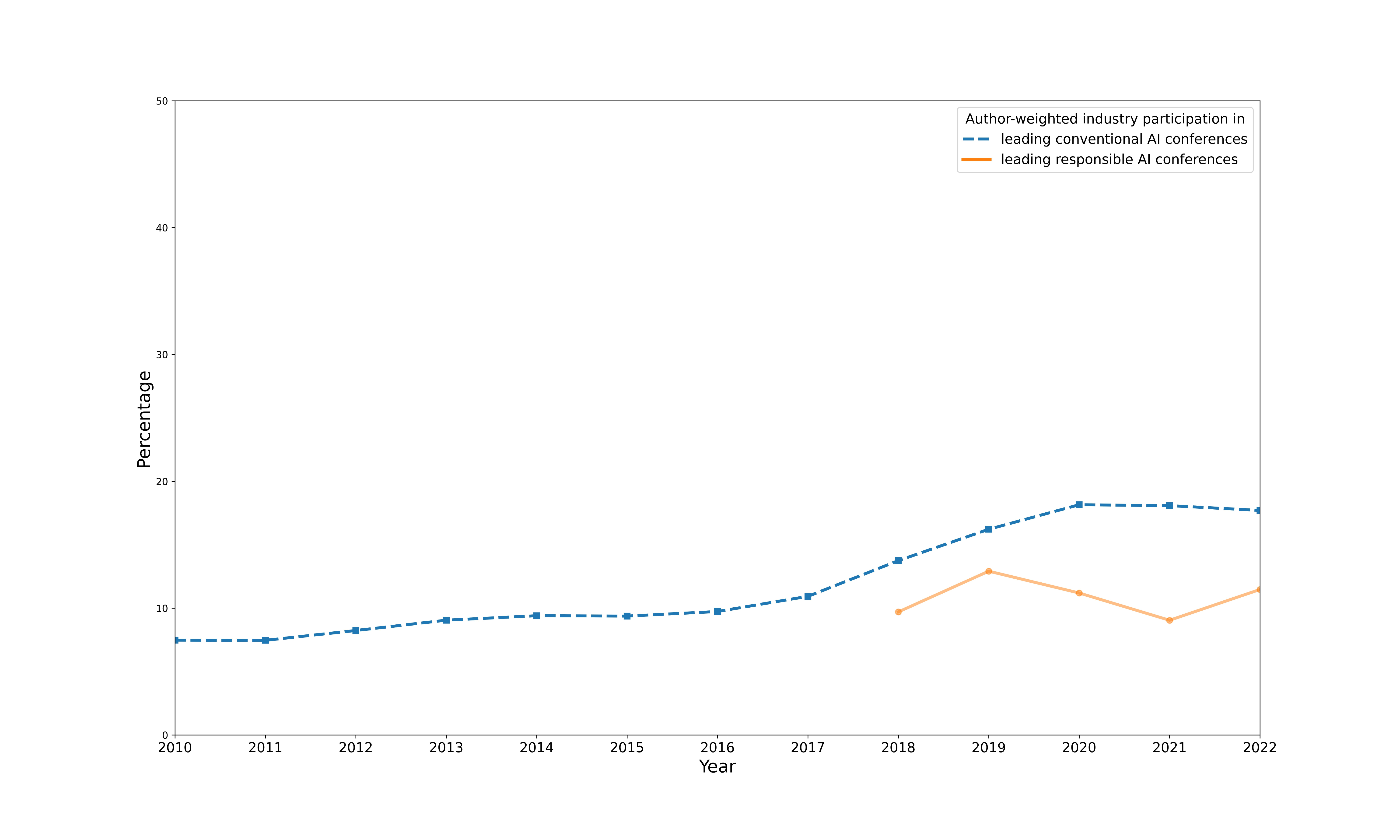}
    \caption{{\fontsize{10}{10}\selectfont The dashed blue line and orange line indicate the author-weighted proportion of papers that have at least one industry co-author in conventional AI research (10 leading conferences; \hyperref[table:3_distinct_data]{dataset 2}) and responsible AI research conferences (three leading conferences; also \hyperref[table:3_distinct_data]{dataset 2}), respectively. }}
    \label{fig:author_weighted}
\end{figure}

\newpage
\begin{figure}[h]
    \centering
    \textbf{Industry’s limited engagement in keywords sampled responsible AI research (author-weighted count)}
    \includegraphics[width=\textwidth]{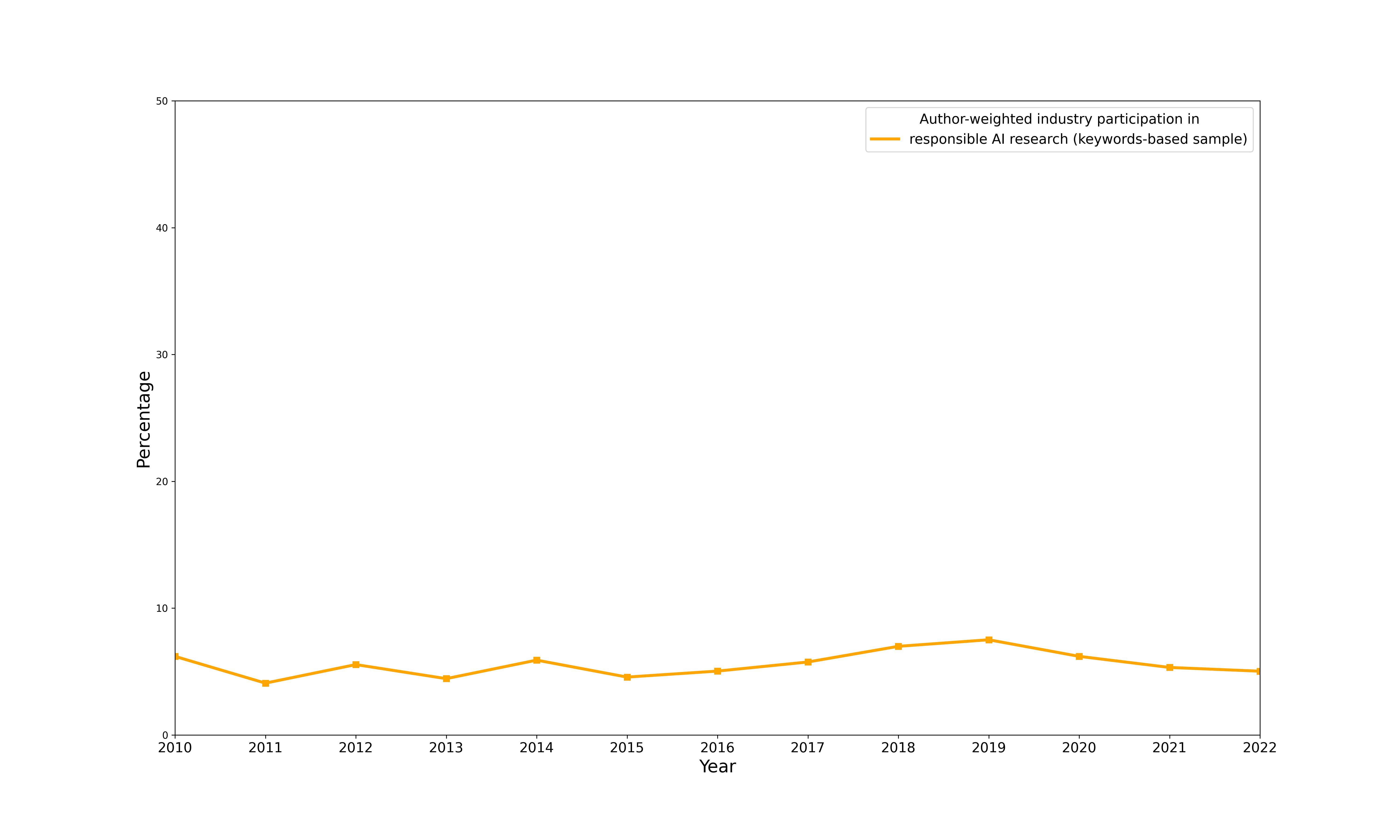}
    \caption{{\fontsize{10}{10}\selectfont The orange line represents the proportion of responsible AI research papers with at least one industry-affiliated author, weighted by author affiliation. The responsible AI research papers are identified using expert-suggested keywords (\hyperref[table:3_distinct_data]{dataset 3}).}}
    \label{fig:author_weighted_Keywords}
\end{figure}

\newpage
\subsection{Validation with Expert Suggested Keywords}
\label{section:keyword-appendix}

\begin{figure}[H]
    \centering
    \textbf{Industry’s limited engagement in responsible AI research (a comparison with academia)}\par\medskip
    \includegraphics[width=\textwidth]{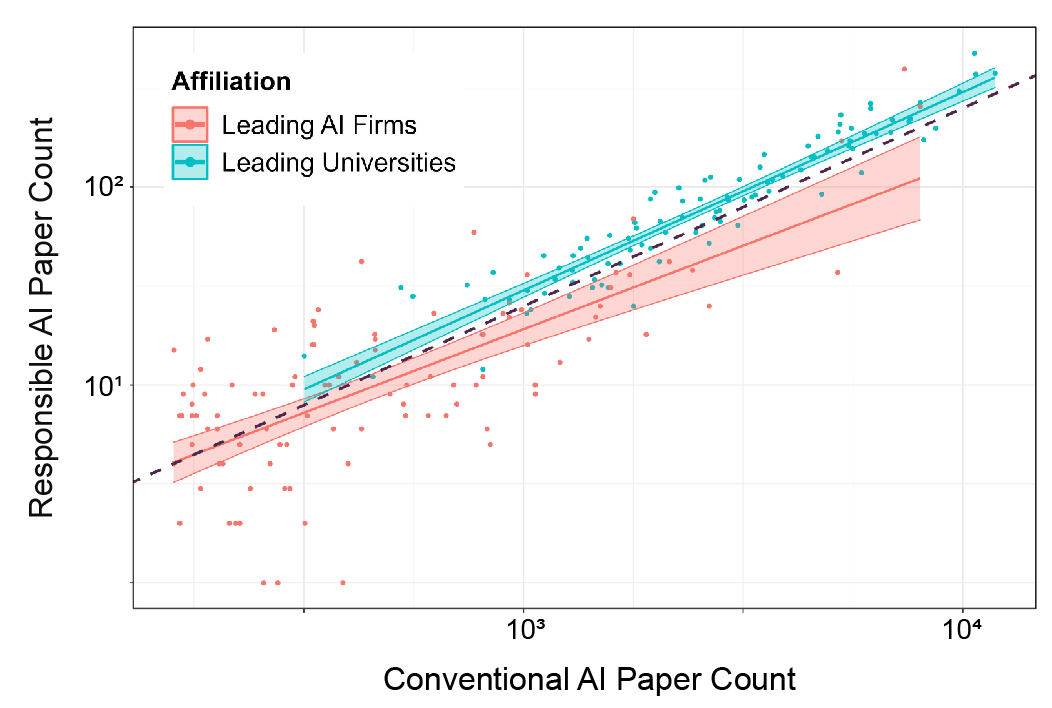}
    \caption{{\fontsize{10}{10}\selectfont Fig. \ref{fig:keyword-appendix} shows the engagement of 1,771 AI firms that have AI research and responsible AI research activities using an expert-recommended keyword list. It showcases a heterogeneity in the leading 100 AI firms’ (n = 506,017 papers) and universities’ (n = 5,265,419 papers) participation in AI research compared to their responsible AI research. Here, individual organizations are symbolized by discrete dots. A trend line reflects the participation trend within each group, and the shaded bands correspond to 95\% confidence intervals. Additionally, the dashed line indicates a reference line where the proportion of responsible AI papers to total AI papers is 2.5\%.}}
    \label{fig:keyword-appendix}
\end{figure}
\newpage

\subsection{Validation with LLM-based Classification}
\label{section:LLM-appendix}
\begin{figure}[H]
\centering
\textbf{Industry's engagement in responsible AI research (a comparison with academia)}\par\medskip
\includegraphics[width=\textwidth]{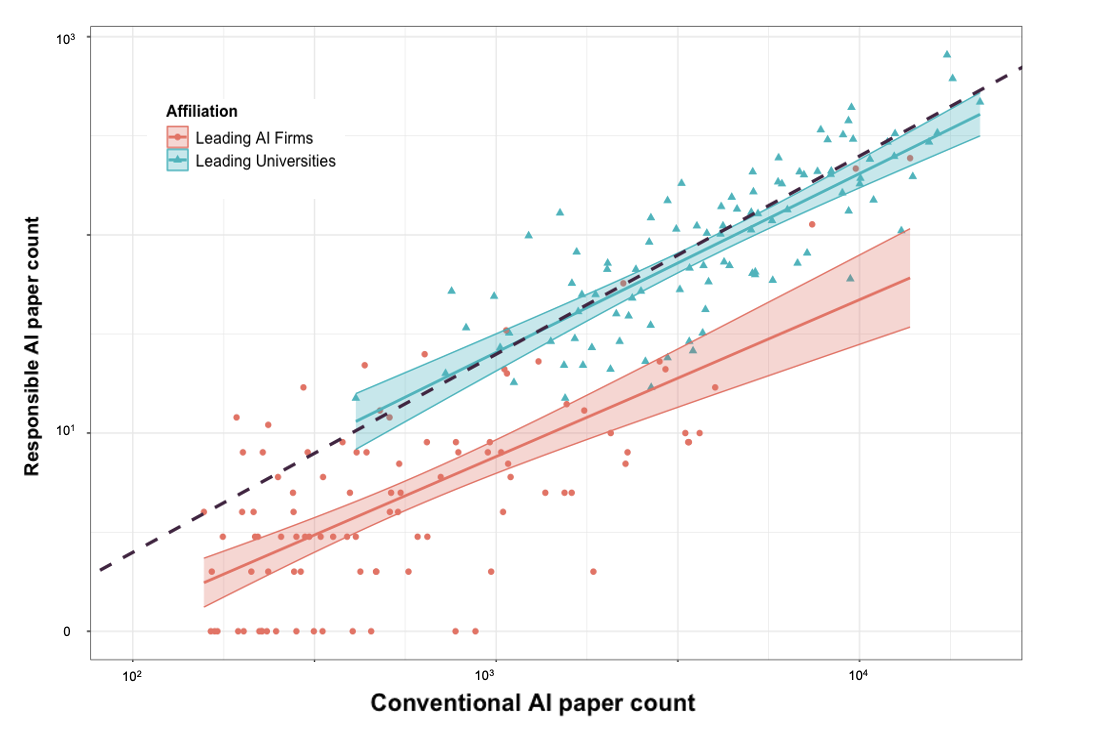}
\caption{{\fontsize{10}{10}\selectfont This figure shows the engagement of AI firms in responsible AI research using LLM-based classification with in-context learning. Fig. \ref{fig:LLM-appendix} presents the distribution of the leading 100 AI firms' (n = 506,017 papers) and universities' (n = 5,265,419 papers) participation in AI research compared to their responsible AI research. Here, individual organizations are symbolized by discrete dots. A trend line reflects the participation trend within each group, and the shaded bands correspond to 95\% confidence intervals. Additionally, the dashed line indicates a reference line where the proportion of responsible AI papers to total AI papers is 2.5\%.}}
\label{fig:LLM-appendix}
\end{figure}

\newpage
\begin{longtable}[H]{ p{3.5cm} p{11cm} }
\caption{Expert-suggested Keywords for Identifying Responsible AI Research Papers \hyperref[table:3_distinct_data]{Dataset 3}}
\label{table:oldA6}
\\ \hline
\textbf{Theme} & \textbf{Keywords/Phrases Covered} \\ \hline
\endfirsthead
\multicolumn{2}{c}%
{{\bfseries Table \thetable\ continued from previous page}} \\
\hline
\textbf{Theme} & \textbf{Keywords/Phrases Covered} \\ \hline
\endhead
\hline \multicolumn{2}{r}{{Continued on next page}} \\ \endfoot
\hline
\endlastfoot
\hline
Ethical \& Moral Implications & ethics, ethical, ethics (in) ai, ethical ai, virtue ethics, transparent, transparency, accountable, accountability, liable, liability, culpability, responsible, value alignment, equality, equity, equitable, moral, morale, morality \\ \hline
Legal \& Governance Issues & justice, ai and law, distributive justice, judicial, governance, audit, auditable ai, gdpr, political bias, de-bias, de-biasing, compliance \\ \hline
Societal \& Social Impacts & human centric, social good, human right(s), human dignity, empowerment, harmony, marginalization, harmony, discriminatory, egalitarian, social implication, societal implication, philosophical aspect, sociotechnical, counteract discrimination, counteracting discrimination, avoid discrimination, avoiding discrimination, mitigate discrimination, mitigating discrimination, address discrimination, addressing discrimination, promote diversity, promoting diversity, improve diversity, improving diversity, increase diversity, increasing diversity \\ \hline
Technical Considerations \& Challenges & interpretable ai, xai, explainable ai, transparent models, increase explainable models, increasing explainable models, enhance explainable models, enhancing explainable models, improve explainable models, improving explainable models, increase interpretable models, increasing interpretable models, enhance interpretable models, enhancing interpretable models, improve interpretable models, improving interpretable models \\ \hline
Environmental Concerns & carbon neutral, carbon neutrality, carbon emission, carbon footprint, carbon intensity, green ai, green computing, decarbonization, climate crisis, climate change, promote sustainable ai, promoting sustainable ai, achieve sustainable ai, achieving sustainable ai \\ \hline
Privacy & private data, preserve privacy, preserving privacy, privacy preserving, privacy enhancing, data protection, digital rights \\ \hline
Risk & risk assessment, harm prevention \\ \hline
Safe & safe ai, responsible ai, robust ai, safe artificial intelligence, responsible artificial intelligence, robust artificial intelligence, ai safety \\ \hline
Toxicity, Hate Speech \& Misinformation & toxicity, hate speech, harmful content, explicit content, toxic content, misinformation detection, disinformation detection, ai generated misinformation, ai generated disinformation \\ \hline
Bias & bias mitigation, mitigating bias, bias in machine learning, bias detection, bias in facial recognition, bias measurement, bias arise, bias arises, bias capturing, bias related, bias (in) automated (systems), bias towards, bias word(s), bias (in) word embedding, implicit bias, gender bias, algorithmic bias, mitigating bias, error (or) bias, source bias, model bias, unintended bias, human bias, reducing bias, reduction (in) bias, racial bias, statistical bias, based bias, potential bias \\ \hline
Fairness & fairness metrics, fairness constraint(s), fairness criteria, fairness criterion, fairness metric(s), fairness measure, fairness measurement(s), fairness perception(s), fairness (in/and) accuracy, fairness definition, fairness notion(s), fairness (and/in) justice, fairness concern(s), fairness context(s), fairness (in) algorithm(s), fairness method(s), fairness (in) ai (models/systems), fairness principle(s), fairness toolkit(s), procedural fairness, definition (of) fairness, improving fairness, measuring fairness, perception (of) fairness, learning fairness, robustness (and) fairness, privacy (and) fairness, fairness trade off(s) \\ \hline
Trust & (public) trust (in) ai, trustworthy/trustable ai, trustworthy/trustable judgement(s), trustworthy/trustable model(s), trustworthy/trustable machine learning (model/system), trustworthy/trustable system(s), user trust, public trust, promoting trust, foster trust, fostering trust \\ \hline
Accountability & accountable framework(s), accountability (of/in) framework(s), accountability (in/and) model(s), accountability process(es), auditable, human oversight, regulatory oversight, stakeholder oversight, public oversight \\ \hline
Accessibility \& Inclusivity & accessible ai, ai for social good, underserved communities, non-discriminatory, disability rights, equal opportunity, equal opportunities, inclusion, inclusive \\ \hline
Model Cards & model card(s) \\ \hline
Misc. & disparate impact, disparate treatment, protected attribute(s), parity, intellectual property, counterfactual explanation, algorithmic decision making, model performance disparities, black box model(s) \\

\end{longtable}

\newpage

\subsection{Keywords for Classifying AI Research Papers}
\label{list:AIkeys}
nlp, opennlp, swarm intelligence, opinion mining, node embeddings, vowpal, opennn, perceptron, monte carlo tree search, automl, rnn, machine translation, probabilistic graphical models, graph convolutional networks, neural network, multi head attention, aggregated model, music generation, active learning, multimodal learning, neural style transfer, regularization, cudnn, neural net, neural model, backpropagation, deep convolutional gan, hardware acceleration neural networks, variational autoencoder, robotics, transfer learning, causal inference, h2o software, lstm, dialogflow, synthetic data, gpt, advances neural information processing systems, deep reinforcement learning, deep network, onnx, deep supervised hashing, paddlepaddle, relu, self attention, transformer xl, catastrophic forgetting, natural language toolkit, data privacy, knowledge graphs commonsense, tokenization, text speech, personalized federated learning, deep linear network, spectral graph theory, siamese networks, time series decomposition, knn, domain adaptation, computational linguistics, sentiment analysis, pixelcnn, max pooling, latent dirichlet allocation, dropconnect, back propagation, gan, bayesian networks, deep metric learning, sentiment classification, local training, natural language understanding, mlpack, stylegan, predictive analytics, attention mechanism, hybrid systems, overfitting, neural arithmetic logic units, model pruning, generative adversarial network, caffe deep learning framework, graspnet, disentanglement, image inpainting, mxnet, pointnet, machine learning, knowledge graph, distribution detection, model deployment, lexical semantics, adversarial examples, multi agent systems, genetic algorithm, ensemble learning, mobilenet, robotic, generative adversarial net, auto regressive model, feature extraction, support vector machine, topic model, quantum algorithms, simulated annealing, arima, deep q learning, early stopping, recommender systems, video summarization, deep encoder decoder, distilbert, markov chain monte carlo, madlib, ensemble methods, data augmentation, affective computing, pre trained models, knowledge discovery data mining, one shot learning, ai, modular audio recognition framework, image captioning, lexalytics, bert, local interpretable model, hierarchical topic modeling, gradient descent, data mining, interpretability ai, concept drift, mode collapse, time series forecasting, nd4j software, chatbot, latent space, sequence sequence, tsne, activation functions, multi modal fusion, gans, asynchronous updates, differential privacy, inverse reinforcement learning, federated averaging, image processing, quantization, neural networks, encoder decoder, edge computing, language model, graph neural networks, image image translation, lda, microsoft cognitive toolkit, continual learning, dataset shift, federated learning, deep deterministic policy gradient, communication efficient learning, r cnn, deberta, self supervised learning, zero knowledge proof, shot learning, genomic data analysis, natural language processing, skip connections, neurosymbolic computing, roberta, decision trees, theano, super resolution, spiking neural network, lasagne, data imputation, text mining, libsvm, inception score, multi armed bandit, model explanation, energy based model, softmax, video analytics, supervised learning, tensor processing unit, synthetic data generation, albert, adversarial networks, dynamic programming, convolutional network, edge ai, semi supervised learning, generative pre trained transformer, feature importance, style transfer, attention mechanisms, bias detection, apertium, object tracking, opencl, optimization algorithms, imagenet, semantic driven subtractive clustering method, secure multi party computation, conditional random field, minilm, antlr, deep embedding, model auditing, quantum neural networks, video object segmentation, resnet, dcgan, radial basis function network, decision tree, frechet inception distance, model accountability, self play, evolutionary algorithms, capsule networks, neural architecture search, tflearn, pixelrnn, neural commonsense knowledge bases, particle swarm optimization, sonnet, ai chatbot, collaborative filtering, pytorch lightning, auto encoder, hill climbing, kernel trick, bidirectional encoder representations, u net, positional encoding, spacy, computer vision, deep generative network, conditional gan, knowledge based systems, nltk, hidden markov model, restricted boltzmann machine, recurrent network, adversarial training, bert variants, gated recurrent unit, dropout, bias ai, model transparency, deep probabilistic model, image recognition, drug discovery ai, pytorch, zero shot learning, deep generative model, differentiable neural computers dnc, cyclegan, data clustering, deep hashing method, long short term memory, anomaly detection, neural language model, ibm watson, federated transfer learning, mahout, information retrieval, proximal policy optimization, deep learning, shapley additive explanations, bagging, vision transformer, deep recurrent network, mlpy, sparse coding, dynet, transformer architectures, neural architecture optimization, tensorflow, natural language learning, audio scene analysis, spectral clustering, depth wise convolution, autoencoder, decentralized training, viterbi algorithm, protein folding ai, variational inference, gluon, word2vec, noise injection, symbolic ai, multi task learning, deeplearning4j, pybrain, deep belief network, deep architecture, k nearest neighbor, q learning, hopfield network, neural turing machine, latent variable, word embedding, automatic speech recognition, federated optimization, nearest neighbor algorithm, xgboost, experts system, feedforward neural networks, split learning, deep representation learning, residual neural network, actor critic, neural architecture, fine tuning, glove, meta learning, model interpretability, text generation, hyperparameter tuning, anomaly detection time series, boltzmann machine, shap, transformer, knowledge distillation, unsupervised learning, ernie, machine vision, reinforcement learning, object recognition, scikit learn, stacked boltzmann, policy gradients, learning representations, interactive learning, graphics processing unit, umap, device learning, naive bayes, keras, allennlp, deep convolutional, conversational agent, artificial intelligence, multilayer perceptron, deep model, hard negative mining, deep autoencoder, hierarchical temporal memory, seq2seq, homomorphic encryption, virtual agents, bidirectional encoder representations transformers, pattern recognition, opencv, random forest, genetic algorithms, alphago, memory networks, word movers distance, speech recognition, convolutional neural networks, residual connections, gradient boosting, sequence model, lexical acquisition, neural processing unit, deep q network, rectifier linear unit, liquid state machine, neural machine translation, cross validation, bayesian neural networks, latent semantic analysis

\newpage
\subsection{Industry’s limited engagement in responsible AI research at conventional AI conferences}
We examined industry engagement in responsible AI at conventional AI conferences. In recent years, conventional AI conferences have also started to focus on responsible AI research \citep*{ai-index-2024}. It is plausible that industry is doing more work in responsible AI but not presenting them at the leading responsible AI conferences. 

\begin{figure}[h]
    \centering
    \includegraphics[width=1\textwidth]{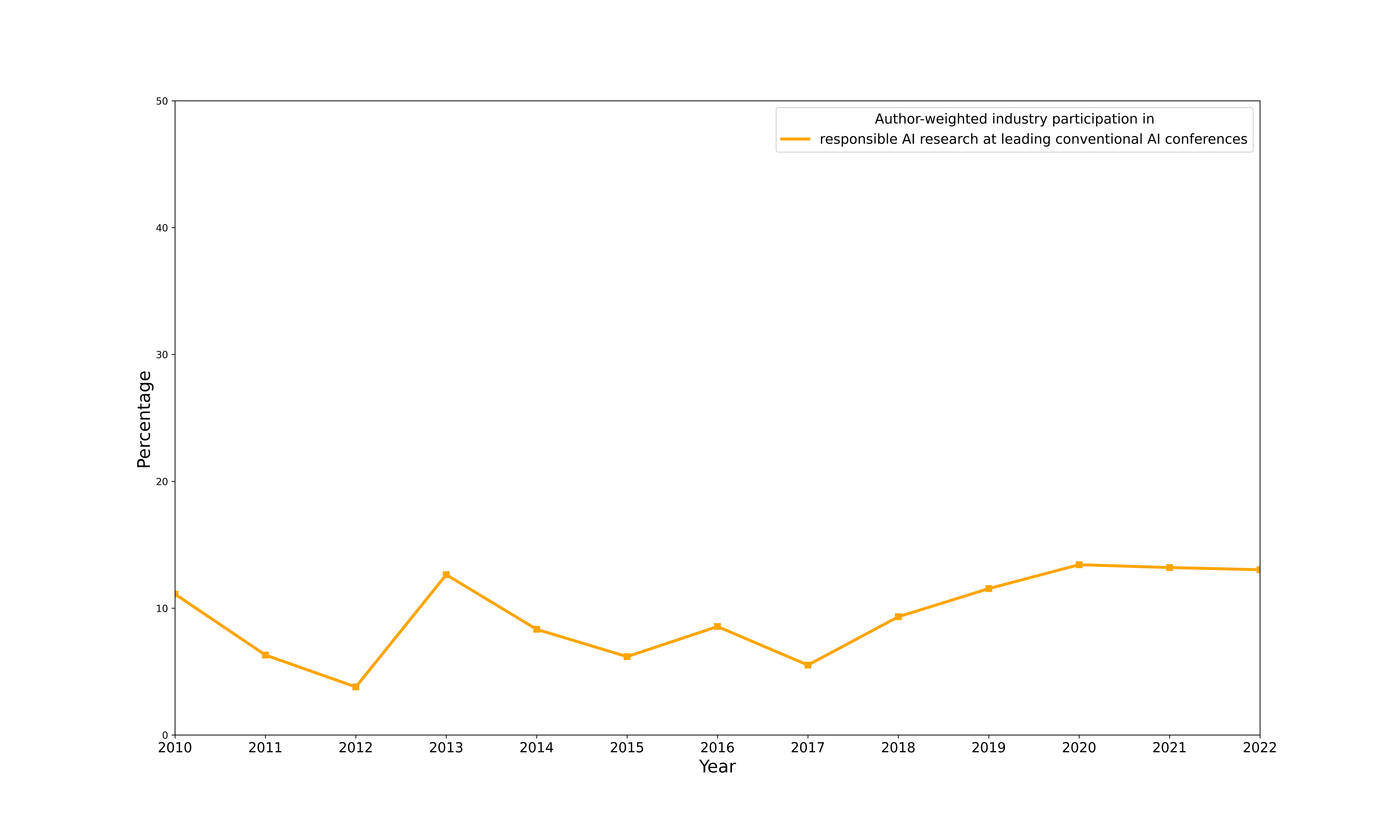}
    \caption{{\fontsize{10}{10}\selectfont 
    This figure illustrates the limited participation of industry researchers in responsible AI research at conventional AI conferences. The graph shows the author affiliation weighted percentage of industry-authored responsible AI papers out of all responsible AI papers presented at 10 leading conventional AI conferences from 2010 to 2022 (n = 1,311; a subset of papers from \hyperref[table:3_distinct_data]{dataset 2}).}}
    \label{fig:industry-rai-paper-cai-conf}
\end{figure}

To investigate this possibility, we classified all papers from the leading 10 conventional AI conferences using our supervised ML model to identify the subset of responsible AI papers. We then calculated the author-weighted percentage of these papers with at least one industry co-author by using \hyperref[eq:awa]{equation 3}. Figure \ref{fig:industry-rai-paper-cai-conf} presents the percentage of responsible AI papers with industry co-authorship at the leading conventional AI conferences. The figure suggests that industry's share of responsible AI research at conventional AI conferences has largely stayed the same in recent years. Additionally, two key issues remain: First, responsible AI research still makes up a small portion of the total research presented at conventional AI conferences. Second, unlike industry's rapid growth in conventional AI (see Figure \ref{fig:fig2}b), the growth in responsible AI research is more gradual and modest in recent years. The majority of these industry-authored responsible AI papers focus on technical issues such as privacy, data bias, and explainable AI. Overall, our conclusion of limited industry engagement in responsible AI research across all publication outlets still holds.

\newpage
\section{Additional Analyses on Linguistic Analysis}
\begin{figure}[H]
     \centering
     \textbf{Linguistic analysis shows industry has a narrower focus on responsible AI research, which broadens over time}\par
     \includegraphics[width=0.65\textwidth]{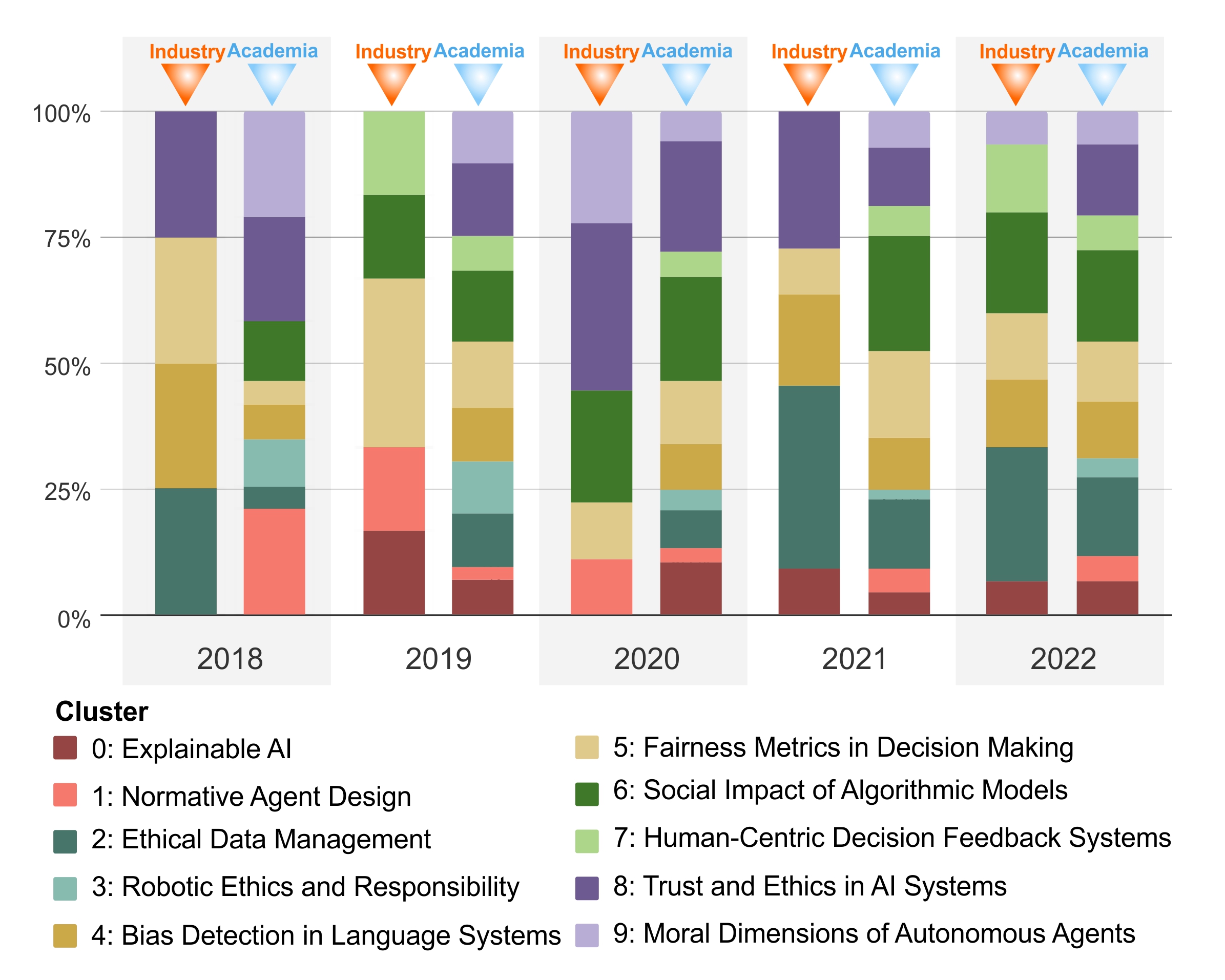}
     \caption{{\fontsize{10}{10}\selectfont This figure shows the clusters of topics in responsible AI (see Table \ref{table:oldA4}) that industry and academia engage in, respectively, as determined by k-means clustering on abstracts (n = 629) from responsible AI conferences (\hyperref[table:3_distinct_data]{dataset 2}). Here, each color represents a specific research cluster. As elaborated in the Methods section, this pattern is robust to changes in the number of clusters or alternative machine learning methods.}}
     \label{fig:ex_fig3}
\end{figure}

\begin{figure}[H]
    \centering
    \textbf{Linguistic analysis shows industry has a narrower focus in responsible AI research but broadens over time (LDA topic modeling, n = 10 topics)}
    \includegraphics[width=0.7\textwidth]{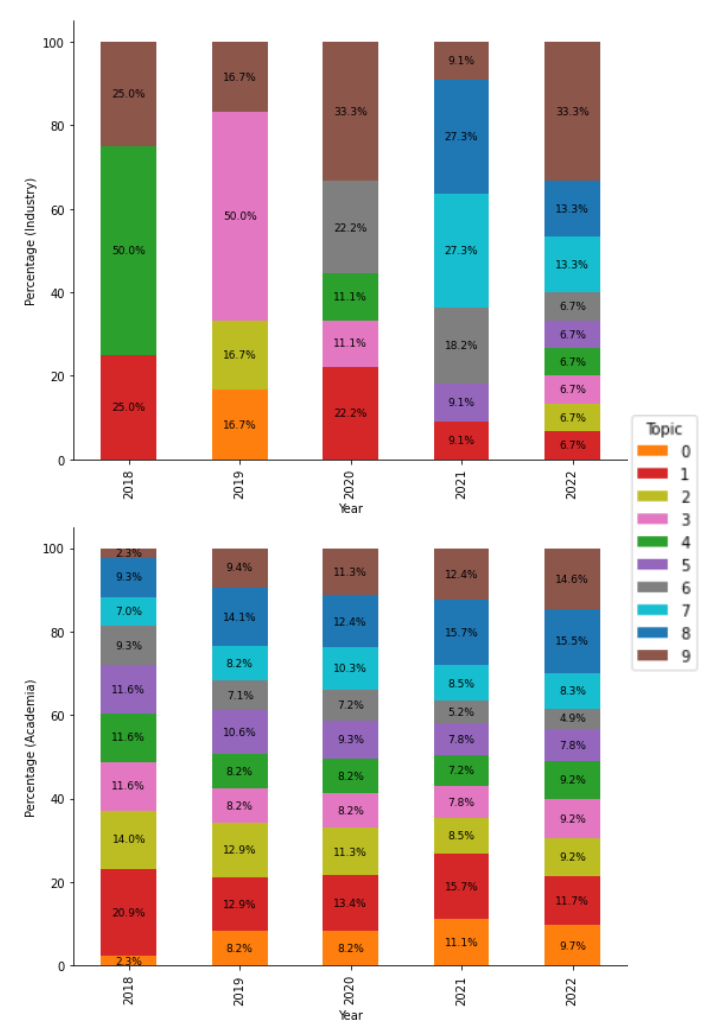}
    \caption{{\fontsize{10}{10}\selectfont This figure shows the clusters of responsible AI topics (see Table \ref{table:oldA8} for keywords) that only industry (top) and only academia (bottom) engage in, respectively, as determined by LDA topic modeling on abstracts (n = 629) from responsible AI conferences (\hyperref[table:3_distinct_data]{dataset 2}). Here, each color represents a specific research cluster. We determined the percentage by dividing the number of papers from each topic by the total number of papers produced by that group for that specific year.}}
    \label{fig:ex11}
\end{figure}

\textbf{Industry has a narrower focus on responsible AI research than academia:} We used publication data from the three leading responsible AI conferences and used the abstracts of those papers to identify key research themes in responsible AI research. Fig. \ref{fig:ex_fig3} shows the research topics that industry and academia explore in their responsible AI research. We used k-means clustering analysis on the abstracts of the responsible AI papers (see keywords list in Table \ref{table:oldA4} in the appendix). After clustering, we labeled each paper based on the cluster and then calculated the percentage of papers associated with that cluster for each year. Here, we present the analysis of 10 clusters on papers from 2018 to 2022 (the results are similar for other clusters, see Fig. \ref{fig:ex9} for 8 clusters \& \ref{fig:ex10} for 6 clusters). Our analysis shows that relative to academia, industry engages in fewer topics (only 4 topics in 2018 compared to academia’s 8 topics). The tendency for industry to have a more limited research scope than academia, as shown across other years, aligns with prior research \citep*{32}. It suggests a consistent pattern of industry’s narrower research focus over time. Moreover, our findings remain consistent even when we employ alternative machine learning methodologies like LDA topic modeling (see Fig. \ref{fig:ex11}) and bigram frequency count (see Fig. \ref{fig:ex12}), or theme-specific analysis (see Fig. \ref{fig:fig3}b).

Furthermore, we found that industry typically began exploring research topics only after academia introduced them. For instance, we observed that industry’s four topics in 2018 increased to seven by the end of 2022 (see Fig. \ref{fig:ex_fig3}). Conversely, academia started with a broader spectrum of topics (8 topics), which eventually increased to 10. Interestingly, we observed that industry research has broadened more over time, underscoring the importance of industry’s consistent engagement in academic venues \citep*{28}. This analysis provides suggestive evidence that research engagement can benefit industry by broadening its understanding of responsible AI. Besides, the expansion of industry’s research portfolio was, in part, likely influenced by the relevance and utility of those additional topics to them. Overall, AI firms have a narrow scope of responsible AI research and appear to trail behind academia in exploring a range of topics within this field.

\newpage

\begin{table}[H]
\centering
\caption{Most associated words for each cluster related to k-means clustering (Fig. \ref{fig:fig3}a)}
\begin{tabular}{l p{10cm}}
\hline
\textbf{Topic name} & \textbf{Top Keywords} \\
\hline
Ethical \& Moral Concerns & moral, foundations, judgments, morality, ethics, ethical, agents, human, theory, social, norms, model, people, reasoning, decision \\
\hline
Bias
 & research, data, bias, policy, equity, systems, racial, use, based, paper, decision, study, social, results, using\\
\hline
Algorithmic Fairness & fairness, fair, learning, algorithms, data, machine, algorithmic, accuracy, algorithm, model, bias, groups, protected, models, group
 \\
\hline
Privacy & privacy, data, differential, private, information, learning, users, model, federated, concerns, differentially, sensitive, protection, user, models \\
\hline
Equitable AI & poverty, inequality, income, economic, poor, racial, wealth, social, households, countries, policy, data, growth, research, pape \\
\hline
Human-AI Interaction & ai, human, intelligence, artificial, systems, humans, research, decision, design, xai, learning, data, use, technologies, making\\
\hline
Explainable AI & models, model, learning, explanations, machine, data, ml, human, explanation, decision, systems, predictions, methods, based, performance\\
\hline
Data Governance & data, research, big, use, science, privacy, collection, information, researchers, sharing, datasets, open, social, access, analysis \\
\hline
Human-cenric AI & social, design, technology, media, research, digital, technologies, hci, ethical, community, users, systems, use, work, platforms \\
\hline
Human-Robot Interaction & robot, robots, trust, human, social, interaction, humans, robotic, participants, hri, team, design, children, study, people\\
\hline
\label{table:3AClusters}

\end{tabular}
\end{table}

\begin{table}[H]
\centering
\caption{Most associated words for each topic on structural topic modeling analysis (Fig. \ref{fig:fig3}b)}
\begin{tabular}{l p{10cm}}
\hline
\textbf{Topic name} & \textbf{Top Keywords} \\
\hline
Human-AI interaction & human, trust, social, people, participants, robot, robots, study, humans, interaction, children, results, behavior, interactions, perceived \\
\hline
Model performance & model, models, learning, data, machine, accuracy, methods, based, explanations, privacy, using, show, training, approach, fairness \\
\hline
Ethical \& moral concerns & moral, theory, ethical, paper, values, argue, social, legal, research, one, ethics, value, framework, making, norms \\
\hline
Societal implication & inequality, poverty, economic, find, population, income, policy, using, market, results, measures, financial, also, level, effects \\
\hline
AI model development & ai, design, systems, technology, intelligence, artificial, technologies, research, learning, work, health, future, machine, care, paper \\
\hline
Algorithmic fairness & fairness, users, systems, system, bias, algorithms, user, algorithmic, online, biases, content, platforms, fair, algorithm, news \\
\hline
Accountability & racial, accountability, black, police, race, gender, use, disparities, public, study, bias, findings, discrimination, crime, white \\
\hline
Decision making & decision, group, performance, team, making, groups, decisions, teams, agents, results \\
\hline
Equitable AI & community, equity, communities, policy, social, digital, urban, public, access, local, justice, housing, cities, paper, research \\
\hline
Privacy & data, research, privacy, social, information, use, researchers, media, analysis, science \\
\hline
\label{table:STMTopics}

\end{tabular}
\end{table}

\begin{table}[H]
\caption{Keywords for constructing Fig. \ref{fig:fig3}b}
\label{tab:topic_compare}
\centering
\begin{tabular}{l p{12cm}}
\hline
\textbf{Category}         & \textbf{Keywords}  \\
\hline
Human Rights              & \begin{tabular}[c]{@{}l@{}}  marginal*, under represent*, human right*, raci* inequal*, \\raci* discrimia*, raci* dispar*, gender discrimia*, human dignity,\\ justice , social value*,global value*, human value*, \\humanity, fundamental right* \end{tabular}\\ \hline

Environment Concerns      &\begin{tabular}[c]{@{}l@{}} environ*, carbon footprint, climate, greenhouse, decarbon*,\\ emission, sustainab*, green ai                                \end{tabular}\\ \hline
Beneficence                   & \begin{tabular}[c]{@{}l@{}} benefi*, well being, peace, social good*, common good*, \\empower*,
                     inclus*, welfare, social value*, people* value*,\\ harness*, public health \end{tabular}     \\ \hline
Non-maleficence         & \begin{tabular}[c]{@{}l@{}} non malef*, security, safety, harm, protect*, \\
                         precaut*, prevent*, integrity , non subver*, oversight \end{tabular} \\ \hline

Explainable AI            & \begin{tabular}[c]{@{}l@{}}explain*, interpret*, xai, transparen*, feature attribution*, \\ feature importance*, reproduci*, human understand*\end{tabular}                                                        \\ \hline

\end{tabular}
\end{table}

\newpage

\newpage

\begin{figure}[H]
    \centering
    \textbf{Linguistic analysis shows industry has a similar focus in conventional AI research (k-means clustering, n = 10 topics)}
    \includegraphics[width=0.7\textwidth]{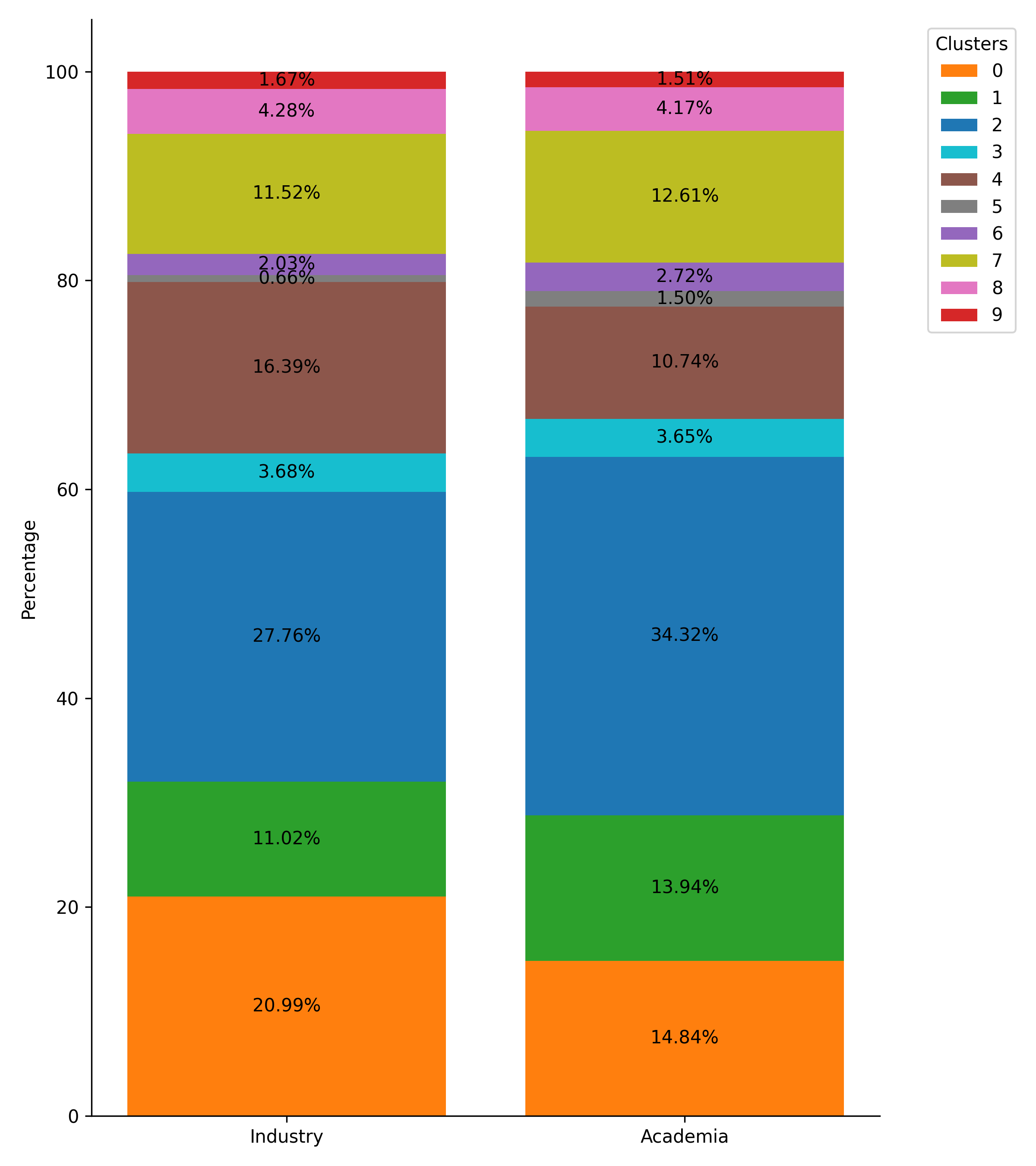}
    \caption{{\fontsize{10}{10}\selectfont This figure shows the clusters of conventional AI topics (see Table \ref{table:CAI-Cluster-keys} for keywords) that only industry and only academia engage in, respectively, as determined by k-means clustering analysis on paper abstracts (n = 79576) from top 10 conventional AI conferences (\hyperref[table:3_distinct_data]{dataset 2}). Here, each color represents a specific research cluster. We determined the percentage by dividing the number of papers from each topic by the total number of papers produced by that group.}}
    \label{fig:CAI_topic_cluster}
\end{figure}

\newpage
\textbf{Linguistic analysis shows industry has a narrower focus in responsible AI research but broadens over time (k-means clustering, 8 clusters)}
\begin{figure}[H]
    \centering
    \includegraphics[width=\textwidth]{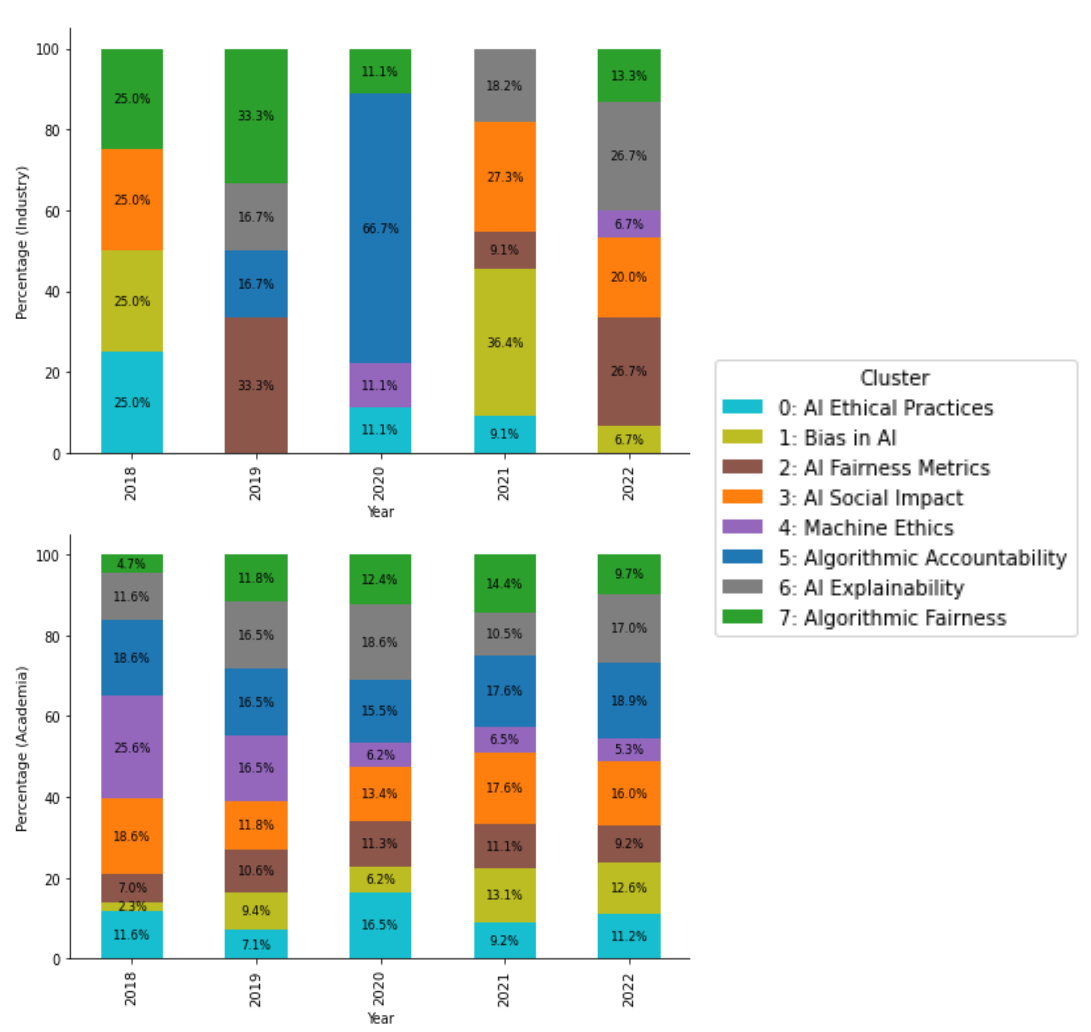}
    \caption{{\fontsize{10}{10}\selectfont This figure shows the clusters of responsible AI topics (see Table \ref{table:oldA5} for keywords) that industry (top) and academia (bottom) engage in, respectively, as determined by k-means clustering on abstracts of papers (n = 629) from three responsible AI conferences. Here, each color represents a specific research cluster. We determine the percentage by dividing the number of papers from each cluster by the total number of papers produced by that group for that specific year. }}
    \label{fig:ex9}
\end{figure} 

\newpage
\textbf{Linguistic analysis shows industry has a narrower focus in responsible AI research but broadens over time (k-means clustering, 6 clusters)}
\begin{figure}[H]
    \centering    \includegraphics[width=\textwidth]{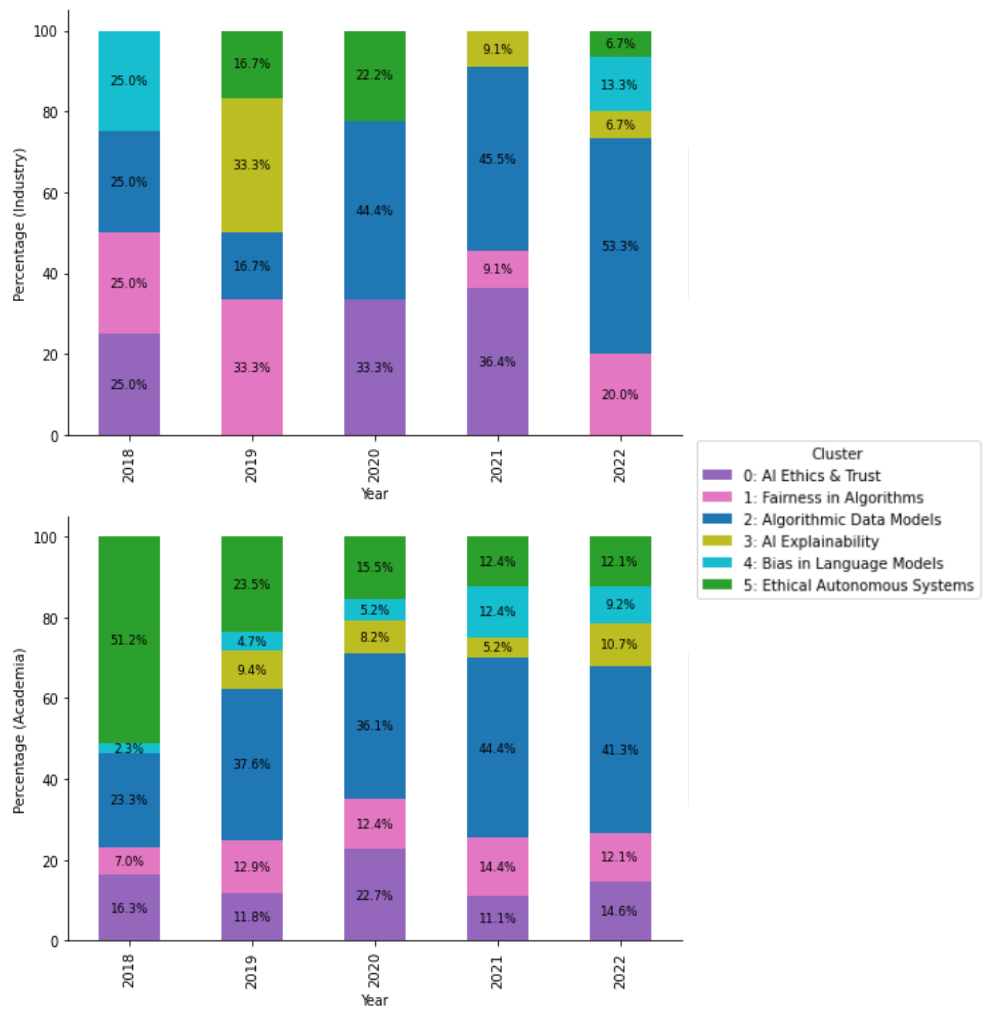}
    \caption{{\fontsize{10}{10}\selectfont This figure shows the clusters of responsible AI topics (see Table \ref{table:oldA5} for keywords) that industry (top) and academia (bottom) engage in, respectively, as determined by k-means clustering on abstracts of papers (n = 629) from three responsible AI conferences. Here, each color represents a specific research cluster. We determined the percentage by dividing the number of papers from each cluster by the total number of papers produced by that group for that specific year. }}
    \label{fig:ex10}
\end{figure} 

\newpage
 \textbf{Linguistic analysis shows industry has a narrower focus in responsible AI research (bigram frequency analysis)}
\begin{figure}[H]
    \centering
    \begin{subfigure}{\textwidth}
        \includegraphics[width=0.7\textwidth]{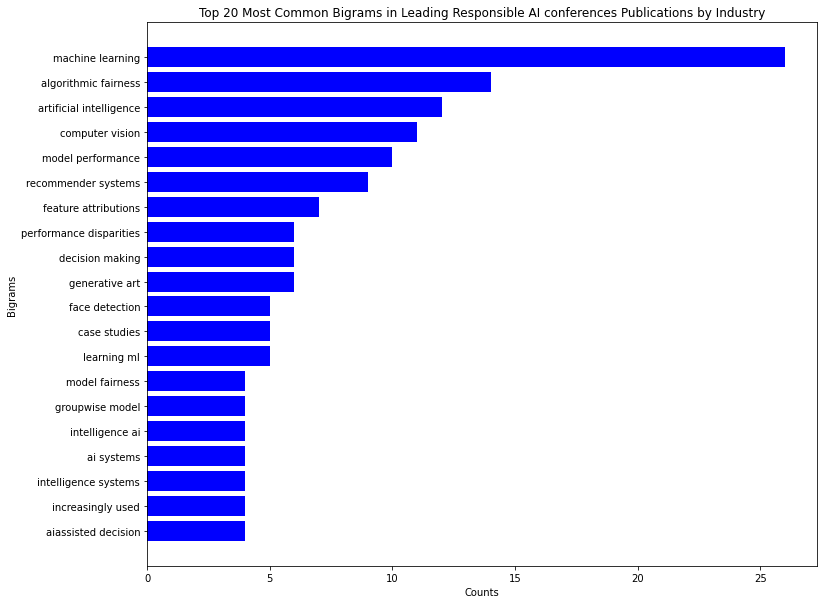}
        \label{fig:sub1}
    \end{subfigure}
    
    \begin{subfigure}{\textwidth}
        \includegraphics[width=0.7\textwidth]{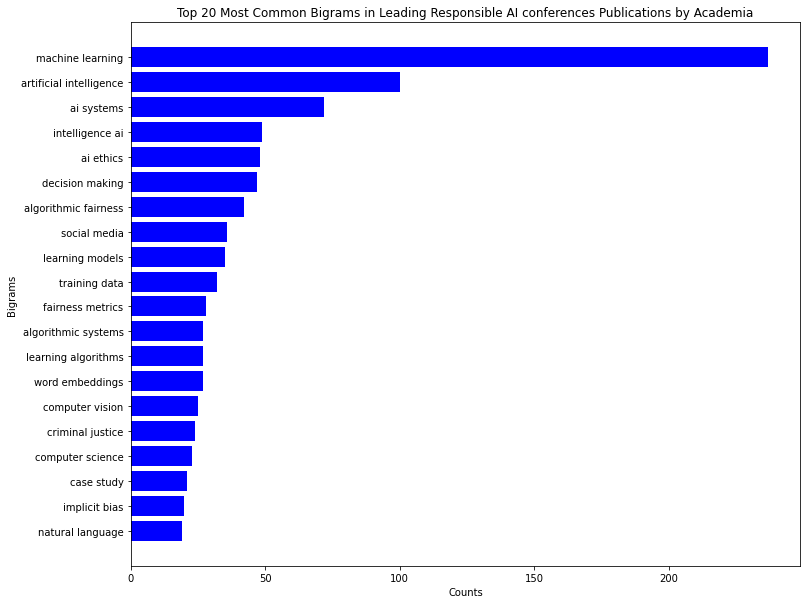}
        \label{fig:sub2}
    \end{subfigure}
    \caption{{\fontsize{10}{10}\selectfont This figure shows the bigram frequency analysis for industry (top) and academia (bottom) based on abstracts of papers from the three leading responsible AI conferences. This analysis suggests that industry research is limited compared to academia. In particular, industry has limited engagement in topics like criminal justice, implicit bias, and fairness metrics.}}
    \label{fig:ex12}
\end{figure}

\begin{table}[H]
\caption{Topic modeling keywords list from Fig. \ref{fig:ex11}}
\begin{tabular}{c p{0.8\textwidth}}
\hline
\multicolumn{1}{l}{Number of topics} & Keywords List \\ \hline
 \multirow{10}{*}{10} & Topic- 0: fairness, individuals, ai, ethics, show, different, social, ethical, machine, research \\ 
 & Topic- 1: data, social, fairness, learning, systems, machine, bias, models, algorithmic, paper \\  
 & Topic- 2: systems, ethical, moral, ai, autonomous, design, ethics, paper, public, agents \\  
 & Topic- 3: learning, ai, systems, machine, fairness, bias, gender, accountability, system, algorithmic \\  
 & Topic- 4: ai, data, model, fairness, systems, learning, models, risk, disparities, decisions \\  
 & Topic- 5: model, systems, models, explanations, ai, algorithmic, transparency, data, information, work \\  
 & Topic- 6: learning, machine, human, ai, fairness, models, moral, social, may, data \\  
 & Topic- 7: fairness, data, ai, algorithmic, work, algorithms, show, impact, fair, systems \\  
 & Topic- 8: ai, data, trust, fairness, models, systems, human, learning, paper, work \\  
 & Topic- 9: fairness, ai, data, model, systems, research, paper, work, fair, analysis \\ \hline 
\label{table:oldA8}
\end{tabular}
\end{table}
\vspace{-4mm}

\begin{table}[H]
\caption{Word clustering keywords of Conventional AI papers list from Fig. \ref{fig:CAI_topic_cluster}}
\begin{tabular}{c p{0.8\textwidth}}
\hline
\multicolumn{1}{l}{Cluster size} & Keywords List \\ \hline
 \multirow{10}{*}{10} & Cluster- 0: learning, data, training, model, network, neural, deep, networks, models, classification, methods, performance, method, tasks, label \\ 
 & Cluster- 1: algorithm, problem, algorithms, optimization, learning, problems, optimal, function, data, method, linear, gradient, time, methods, matrix \\  
 & Cluster- 2: data, model, based, paper, information, social, using, time, user, results, approach, method, systems, users, used \\  
 & Cluster- 3: video, action, temporal, videos, motion, frames, frame, recognition, human, actions, model, method, features, based, dataset \\  
 & Cluster- 4: language, word, translation, models, model, text, task, languages, sentence, words, english, semantic, based, tasks, embeddings \\  
 & Cluster- 5: clustering, data, clusters, cluster, algorithm, means, method, subspace, algorithms, view, based, proposed, spectral, methods, matrix \\  
 & Cluster- 6: graph, graphs, node, nodes, network, networks, learning, knowledge, based, model, information, structure, methods, embedding, gnns \\  
 & Cluster- 7: image, images, method, face, depth, based, network, proposed, segmentation, features, pose, methods, model, feature, resolution \\  
 & Cluster- 8: object, objects, detection, tracking, segmentation, image, method, visual, based, scene, model, approach, images, dataset, pose \\  
 & Cluster- 9: domain, adaptation, target, source, domains, data, transfer, learning, cross, model, training, method, unsupervised, labeled, knowledge \\ \hline 
\label{table:CAI-Cluster-keys}
\end{tabular}
\end{table}

\newcolumntype{L}[1]{>{\raggedright\arraybackslash}p{#1}}
\newcolumntype{C}[1]{>{\centering\arraybackslash}p{#1}}

\begin{table}[H]
\caption{Top 10 keywords per cluster in K-means clustering}
\footnotesize 
\begin{tabular}{ c p{3cm} p{9cm} }
\hline
\textbf{Cluster size} & \textbf{Cluster name} & \textbf{Top keywords} \\ \hline
\multirow{10}{*}{10} & Explainable AI & Cluster 0: explanations, explanation, model, counterfactual, models, recourse, feature, black, box, explainability \\ \cline{2-3} 
& Normative Agent Design & Cluster 1: systems, design, agent, learning, reward, norm, value, agents, human, norms \\ \cline{2-3} 
& Ethical Data Management & Cluster 2: data, privacy, learning, dataset, access, ml, ethical, public, machine, training \\ \cline{2-3} 
& Robotic Ethics and Responsibility & Cluster 3: robots, human, responsibility, robot, ethical, norms, rights, moral, autonomous, machine \\ \cline{2-3} 
& Bias Detection in Language Systems & Cluster 4: bias, biases, gender, language, word, content, detection, systems, embeddings, social \\ \cline{2-3} 
& Fairness Metrics in Decision Making & Cluster 5: fairness, fair, group, decision, groups, metrics, learning, parity, individual, model \\ \cline{2-3} 
& Social Impact of Algorithmic Models & Cluster 6: algorithmic, social, model, fairness, systems, work, algorithms, models, learning, use \\ \cline{2-3} 
& Human-Centric Decision Feedback Systems & Cluster 7: decision, making, feedback, assessments, fairness, algorithmic, human, model, decisions, ai \\ \cline{2-3} 
& Trust and Ethics in AI Systems & Cluster 8: ai, ethics, intelligence, ethical, artificial, systems, trust, human, research, technologies \\ \cline{2-3} 
& Moral Dimensions of Autonomous Agents & Cluster 9: moral, ethical, autonomous, agents, machine, ethics, systems, vehicles, human, learning \\ \hline
\end{tabular}
\label{table:oldA4}
\end{table}

\begin{table}[H]
\caption{Top 10 keywords per cluster in K-means clustering}
\begin{tabular}{ c p{3cm} p{9cm} }
\hline 
\textbf{Cluster size} & \textbf{Cluster name} & \textbf{Top keywords} \\ \hline
\multirow{8}{*}{8} & AI Ethical Practices & Cluster 0: ai, ethics, intelligence, ethical, artificial, systems, research, human, trust, technologies \\ \cline{2-3} 
& Bias in AI & Cluster 1: bias, biases, racial, gender, language, word, embeddings, social, datasets, images \\ \cline{2-3} 
& AI Fairness Metrics & Cluster 2: fairness, group, decision, model, groups, utility, parity, outcomes, based, individuals \\ \cline{2-3} 
& AI Social Impact & Cluster 3: data, ai, systems, social, design, public, diversity, research, communities, challenges \\ \cline{2-3} 
& Machine Ethics & Cluster 4: moral, machine, ethical, learning, human, robots, autonomous, robot, trust, norms \\ \cline{2-3} 
& Algorithmic Accountability & Cluster 5: algorithmic, decision, systems, making, accountability, ethical, decisions, information, risk, ai \\ \cline{2-3} 
& AI Explainability & Cluster 6: explanations, models, data, model, learning, explanation, machine, xai, human, transparency \\ \cline{2-3} 
& Algorithmic Fairness & Cluster 7: fairness, fair, data, learning, algorithms, attributes, sensitive, model, algorithmic, notions \\ \hline

\multirow{6}{*}{6} & AI Ethics \& Trust & Cluster 0: ai, ethics, intelligence, systems, ethical, artificial, human, research, trust, public \\ \cline{2-3} 
& Fairness in Algorithms & Cluster 1: fairness, fair, group, groups, algorithms, parity, metrics, notions, learning, decision \\ \cline{2-3} 
& Algorithmic Data Models & Cluster 2: data, algorithmic, model, fairness, learning, decision, systems, models, work, social \\ \cline{2-3} 
& AI Explainability & Cluster 3: explanations, explanation, model, xai, models, trust, explainability, user, explainable, decision \\ \cline{2-3} 
& Bias in Language Models & Cluster 4: bias, biases, gender, language, word, embeddings, social, nlp, models, english \\ \cline{2-3} 
& Ethical Autonomous Systems & Cluster 5: ethical, moral, human, autonomous, systems, agents, ethics, agent, machine, accountability \\ \hline
\end{tabular}
\label{table:oldA5}
\end{table}

\newpage
\section{Additional Analyses on Citation Analysis}
 
\begin{figure}[h]
    \centering
    \textbf{Process for constructing Fig. \ref{fig:patent-analysis}a (examining industry’s reliance on responsible AI research by industry)} \\
    \includegraphics[width=.95\textwidth]{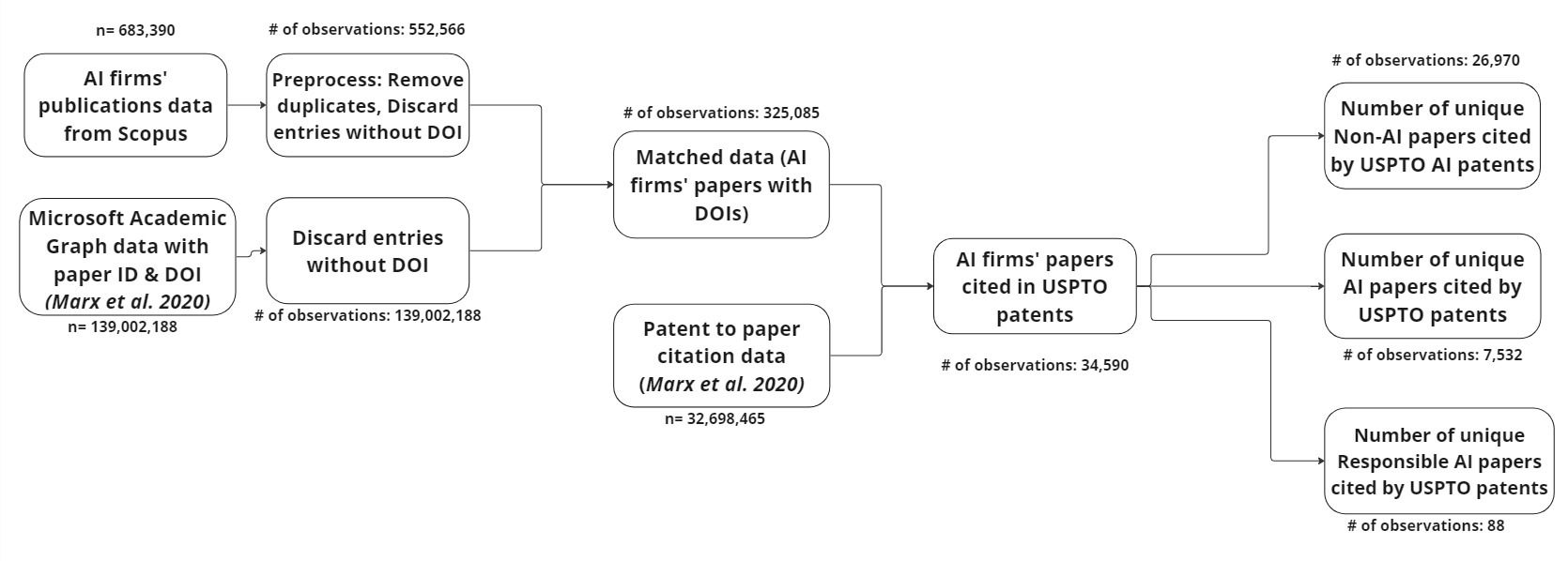}
    \caption{{\fontsize{10}{10}\selectfont This figure documents the process of how we combined multiple datasets to examine USPTO patents’ reliance on responsible AI research. }}
    \label{fig:flowchart1}
\end{figure}

\textbf{Process for constructing Fig. \ref{fig:patent-analysis}b (examining industry’s reliance on responsible AI research by industry)} \\ 

\begin{figure}[h]
    \centering
    \includegraphics[width=.95\textwidth]{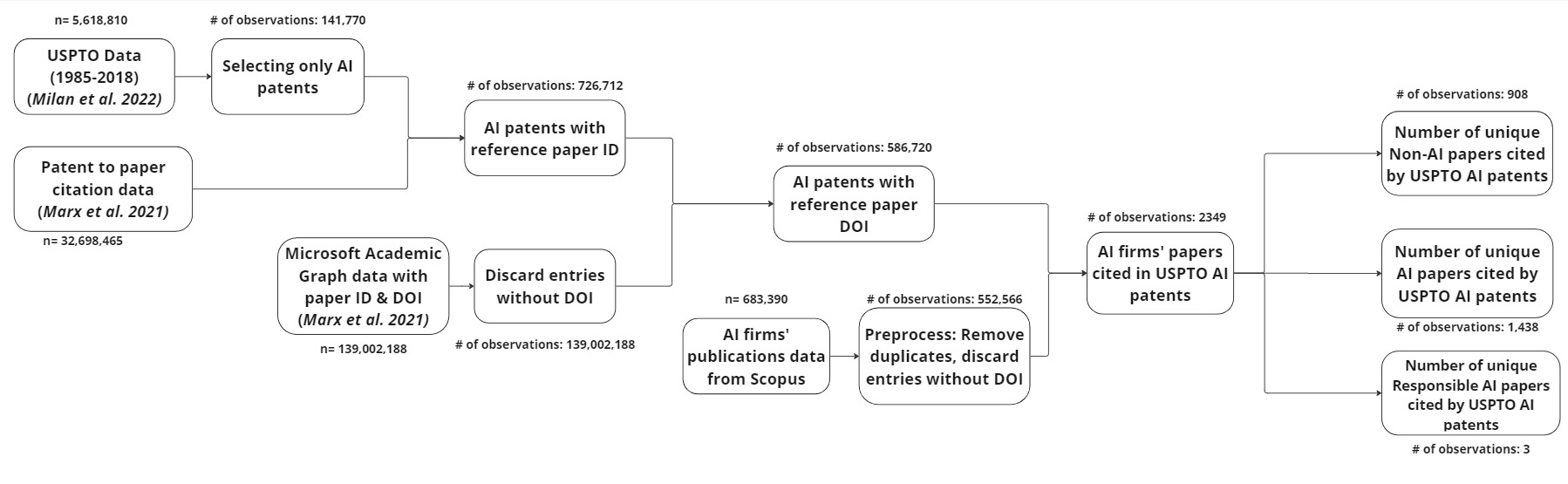}
    \caption{{\fontsize{10}{10}\selectfont This figure documents the process of how we combined multiple datasets to examine AI patents’ reliance on responsible AI research.  }}
    \label{fig:flowchart2}
\end{figure}

\newpage
\textbf{Process for constructing Fig. \ref{fig:patent-analysis}c (examining industry’s reliance on responsible AI research by academia)}  
\begin{figure}[h]
    \centering
    \includegraphics[width=.95\textwidth]{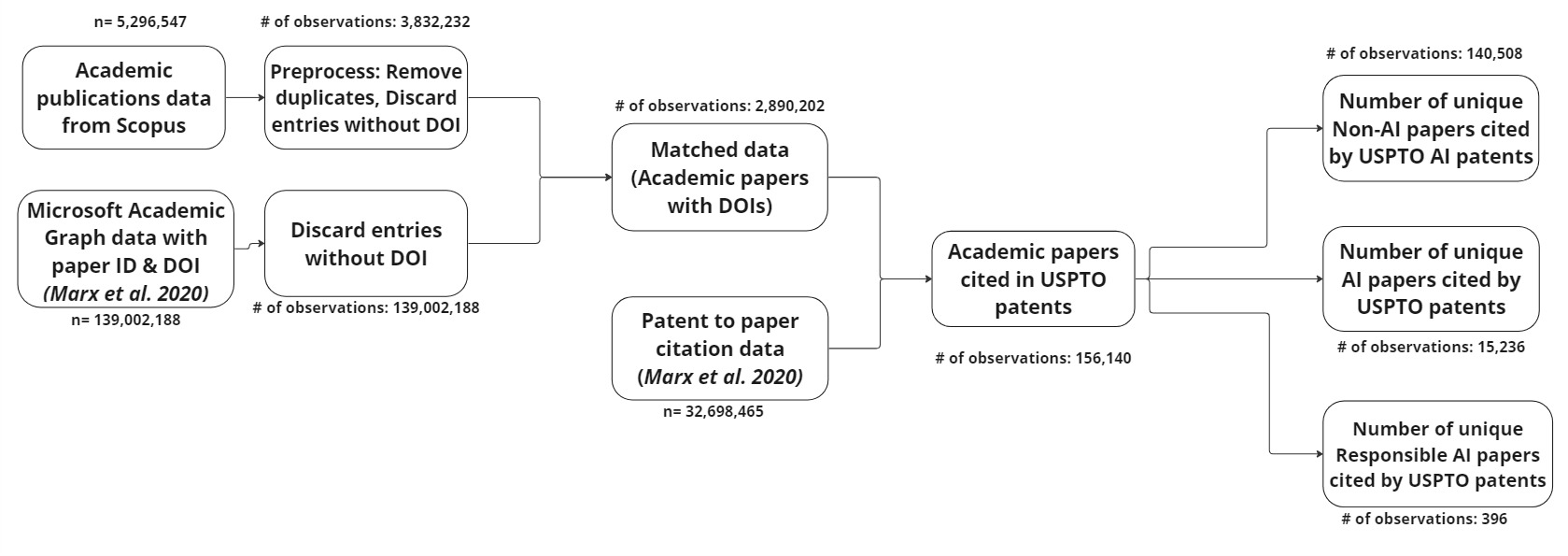}
    \caption{{\fontsize{10}{10}\selectfont This figure documents the process of how we combined multiple datasets to examine USPTO patents’ reliance on academic responsible AI research. }}
    \label{fig:flowchart3}
\end{figure}

\begin{figure}[h]
    \centering
    \textbf{Process for constructing Fig. \ref{fig:patent-analysis}d (examining industry’s reliance on academic responsible AI research)}
    \includegraphics[width=.95\textwidth]{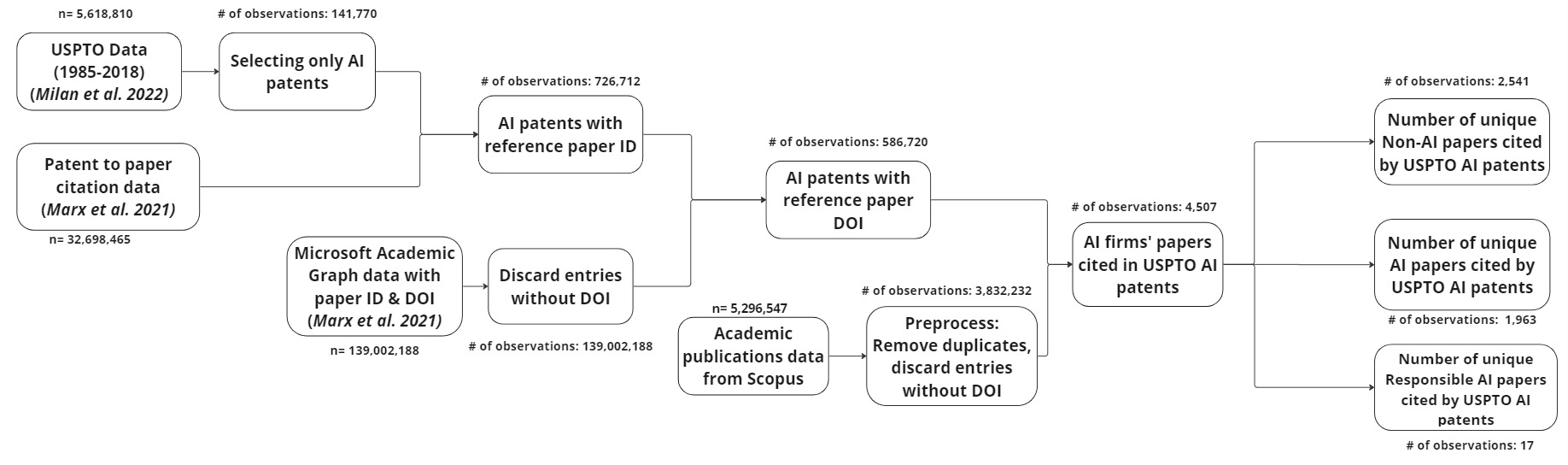}
    \caption{{\fontsize{10}{10}\selectfont This figure documents the process of how we combined multiple datasets to examine AI patents’ reliance on academic responsible AI research.  }}
    \label{fig:flowchart4}
\end{figure}

\newpage
\section{Industry's Academic Research builds on Academia’s Responsible AI Research}

We analyzed citation data from the three leading responsible AI conferences. First, we took all of the citations from industry co-authored paper titles and matched those with academic paper titles using a cosine similarity, with a 90\% threshold. Results were similar when we used ``exact matching.'' We counted the unique number of industry papers that cited one or more academic papers. Then, we counted the total number of papers co-authored by industry for that specific year. This process allowed us to compute the percentage of industry papers that cite academic works. We followed the same procedure to examine responsible AI papers co-authored by academia that cite responsible AI papers co-authored by industry.

\begin{figure}[h]
    \centering
    \textbf{Citation analysis suggests that industry builds upon academic responsible AI research}\par\medskip
    \includegraphics[width=0.75\textwidth]{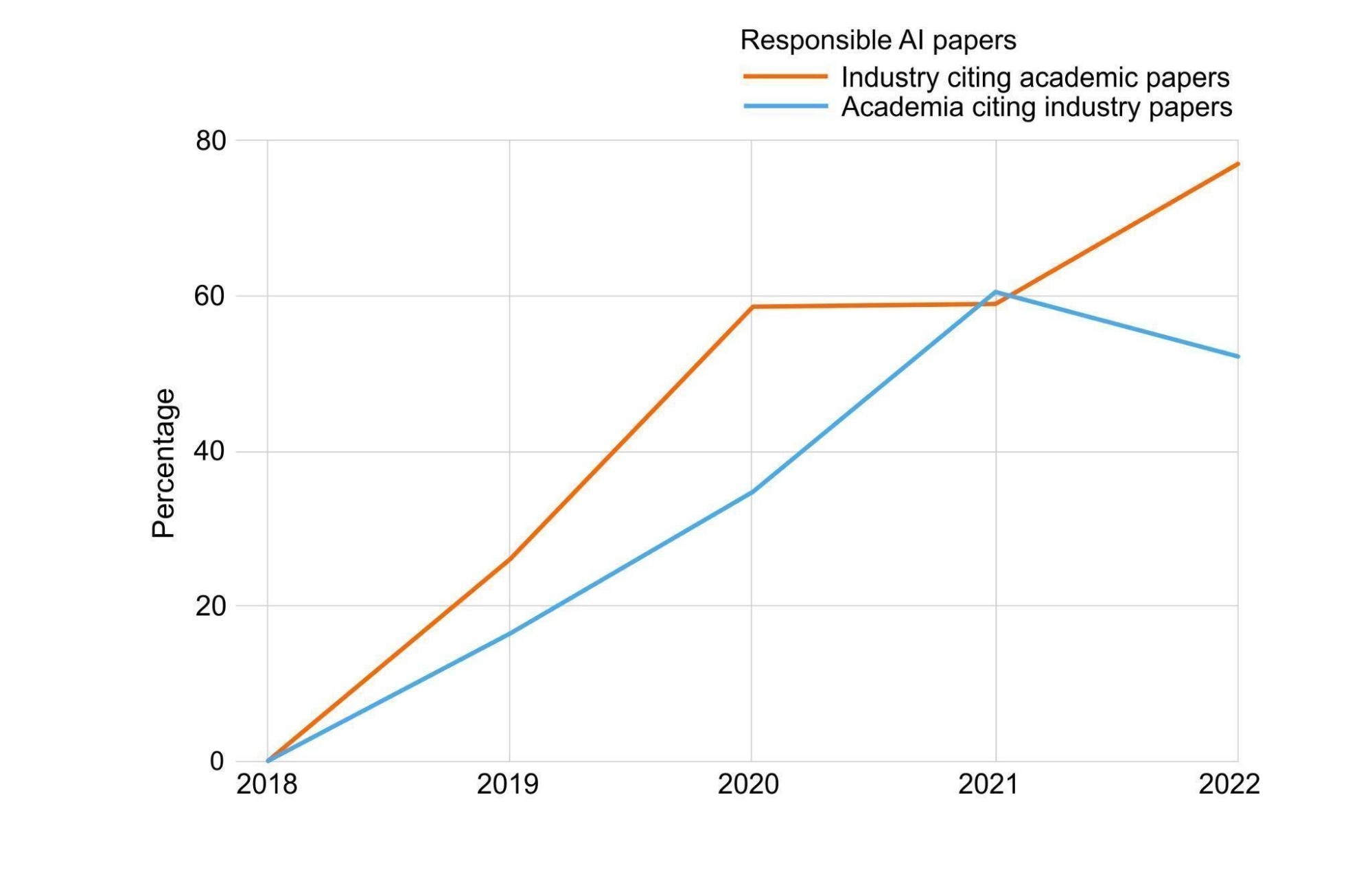}
    \caption{{\fontsize{10}{10}\selectfont This figure shows the percentage of industry co-authored responsible papers (n = 851) that cite academic papers, and academia co-authored papers that cite industry papers at the three leading responsible AI conferences (2018-2022; \hyperref[table:3_distinct_data]{dataset 2}). The yearly percentages are computed by dividing the number of industry-authored papers that cite one or more academic papers by the total number of industry-authored papers for that same year. A similar calculation is performed for academic-authored papers that cite industry papers.}}
    \label{fig:citation}
\end{figure}

\newpage
\begin{figure}[H]
    \centering
    \textbf{Patent citation analyses: cumulative count over the years} \par\medskip
    \includegraphics[width=0.8\textwidth]{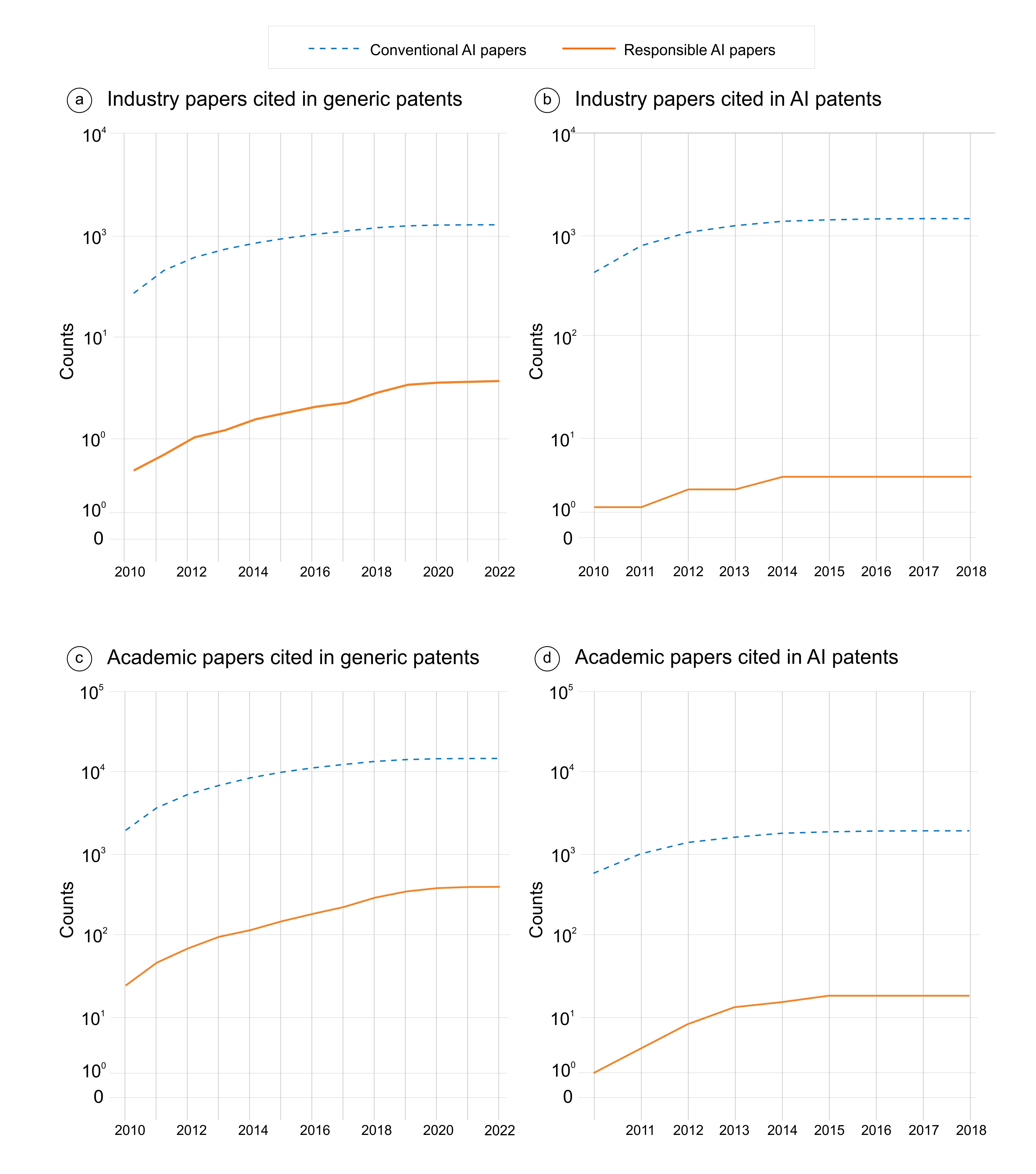}
    \caption{{\fontsize{10}{10} \selectfont This figure shows industry and academic papers’ cumulative citations in USPTO patents. By matching against a comprehensive patent-paper citation data (n = 32,698,465; 1947-2022), we show in Fig. \ref{fig:citation-cumulative}a and \ref{fig:citation-cumulative}c that, only 88 and 396 responsible AI papers from industry and academia, respectively, have been cited in generic patents, while 15,236 academia and 7,532 industry conventional AI and non-AI papers have been cited between 2010-22. Using a separate dataset of AI patents (n = 141,770; 1985-2018), Fig. \ref{fig:citation-cumulative}b and \ref{fig:citation-cumulative}d further documents, 3 industry and 17 academic responsible AI papers have been cited in any AI patents between 2010-22, in stark contrast to the incorporation of conventional AI and non-AI research. }}
    \label{fig:citation-cumulative}
\end{figure}

\newpage
\begin{longtable}[c]{ p{1cm}  p{6cm}  p{1cm}  p{6cm} }
\caption{List of leading 100 universities: \hyperref[table:3_distinct_data]{Dataset 1} (Based on csranking.org’s list, which considers the number of publications at leading conferences)} \label{table:oldA2} \\
\hline
Rank & Institution & Rank & Institution \\
\hline
\endfirsthead

\multicolumn{4}{c}%
{\tablename\ \thetable\ -- \textit{Continued from previous page}} \\
\hline
Rank & Institution & Rank & Institution \\
\hline
\endhead

\hline \multicolumn{4}{r}{\textit{Continued on next page}} \\
\endfoot

\hline
\endlastfoot

1 & Massachusetts Institute of Technology & 51 & University of Utah \\
2 & Carnegie Mellon University & 52 & Boston University \\
3 & Georgia Institute of Technology & 53 & George Mason University \\
4 & Stanford University & 54 & University of Massachusetts \\
5 & University of California, Berkeley & 55 & Yale University \\
6 & University of Illinois Urbana-Champaign & 56 & University of Connecticut \\
7 & University of California, Santa Barbara & 57 & University of Texas at Arlington \\
8 & University of Michigan & 58 & City University of New York \\
\end{longtable}

\newpage
\begin{longtable}[c]{p{1cm}  p{6cm}  p{1cm}  p{6cm}}
\caption{List of leading 100 AI firms (based on the number of AI papers)} \\ \hline
Rank & Company & Rank & Company \\ \hline
\endfirsthead

\multicolumn{4}{c}
{\tablename\ \thetable\ -- \textit{Continued from previous page}} \\
\hline
Rank & Company & Rank & Company \\
\hline
\endhead

\hline \multicolumn{4}{r}{\textit{Continued on next page}} \\
\endfoot

\hline
\endlastfoot
\hline
1 & Microsoft & 51 & NVIDIA \\
2 & IBM & 52 & STMicroelectronics \\
3 & Google & 53 & Honda R\&D Co \\
4 & Nokia Corporation & 54 & Thales Group \\
5 & Huawei Technologies & 55 & Ericsson Sweden \\
6 & Nippon Telegraph and Telephone Corporation & 56 & Bytedance \\
7 & Tencent & 57 & Airbus Defence and Space \\
8 & Siemens AG & 58 & BAE Systems plc \\
\end{longtable}

\newpage

\begin{longtable}[H]{p{0.4\textwidth}p{0.1\textwidth}p{0.2\textwidth}p{0.2\textwidth}}
\caption{Details of the training data for the supervised machine learning model to classify responsible AI papers}\\
\hline
\textbf{Data Source Name} & \textbf{Paper Count} & \textbf{Paper Type} & \textbf{Data Source Type} \\
\hline
\endfirsthead
\hline
\textbf{Data Source Name} & \textbf{Paper Count} & \textbf{Paper Type} & \textbf{Data Source Type} \\
\hline
\endhead
AAAI & 132 & Other papers\footnote{Consists of conventional AI papers \& Non-AI papers} & Conference \\
AAAI Conference on Human Computation and Crowdsourcing (HCOMP) & 1 & Other papers & Conference \\
AAAI Spring Symposium - Technical Report & 1 & Other papers & Conference \\
AAAI Workshop - Technical Report & 1 & Other papers & Conference \\
ACL & 241 & Other papers & Conference \\
ACM Conference on Computer Supported Cooperative Work and Social Computing (CSCW) & 1 & Other papers & Conference \\
ACM Conference on Human Factors in Computing Systems & 23 & Responsible AI & Conference \\
ACM Conference on Human Factors in Computing Systems & 8 & Other papers & Conference \\
\end{longtable}

\newpage
\begin{longtable}[H]
{p{0.4\textwidth}p{0.1\textwidth}p{0.2\textwidth}p{0.2\textwidth}}
\caption{Details of the training data for the supervised machine learning model to classify conventional AI papers}\\
\hline
\textbf{Data Source Name} & \textbf{Paper Count} & \textbf{Paper Type} & \textbf{Data Source Type} \\
\hline
\endfirsthead

\hline
\textbf{Data Source Name} & \textbf{Paper Count} & \textbf{Paper Type} & \textbf{Data Source Type} \\
\hline
\endhead
 AAAI                                                                                                                                            &           132 & AI           & Conference         \\
 ACL                                                                                                                                             &           241 & AI           & Conference         \\
 ACM Conference on Human Factors in Computing Systems                                                                                            &            23 & AI           & Conference         \\
 ACM International WSDM Conference                                                                                                               &             4 & AI           & Conference         \\
 AIES                                                                                                                                            &           362 & AI           & Conference         \\
 American Control Conference (ACC)                                                                                                               &             1 & Non-AI       & Conference         \\
 Annual IEEE/ACM International Symposium on Microarchitecture (MICRO)                                                                            &            22 & Non-AI       & Conference         \\
 Asia-Pacific Power and Energy Engineering Conference (APPEEC)                                                                                   &            10 & Non-AI       & Conference         \\
\end{longtable}

\newpage

\begin{longtable}{llll}
\caption{Deatils of the conference only training data to classify conventional AI papers}\\
\hline
\textbf{Data Source Name} & \textbf{Paper Count} & \textbf{Paper Type} & \textbf{Data Source Type} \\
\hline
\endfirsthead

AIES & 144 & AI & Conference \\
FaccT & 142 & AI & Conference \\
ACL & 65 & AI & Conference \\
NeurIPS & 57 & AI & Conference \\
ICML & 48 & AI & Conference \\
AAAI & 46 & AI & Conference \\
CVPR & 45 & AI & Conference \\
IJCAI & 37 & AI & Conference \\
ECCV & 29 & AI & Conference \\
EMNLP & 28 & AI & Conference \\
ICCV & 23 & AI & Conference \\
KDD & 22 & AI & Conference \\
EAAMO & 14 & AI & Conference \\
ICSE & 137 & Non-AI & Conference \\
DAC & 124 & Non-AI & Conference \\
USENIXSecurity & 83 & Non-AI & Conference \\
FOCS & 36 & Non-AI & Conference \\
STOC & 35 & Non-AI & Conference \\
MOBISYS & 32 & Non-AI & Conference \\
MICRO & 27 & Non-AI & Conference \\
NSDI & 13 & Non-AI & Conference \\
RTAS & 13 & Non-AI & Conference \\
\hline
\label{table: Confonly_data}

\end{longtable}

\newpage
\begin{longtable}{llll}
\caption{Deatils of the conference only responsible AI training data}\\
\hline
\textbf{Data Source Name} & \textbf{Paper Count} & \textbf{Paper Type} & \textbf{Data Source Type} \\
\hline
\endfirsthead

AIES & 144 & Responsible AI & Conference \\
FaccT & 142 & Responsible AI & Conference \\
ACL & 65 & Other papers & Conference \\
NeurIPS & 57 & Other papers & Conference \\
ICML & 48 & Other papers & Conference \\
AAAI & 46 & Other papers & Conference \\
CVPR & 45 & Other papers & Conference \\
IJCAI & 37 & Other papers & Conference \\
ECCV & 29 & Other papers & Conference \\
EMNLP & 28 & Other papers & Conference \\
ICCV & 23 & Other papers & Conference \\
KDD & 22 & Other papers & Conference \\
EAAMO & 14 & Responsible AI & Conference \\
ICSE & 137 & Other papers & Conference \\
DAC & 124 & Other papers & Conference \\
USENIXSecurity & 83 & Other papers & Conference \\
FOCS & 36 & Other papers & Conference \\
STOC & 35 & Other papers & Conference \\
MOBISYS & 32 & Other papers & Conference \\
MICRO & 27 & Other papers & Conference \\
NSDI & 13 & Other papers & Conference \\
RTAS & 13 & Other papers & Conference \\
\hline
\label{table: Facct_Confonly_data}

\end{longtable}

\end{document}